\documentclass[11pt, a4paper]{article}
\usepackage[utf8]{inputenc}
\usepackage{geometry}
\usepackage{authblk}
\usepackage{setspace}
\usepackage{mathptmx}
\usepackage{fancyhdr}
\usepackage{hyperref}
\usepackage{amsmath}
\usepackage{amssymb}
\usepackage{graphicx} 
\usepackage{float}
\usepackage{booktabs}
\usepackage{longtable}
\usepackage{tabularx}
\usepackage{natbib}
\usepackage{parskip}
\usepackage[table]{xcolor}

\newcommand{\TableReceivedTrainingSubsector}{%
  \begin{longtable}{>{\raggedright\arraybackslash}p{0.15\textwidth}rrrrrrrr}
  \caption{Received Formal Training (Subsector)} \label{table:received_training_subsector} \\
  \toprule
   & $\text{Incidence}_g$ & $\text{Intensity}_g$ & $\overline{I_n}$ & $\overline{I^W_n}$ & $\overline{I^C_n}$ & $\overline{I^M_n}$ & Count & \% w/ Index \\
  \midrule
  \endfirsthead
  \caption[]{Received Formal Training (Subsector)} \\
  \toprule
   & $\text{Incidence}_g$ & $\text{Intensity}_g$ & $\overline{I_n}$ & $\overline{I^W_n}$ & $\overline{I^C_n}$ & $\overline{I^M_n}$ & Count & \% w/ Index \\
  \midrule
  \endhead
  \midrule
  \multicolumn{9}{r}{Continued on next page} \\
  \midrule
  \endfoot
  \bottomrule
  \endlastfoot
  No & 0.48 & 0.02 & -0.01 & -0.02 & -0.00 & -0.00 & 22,681,874 & 48.1\% \\
  Yes & 0.58 & 0.02 & 0.01 & 0.04 & 0.02 & 0.02 & 1,080,810 & 43.8\% \\
  \end{longtable}
}

\newcommand{\TableReceivedTrainingOccupationIpw}{%
  \begin{longtable}{>{\raggedright\arraybackslash}p{0.15\textwidth}rrrrrrrr}
  \caption{Received Formal Training (Occupation IPW)} \label{table:received_training_occupation_ipw} \\
  \toprule
   & $\text{Incidence}_g$ & $\text{Intensity}_g$ & $\overline{I_n}$ & $\overline{I^W_n}$ & $\overline{I^C_n}$ & $\overline{I^M_n}$ & Count & \% w/ Index \\
  \midrule
  \endfirsthead
  \caption[]{Received Formal Training (Occupation IPW)} \\
  \toprule
   & $\text{Incidence}_g$ & $\text{Intensity}_g$ & $\overline{I_n}$ & $\overline{I^W_n}$ & $\overline{I^C_n}$ & $\overline{I^M_n}$ & Count & \% w/ Index \\
  \midrule
  \endhead
  \midrule
  \multicolumn{9}{r}{Continued on next page} \\
  \midrule
  \endfoot
  \bottomrule
  \endlastfoot
  No & 0.58 & 0.07 & 0.04 & 0.10 & 0.01 & 0.01 & 22,681,874 & 0.7\% \\
  Yes & 0.61 & 0.10 & 0.06 & 0.17 & 0.02 & 0.06 & 1,080,810 & 7.6\% \\
  \end{longtable}
}

\newcommand{\TableEmploymentStatusSubsector}{%
  \begin{longtable}{>{\raggedright\arraybackslash}p{0.15\textwidth}rrrrrrrr}
  \caption{Employment Status at Entry (Subsector)} \label{table:employment_status_subsector} \\
  \toprule
   & $\text{Incidence}_g$ & $\text{Intensity}_g$ & $\overline{I_n}$ & $\overline{I^W_n}$ & $\overline{I^C_n}$ & $\overline{I^M_n}$ & Count & \% w/ Index \\
  \midrule
  \endfirsthead
  \caption[]{Employment Status at Entry (Subsector)} \\
  \toprule
   & $\text{Incidence}_g$ & $\text{Intensity}_g$ & $\overline{I_n}$ & $\overline{I^W_n}$ & $\overline{I^C_n}$ & $\overline{I^M_n}$ & Count & \% w/ Index \\
  \midrule
  \endhead
  \midrule
  \multicolumn{9}{r}{Continued on next page} \\
  \midrule
  \endfoot
  \bottomrule
  \endlastfoot
  Unemployed & 0.47 & 0.02 & -0.01 & -0.02 & -0.00 & -0.00 & 19,952,496 & 46.4\% \\
  Employed & 0.54 & 0.02 & 0.00 & 0.01 & 0.01 & 0.00 & 3,542,893 & 56.7\% \\
  Employed, but Notice of Termination & 0.47 & 0.02 & -0.01 & -0.02 & -0.00 & -0.01 & 210,759 & 48.0\% \\
  Not in labor force & 0.50 & 0.01 & -0.00 & 0.01 & 0.02 & 0.01 & 56,536 & 13.4\% \\
  \end{longtable}
}

\newcommand{\TableEmploymentStatusOccupationIpw}{%
  \begin{longtable}{>{\raggedright\arraybackslash}p{0.15\textwidth}rrrrrrrr}
  \caption{Employment Status at Entry (Occupation IPW)} \label{table:employment_status_occupation_ipw} \\
  \toprule
   & $\text{Incidence}_g$ & $\text{Intensity}_g$ & $\overline{I_n}$ & $\overline{I^W_n}$ & $\overline{I^C_n}$ & $\overline{I^M_n}$ & Count & \% w/ Index \\
  \midrule
  \endfirsthead
  \caption[]{Employment Status at Entry (Occupation IPW)} \\
  \toprule
   & $\text{Incidence}_g$ & $\text{Intensity}_g$ & $\overline{I_n}$ & $\overline{I^W_n}$ & $\overline{I^C_n}$ & $\overline{I^M_n}$ & Count & \% w/ Index \\
  \midrule
  \endhead
  \midrule
  \multicolumn{9}{r}{Continued on next page} \\
  \midrule
  \endfoot
  \bottomrule
  \endlastfoot
  Unemployed & 0.57 & 0.08 & 0.05 & 0.11 & 0.01 & 0.01 & 19,952,496 & 0.8\% \\
  Employed & 0.61 & 0.06 & 0.04 & 0.08 & 0.01 & 0.01 & 3,542,893 & 1.7\% \\
  Employed, but Notice of Termination & 0.49 & 0.04 & 0.01 & 0.03 & 0.01 & 0.01 & 210,759 & 0.9\% \\
  Not in labor force & 0.52 & 0.04 & 0.01 & 0.01 & -0.01 & -0.01 & 56,536 & 3.8\% \\
  \end{longtable}
}

\newcommand{\TableProgramYearSubsector}{%
  \begin{longtable}{>{\raggedright\arraybackslash}p{0.15\textwidth}rrrrrrrr}
  \caption{Program Year (Subsector)} \label{table:program_year_subsector} \\
  \toprule
   & $\text{Incidence}_g$ & $\text{Intensity}_g$ & $\overline{I_n}$ & $\overline{I^W_n}$ & $\overline{I^C_n}$ & $\overline{I^M_n}$ & Count & \% w/ Index \\
  \midrule
  \endfirsthead
  \caption[]{Program Year (Subsector)} \\
  \toprule
   & $\text{Incidence}_g$ & $\text{Intensity}_g$ & $\overline{I_n}$ & $\overline{I^W_n}$ & $\overline{I^C_n}$ & $\overline{I^M_n}$ & Count & \% w/ Index \\
  \midrule
  \endhead
  \midrule
  \multicolumn{9}{r}{Continued on next page} \\
  \midrule
  \endfoot
  \bottomrule
  \endlastfoot
  2016 & 0.48 & 0.02 & -0.01 & -0.01 & 0.00 & 0.00 & 4,401,222 & 44.8\% \\
  2017 & 0.48 & 0.02 & -0.01 & -0.01 & 0.00 & 0.00 & 3,876,884 & 47.5\% \\
  2018 & 0.46 & 0.02 & -0.01 & -0.02 & 0.00 & 0.00 & 3,403,551 & 51.9\% \\
  2019 & 0.42 & 0.01 & -0.02 & -0.03 & 0.01 & -0.00 & 3,092,842 & 49.9\% \\
  2020 & 0.55 & 0.02 & 0.00 & 0.00 & 0.01 & -0.00 & 2,250,230 & 48.3\% \\
  2021 & 0.54 & 0.02 & 0.00 & 0.00 & -0.00 & -0.00 & 2,307,662 & 46.4\% \\
  2022 & 0.47 & 0.02 & -0.01 & -0.02 & -0.01 & -0.00 & 2,205,751 & 55.0\% \\
  2023 & 0.46 & 0.01 & -0.01 & -0.03 & -0.01 & -0.01 & 2,224,542 & 39.7\% \\
  \end{longtable}
}

\newcommand{\TableProgramYearOccupationIpw}{%
  \begin{longtable}{>{\raggedright\arraybackslash}p{0.15\textwidth}rrrrrrrr}
  \caption{Program Year (Occupation IPW)} \label{table:program_year_occupation_ipw} \\
  \toprule
   & $\text{Incidence}_g$ & $\text{Intensity}_g$ & $\overline{I_n}$ & $\overline{I^W_n}$ & $\overline{I^C_n}$ & $\overline{I^M_n}$ & Count & \% w/ Index \\
  \midrule
  \endfirsthead
  \caption[]{Program Year (Occupation IPW)} \\
  \toprule
   & $\text{Incidence}_g$ & $\text{Intensity}_g$ & $\overline{I_n}$ & $\overline{I^W_n}$ & $\overline{I^C_n}$ & $\overline{I^M_n}$ & Count & \% w/ Index \\
  \midrule
  \endhead
  \midrule
  \multicolumn{9}{r}{Continued on next page} \\
  \midrule
  \endfoot
  \bottomrule
  \endlastfoot
  2016 & 0.57 & 0.08 & 0.05 & 0.11 & 0.01 & 0.01 & 4,401,222 & 1.0\% \\
  2017 & 0.63 & 0.09 & 0.07 & 0.15 & 0.01 & 0.01 & 3,876,884 & 1.0\% \\
  2018 & 0.63 & 0.10 & 0.07 & 0.15 & 0.01 & 0.01 & 3,403,551 & 1.1\% \\
  2019 & 0.58 & 0.08 & 0.05 & 0.12 & 0.01 & 0.01 & 3,092,842 & 0.9\% \\
  2020 & 0.65 & 0.09 & 0.07 & 0.15 & 0.01 & 0.02 & 2,250,230 & 0.9\% \\
  2021 & 0.60 & 0.07 & 0.04 & 0.09 & 0.00 & 0.01 & 2,307,662 & 1.0\% \\
  2022 & 0.50 & 0.04 & 0.01 & 0.02 & 0.01 & 0.01 & 2,205,751 & 0.9\% \\
  2023 & 0.48 & 0.03 & -0.00 & 0.01 & 0.01 & 0.02 & 2,224,542 & 0.8\% \\
  \end{longtable}
}

\newcommand{\TableFundingStreamSubsector}{%
  \begin{longtable}{>{\raggedright\arraybackslash}p{0.15\textwidth}rrrrrrrr}
  \caption{Funding Stream (Subsector)} \label{table:funding_stream_subsector} \\
  \toprule
   & $\text{Incidence}_g$ & $\text{Intensity}_g$ & $\overline{I_n}$ & $\overline{I^W_n}$ & $\overline{I^C_n}$ & $\overline{I^M_n}$ & Count & \% w/ Index \\
  \midrule
  \endfirsthead
  \caption[]{Funding Stream (Subsector)} \\
  \toprule
   & $\text{Incidence}_g$ & $\text{Intensity}_g$ & $\overline{I_n}$ & $\overline{I^W_n}$ & $\overline{I^C_n}$ & $\overline{I^M_n}$ & Count & \% w/ Index \\
  \midrule
  \endhead
  \midrule
  \multicolumn{9}{r}{Continued on next page} \\
  \midrule
  \endfoot
  \bottomrule
  \endlastfoot
  Adult & 0.54 & 0.02 & 0.00 & 0.02 & 0.01 & 0.00 & 2,216,798 & 44.5\% \\
  Adult, Dislocated worker, or Youth & 0.45 & 0.02 & -0.01 & -0.02 & 0.00 & -0.00 & 458,334 & 52.0\% \\
  Dislocated Worker & 0.45 & 0.02 & -0.01 & -0.02 & -0.00 & 0.00 & 1,397,789 & 47.7\% \\
  Wagner-Peyser & 0.48 & 0.02 & -0.01 & -0.02 & -0.00 & -0.00 & 19,124,612 & 49.3\% \\
  Youth & 0.61 & 0.01 & 0.03 & 0.06 & 0.03 & -0.01 & 565,151 & 11.7\% \\
  \end{longtable}
}

\newcommand{\TableFundingStreamOccupationIpw}{%
  \begin{longtable}{>{\raggedright\arraybackslash}p{0.15\textwidth}rrrrrrrr}
  \caption{Funding Stream (Occupation IPW)} \label{table:funding_stream_occupation_ipw} \\
  \toprule
   & $\text{Incidence}_g$ & $\text{Intensity}_g$ & $\overline{I_n}$ & $\overline{I^W_n}$ & $\overline{I^C_n}$ & $\overline{I^M_n}$ & Count & \% w/ Index \\
  \midrule
  \endfirsthead
  \caption[]{Funding Stream (Occupation IPW)} \\
  \toprule
   & $\text{Incidence}_g$ & $\text{Intensity}_g$ & $\overline{I_n}$ & $\overline{I^W_n}$ & $\overline{I^C_n}$ & $\overline{I^M_n}$ & Count & \% w/ Index \\
  \midrule
  \endhead
  \midrule
  \multicolumn{9}{r}{Continued on next page} \\
  \midrule
  \endfoot
  \bottomrule
  \endlastfoot
  Adult & 0.66 & 0.12 & 0.09 & 0.20 & 0.02 & 0.03 & 2,216,798 & 3.7\% \\
  Adult, Dislocated worker, or Youth & 0.53 & 0.07 & 0.03 & 0.09 & 0.02 & 0.02 & 458,334 & 1.1\% \\
  Dislocated Worker & 0.49 & 0.06 & 0.01 & 0.06 & 0.03 & 0.03 & 1,397,789 & 3.5\% \\
  Wagner-Peyser & 0.56 & 0.06 & 0.04 & 0.08 & 0.01 & 0.01 & 19,124,612 & 0.5\% \\
  Youth & 0.75 & 0.15 & 0.13 & 0.27 & 0.01 & 0.01 & 565,151 & 1.3\% \\
  \end{longtable}
}

\newcommand{\TableTrainingServiceSubsector}{%
  \begin{longtable}{>{\raggedright\arraybackslash}p{0.15\textwidth}rrrrrrrr}
  \caption{Training Service Type (Subsector)} \label{table:training_service_subsector} \\
  \toprule
   & $\text{Incidence}_g$ & $\text{Intensity}_g$ & $\overline{I_n}$ & $\overline{I^W_n}$ & $\overline{I^C_n}$ & $\overline{I^M_n}$ & Count & \% w/ Index \\
  \midrule
  \endfirsthead
  \caption[]{Training Service Type (Subsector)} \\
  \toprule
   & $\text{Incidence}_g$ & $\text{Intensity}_g$ & $\overline{I_n}$ & $\overline{I^W_n}$ & $\overline{I^C_n}$ & $\overline{I^M_n}$ & Count & \% w/ Index \\
  \midrule
  \endhead
  \midrule
  \multicolumn{9}{r}{Continued on next page} \\
  \midrule
  \endfoot
  \bottomrule
  \endlastfoot
  Not available & 0.48 & 0.01 & -0.01 & -0.02 & 0.00 & -0.00 & 1,692,474 & 42.2\% \\
  No Training Service & 0.48 & 0.02 & -0.01 & -0.02 & -0.00 & -0.00 & 21,021,079 & 48.6\% \\
  On the Job Training & 0.53 & 0.03 & 0.01 & 0.04 & 0.02 & 0.04 & 100,196 & 53.2\% \\
  Skill Upgrading & 0.58 & 0.03 & 0.01 & 0.04 & 0.01 & 0.02 & 133,589 & 53.3\% \\
  Entrepreneurial Training & 0.53 & 0.01 & -0.00 & 0.01 & 0.02 & 0.01 & 3,034 & 17.3\% \\
  ABE or ESL w/ Training & 0.59 & 0.02 & 0.02 & 0.04 & 0.01 & -0.01 & 13,025 & 28.8\% \\
  Customized Training & 0.67 & 0.01 & 0.01 & 0.03 & 0.01 & 0.00 & 24,710 & 52.0\% \\
  Occupational Skills Training & 0.57 & 0.02 & 0.01 & 0.04 & 0.02 & 0.02 & 609,007 & 46.6\% \\
  ABE or ESL w/out Training & 0.60 & 0.02 & 0.03 & 0.00 & -0.02 & -0.08 & 5,259 & 41.1\% \\
  Prerequisite Training & 0.57 & 0.02 & 0.02 & 0.03 & -0.01 & -0.00 & 1,841 & 40.1\% \\
  Registered Apprenticeship & 0.69 & 0.02 & 0.02 & 0.05 & -0.01 & 0.02 & 12,893 & 56.8\% \\
  Youth Occupational Skills Training & 0.62 & 0.01 & 0.03 & 0.07 & 0.03 & -0.01 & 131,577 & 15.8\% \\
  Non-Occupational-Skills Training & 0.57 & 0.02 & 0.02 & 0.04 & 0.03 & -0.02 & 11,188 & 27.4\% \\
  Job Readiness Training & 0.61 & 0.02 & 0.04 & 0.05 & -0.03 & -0.03 & 2,812 & 24.3\% \\
  \end{longtable}
}

\newcommand{\TableTrainingServiceOccupationIpw}{%
  \begingroup 
  \small 
  \setlength{\tabcolsep}{4pt} 
  \begin{longtable}{>{\raggedright\arraybackslash}p{0.24\textwidth}rrrrrrrr}
  \caption{Training Service Type (Occupation IPW)} \label{table:training_service_occupation_ipw} \\
  \toprule
   & $\text{Incidence}_g$ & $\text{Intensity}_g$ & $\overline{I_n}$ & $\overline{I^W_n}$ & $\overline{I^C_n}$ & $\overline{I^M_n}$ & Count & \% w/ Index \\
  \midrule
  \endfirsthead
  \caption[]{Training Service Type (Occupation IPW)} \\
  \toprule
   & $\text{Incidence}_g$ & $\text{Intensity}_g$ & $\overline{I_n}$ & $\overline{I^W_n}$ & $\overline{I^C_n}$ & $\overline{I^M_n}$ & Count & \% w/ Index \\
  \midrule
  \endhead
  \midrule
  \multicolumn{9}{r}{Continued on next page} \\
  \midrule
  \endfoot
  \bottomrule
  \endlastfoot
  No Training Service & 0.58 & 0.07 & 0.04 & 0.10 & 0.01 & 0.01 & 21,021,079 & 0.6\% \\
  On the Job Training & 0.61 & 0.09 & 0.06 & 0.16 & 0.04 & 0.04 & 100,196 & 14.8\% \\
  Skill Upgrading & 0.61 & 0.11 & 0.07 & 0.19 & 0.01 & 0.08 & 133,589 & 9.7\% \\
  Entrepreneurial Training & 0.61 & 0.11 & 0.08 & 0.18 & -0.03 & 0.04 & 3,034 & 3.2\% \\
  ABE or ESL w/ Training & 0.67 & 0.12 & 0.09 & 0.21 & 0.02 & 0.02 & 13,025 & 5.2\% \\
  Customized Training & 0.71 & 0.10 & 0.08 & 0.17 & 0.02 & 0.00 & 24,710 & 9.6\% \\
  Occupational Skills Training & 0.59 & 0.09 & 0.06 & 0.16 & 0.01 & 0.08 & 609,007 & 7.6\% \\
  ABE or ESL w/out Training & 0.57 & 0.08 & 0.03 & 0.05 & 0.05 & -0.05 & 5,259 & 4.4\% \\
  Prerequisite Training & 0.56 & 0.07 & 0.02 & 0.12 & 0.06 & 0.09 & 1,841 & 4.1\% \\
  Registered Apprenticeship & 0.71 & 0.07 & 0.05 & 0.10 & 0.00 & 0.00 & 12,893 & 14.9\% \\
  Youth Occupational Skills Training & 0.69 & 0.13 & 0.10 & 0.22 & 0.01 & 0.02 & 131,577 & 1.7\% \\
  Non-Occupational-Skills Training & 0.59 & 0.08 & 0.05 & 0.10 & 0.01 & -0.03 & 11,188 & 3.6\% \\
  Job Readiness Training & 0.73 & 0.16 & 0.13 & 0.31 & 0.08 & 0.02 & 2,812 & 2.0\% \\
  \end{longtable}
  \endgroup
}

\newcommand{\TableAgeOccupationIpw}{%
  \begin{longtable}{>{\raggedright\arraybackslash}p{0.15\textwidth}rrrrrrrr}
  \caption{Participant Age (Occupation IPW)} \label{table:age_occupation_ipw} \\
  \toprule
   & $\text{Incidence}_g$ & $\text{Intensity}_g$ & $\overline{I_n}$ & $\overline{I^W_n}$ & $\overline{I^C_n}$ & $\overline{I^M_n}$ & Count & \% w/ Index \\
  \midrule
  \endfirsthead
  \caption[]{Participant Age (Occupation IPW)} \\
  \toprule
   & $\text{Incidence}_g$ & $\text{Intensity}_g$ & $\overline{I_n}$ & $\overline{I^W_n}$ & $\overline{I^C_n}$ & $\overline{I^M_n}$ & Count & \% w/ Index \\
  \midrule
  \endhead
  \midrule
  \multicolumn{9}{r}{Continued on next page} \\
  \midrule
  \endfoot
  \bottomrule
  \endlastfoot
  15–19 & 0.75 & 0.14 & 0.12 & 0.25 & 0.01 & 0.01 & 998,011 & 1.2\% \\
  20–24 & 0.65 & 0.08 & 0.06 & 0.13 & 0.01 & 0.01 & 2,606,163 & 1.5\% \\
  25–29 & 0.61 & 0.08 & 0.05 & 0.11 & 0.01 & 0.01 & 2,989,862 & 1.3\% \\
  30–34 & 0.59 & 0.07 & 0.04 & 0.10 & 0.01 & 0.01 & 2,900,055 & 1.0\% \\
  35–39 & 0.59 & 0.07 & 0.05 & 0.10 & 0.01 & 0.01 & 2,657,372 & 1.0\% \\
  40–44 & 0.56 & 0.07 & 0.04 & 0.09 & 0.01 & 0.02 & 2,367,387 & 0.9\% \\
  45–49 & 0.55 & 0.07 & 0.04 & 0.09 & 0.01 & 0.02 & 2,286,281 & 0.9\% \\
  50–54 & 0.53 & 0.06 & 0.03 & 0.07 & 0.01 & 0.02 & 2,266,549 & 0.8\% \\
  55–59 & 0.50 & 0.06 & 0.02 & 0.06 & 0.02 & 0.01 & 2,147,074 & 0.7\% \\
  60–64 & 0.47 & 0.05 & 0.02 & 0.04 & 0.01 & 0.01 & 1,480,495 & 0.5\% \\
  65–69 & 0.50 & 0.06 & 0.03 & 0.08 & 0.01 & 0.00 & 651,361 & 0.4\% \\
  70–74 & 0.47 & 0.06 & 0.04 & 0.07 & 0.01 & -0.01 & 267,389 & 0.3\% \\
  75–79 & 0.47 & 0.07 & 0.04 & 0.08 & 0.02 & -0.02 & 96,371 & 0.3\% \\
  \end{longtable}
}

\newcommand{\TableDemographicsSubsector}{%
  \begin{longtable}{>{\raggedright\arraybackslash}p{0.15\textwidth}rrrrrrrr}
  \caption{Demographics (Subsector)} \label{table:demographics_subsector} \\
  \toprule
   & $\text{Incidence}_g$ & $\text{Intensity}_g$ & $\overline{I_n}$ & $\overline{I^W_n}$ & $\overline{I^C_n}$ & $\overline{I^M_n}$ & Count & \% w/ Index \\
  \midrule
  \endfirsthead
  \caption[]{Demographics (Subsector)} \\
  \toprule
   & $\text{Incidence}_g$ & $\text{Intensity}_g$ & $\overline{I_n}$ & $\overline{I^W_n}$ & $\overline{I^C_n}$ & $\overline{I^M_n}$ & Count & \% w/ Index \\
  \midrule
  \endhead
  \midrule
  \multicolumn{9}{r}{Continued on next page} \\
  \midrule
  \endfoot
  \bottomrule
  \endlastfoot
  \multicolumn{9}{l}{\textit{Sex}} \\
  Male & 0.48 & 0.02 & -0.01 & -0.01 & 0.00 & -0.00 & 12,203,827 & 48.6\% \\
  Female & 0.48 & 0.02 & -0.01 & -0.01 & -0.00 & -0.00 & 11,231,720 & 47.3\% \\
  Participant did not self-identify & 0.50 & 0.01 & -0.01 & -0.01 & -0.00 & -0.00 & 327,137 & 40.8\% \\
  \midrule
  \multicolumn{9}{l}{\textit{Race/Ethnicity}} \\
  Not available & 0.48 & 0.01 & -0.01 & -0.02 & -0.00 & 0.00 & 1,966,695 & 42.9\% \\
  Hispanic & 0.49 & 0.02 & -0.00 & -0.01 & 0.00 & -0.00 & 4,404,775 & 47.0\% \\
  Asian & 0.46 & 0.01 & -0.01 & -0.02 & -0.00 & 0.00 & 565,591 & 42.1\% \\
  Black & 0.50 & 0.02 & -0.00 & -0.01 & 0.00 & 0.00 & 5,744,866 & 51.7\% \\
  Native Hawaiian or Pacific Islander & 0.50 & 0.01 & -0.00 & -0.01 & -0.00 & -0.00 & 86,233 & 40.7\% \\
  American Indian or Alaska Native & 0.48 & 0.01 & -0.01 & -0.01 & 0.00 & -0.00 & 268,389 & 42.0\% \\
  White & 0.47 & 0.01 & -0.01 & -0.02 & -0.00 & -0.00 & 10,285,175 & 47.8\% \\
  Multiple Race & 0.49 & 0.02 & -0.01 & -0.01 & 0.00 & 0.00 & 440,960 & 45.4\% \\
  \midrule
  \multicolumn{9}{l}{\textit{Low Income Status}} \\
  No & 0.47 & 0.02 & -0.01 & -0.02 & -0.00 & -0.00 & 18,499,333 & 49.8\% \\
  Yes & 0.51 & 0.02 & 0.00 & 0.00 & 0.00 & -0.00 & 5,263,351 & 41.1\% \\
  \end{longtable}
}

\newcommand{\TableDemographicsOccupationIpw}{%
  \begin{longtable}{>{\raggedright\arraybackslash}p{0.15\textwidth}rrrrrrrr}
  \caption{Demographics (Occupation IPW)} \label{table:demographics_occupation_ipw} \\
  \toprule
   & $\text{Incidence}_g$ & $\text{Intensity}_g$ & $\overline{I_n}$ & $\overline{I^W_n}$ & $\overline{I^C_n}$ & $\overline{I^M_n}$ & Count & \% w/ Index \\
  \midrule
  \endfirsthead
  \caption[]{Demographics (Occupation IPW)} \\
  \toprule
   & $\text{Incidence}_g$ & $\text{Intensity}_g$ & $\overline{I_n}$ & $\overline{I^W_n}$ & $\overline{I^C_n}$ & $\overline{I^M_n}$ & Count & \% w/ Index \\
  \midrule
  \endhead
  \midrule
  \multicolumn{9}{r}{Continued on next page} \\
  \midrule
  \endfoot
  \bottomrule
  \endlastfoot
  \multicolumn{9}{l}{\textit{Sex}} \\
  Male & 0.58 & 0.07 & 0.05 & 0.11 & 0.01 & 0.02 & 12,203,827 & 1.0\% \\
  Female & 0.58 & 0.07 & 0.04 & 0.10 & 0.01 & 0.01 & 11,231,720 & 1.0\% \\
  Participant did not self-identify & 0.67 & 0.11 & 0.08 & 0.18 & 0.01 & 0.01 & 327,137 & 0.3\% \\
  \midrule
  \multicolumn{9}{l}{\textit{Race/Ethnicity}} \\
  Not available & 0.57 & 0.08 & 0.05 & 0.11 & 0.01 & 0.01 & 1,966,695 & 0.5\% \\
  Hispanic & 0.56 & 0.08 & 0.05 & 0.11 & 0.01 & 0.01 & 4,404,775 & 0.7\% \\
  Asian & 0.55 & 0.07 & 0.04 & 0.08 & 0.01 & 0.01 & 565,591 & 0.7\% \\
  Black & 0.62 & 0.08 & 0.05 & 0.11 & 0.01 & 0.01 & 5,744,866 & 1.6\% \\
  Native Hawaiian or Pacific Islander & 0.54 & 0.05 & 0.02 & 0.03 & -0.01 & 0.00 & 86,233 & 0.9\% \\
  American Indian or Alaska Native & 0.50 & 0.05 & 0.01 & 0.03 & -0.00 & 0.01 & 268,389 & 1.5\% \\
  White & 0.56 & 0.07 & 0.04 & 0.10 & 0.02 & 0.02 & 10,285,175 & 0.8\% \\
  Multiple Race & 0.53 & 0.06 & 0.03 & 0.06 & 0.01 & 0.01 & 440,960 & 0.9\% \\
  \midrule
  \multicolumn{9}{l}{\textit{Low Income Status}} \\
  No & 0.56 & 0.06 & 0.03 & 0.08 & 0.01 & 0.01 & 18,499,333 & 0.7\% \\
  Yes & 0.63 & 0.10 & 0.08 & 0.16 & 0.01 & 0.01 & 5,263,351 & 2.0\% \\
  \end{longtable}
}
\newcommand{\TableHighestEducationalLevelSubsector}{%
  \begin{longtable}{>{\raggedright\arraybackslash}p{0.15\textwidth}rrrrrrrr}
  \caption{Highest Educational Level (Subsector)} \label{table:highest_educational_level_subsector} \\
  \toprule
   & $\text{Incidence}_g$ & $\text{Intensity}_g$ & $\overline{I_n}$ & $\overline{I^W_n}$ & $\overline{I^C_n}$ & $\overline{I^M_n}$ & Count & \% w/ Index \\
  \midrule
  \endfirsthead
  \caption[]{Highest Educational Level (Subsector)} \\
  \toprule
   & $\text{Incidence}_g$ & $\text{Intensity}_g$ & $\overline{I_n}$ & $\overline{I^W_n}$ & $\overline{I^C_n}$ & $\overline{I^M_n}$ & Count & \% w/ Index \\
  \midrule
  \endhead
  \midrule
  \multicolumn{9}{r}{Continued on next page} \\
  \midrule
  \endfoot
  \bottomrule
  \endlastfoot
  No Educational Level Completed & 0.49 & 0.01 & -0.00 & -0.01 & 0.00 & -0.01 & 4,011,343 & 45.2\% \\
  Secondary School Diploma & 0.49 & 0.02 & -0.00 & -0.01 & 0.00 & -0.00 & 8,319,493 & 49.3\% \\
  Secondary School Equivalency & 0.49 & 0.02 & -0.00 & -0.01 & 0.00 & -0.00 & 1,916,584 & 47.6\% \\
  +1 Years of Postsecondary Education & 0.48 & 0.02 & -0.01 & -0.02 & -0.00 & 0.00 & 3,612,740 & 49.1\% \\
  Postsecondary Technical/Vocational Certificate & 0.48 & 0.02 & -0.01 & -0.01 & -0.00 & 0.00 & 703,887 & 50.2\% \\
  Associate's Degree & 0.47 & 0.02 & -0.01 & -0.02 & -0.00 & 0.00 & 1,504,002 & 48.5\% \\
  Bachelor's Degree & 0.45 & 0.02 & -0.01 & -0.02 & -0.00 & 0.01 & 2,680,297 & 46.5\% \\
  Advanced Degree & 0.43 & 0.01 & -0.02 & -0.03 & -0.00 & 0.01 & 983,681 & 45.0\% \\
  \end{longtable}
}

\newcommand{\TableHighestEducationalLevelOccupationIpw}{%
  \begin{longtable}{>{\raggedright\arraybackslash}p{0.15\textwidth}rrrrrrrr}
  \caption{Highest Educational Level (Occupation IPW)} \label{table:highest_educational_level_occupation_ipw} \\
  \toprule
   & $\text{Incidence}_g$ & $\text{Intensity}_g$ & $\overline{I_n}$ & $\overline{I^W_n}$ & $\overline{I^C_n}$ & $\overline{I^M_n}$ & Count & \% w/ Index \\
  \midrule
  \endfirsthead
  \caption[]{Highest Educational Level (Occupation IPW)} \\
  \toprule
   & $\text{Incidence}_g$ & $\text{Intensity}_g$ & $\overline{I_n}$ & $\overline{I^W_n}$ & $\overline{I^C_n}$ & $\overline{I^M_n}$ & Count & \% w/ Index \\
  \midrule
  \endhead
  \midrule
  \multicolumn{9}{r}{Continued on next page} \\
  \midrule
  \endfoot
  \bottomrule
  \endlastfoot
  No Educational Level Completed & 0.60 & 0.08 & 0.06 & 0.12 & 0.01 & 0.00 & 4,011,343 & 0.5\% \\
  Secondary School Diploma & 0.60 & 0.08 & 0.05 & 0.11 & 0.01 & 0.01 & 8,319,493 & 1.3\% \\
  Secondary School Equivalency & 0.60 & 0.08 & 0.05 & 0.11 & 0.00 & 0.01 & 1,916,584 & 1.3\% \\
  +1 Years of Postsecondary Education & 0.57 & 0.07 & 0.04 & 0.09 & 0.01 & 0.02 & 3,612,740 & 0.9\% \\
  Postsecondary Technical/Vocational Certificate & 0.55 & 0.07 & 0.04 & 0.08 & 0.02 & 0.01 & 703,887 & 1.0\% \\
  Associate's Degree & 0.55 & 0.07 & 0.03 & 0.08 & 0.02 & 0.02 & 1,504,002 & 0.9\% \\
  Bachelor's Degree & 0.51 & 0.06 & 0.02 & 0.07 & 0.03 & 0.03 & 2,680,297 & 0.7\% \\
  Advanced Degree & 0.49 & 0.06 & 0.02 & 0.07 & 0.03 & 0.02 & 983,681 & 0.5\% \\
  \end{longtable}
}
\newcommand{\TableStateSubsector}{%
  \begin{longtable}{>{\raggedright\arraybackslash}p{0.15\textwidth}rrrrrrrr}
  \caption{State (Subsector)} \label{table:state_subsector} \\
  \toprule
   & $\text{Incidence}_g$ & $\text{Intensity}_g$ & $\overline{I_n}$ & $\overline{I^W_n}$ & $\overline{I^C_n}$ & $\overline{I^M_n}$ & Count & \% w/ Index \\
  \midrule
  \endfirsthead
  \caption[]{State (Subsector)} \\
  \toprule
   & $\text{Incidence}_g$ & $\text{Intensity}_g$ & $\overline{I_n}$ & $\overline{I^W_n}$ & $\overline{I^C_n}$ & $\overline{I^M_n}$ & Count & \% w/ Index \\
  \midrule
  \endhead
  \midrule
  \multicolumn{9}{r}{Continued on next page} \\
  \midrule
  \endfoot
  \bottomrule
  \endlastfoot
  AK & 0.47 & 0.01 & -0.01 & -0.01 & 0.00 & 0.01 & 66,858 & 32.6\% \\
  AL & 0.52 & 0.02 & 0.00 & 0.01 & 0.00 & -0.00 & 409,383 & 60.3\% \\
  AR & 0.48 & 0.02 & -0.01 & -0.02 & -0.00 & -0.01 & 535,895 & 65.0\% \\
  AZ & 0.47 & 0.02 & -0.01 & -0.01 & -0.00 & 0.00 & 312,343 & 53.2\% \\
  CA & 0.43 & 0.02 & -0.01 & -0.03 & -0.00 & -0.00 & 1,528,270 & 48.7\% \\
  CO & 0.48 & 0.02 & -0.01 & -0.01 & 0.00 & -0.00 & 482,724 & 49.2\% \\
  CT & 0.44 & 0.01 & -0.01 & -0.03 & -0.00 & 0.00 & 242,953 & 46.7\% \\
  DC & 0.48 & 0.00 & -0.01 & -0.02 & -0.01 & -0.00 & 50,700 & 5.8\% \\
  DE & 0.47 & 0.02 & -0.01 & -0.02 & -0.01 & -0.01 & 75,127 & 57.5\% \\
  FL & 0.51 & 0.01 & -0.00 & 0.00 & 0.00 & 0.00 & 1,490,980 & 36.4\% \\
  GA & 0.52 & 0.02 & -0.00 & -0.01 & -0.00 & -0.00 & 1,053,080 & 43.9\% \\
  IA & 0.48 & 0.01 & -0.01 & -0.02 & -0.00 & -0.01 & 368,611 & 39.2\% \\
  ID & 0.47 & 0.02 & -0.01 & -0.02 & 0.00 & 0.00 & 95,327 & 62.3\% \\
  IL & 0.47 & 0.02 & -0.01 & -0.02 & -0.01 & 0.00 & 271,329 & 55.4\% \\
  IN & 0.49 & 0.02 & -0.01 & -0.02 & 0.00 & -0.01 & 540,814 & 66.9\% \\
  KS & 0.52 & 0.02 & -0.00 & 0.00 & 0.00 & 0.00 & 144,026 & 56.2\% \\
  KY & 0.49 & 0.02 & -0.00 & -0.02 & -0.00 & -0.01 & 390,370 & 48.3\% \\
  LA & 0.49 & 0.02 & -0.00 & -0.01 & -0.00 & -0.00 & 300,414 & 47.7\% \\
  MA & 0.42 & 0.01 & -0.02 & -0.03 & 0.00 & 0.01 & 720,441 & 34.4\% \\
  MD & 0.43 & 0.02 & -0.02 & -0.03 & 0.00 & 0.00 & 288,332 & 58.2\% \\
  ME & 0.47 & 0.02 & -0.00 & -0.01 & 0.00 & -0.01 & 40,158 & 53.4\% \\
  MI & 0.45 & 0.01 & -0.01 & -0.02 & -0.00 & -0.00 & 1,125,769 & 56.4\% \\
  MN & 0.45 & 0.02 & -0.01 & -0.03 & -0.00 & 0.00 & 208,683 & 53.5\% \\
  MO & 0.47 & 0.02 & -0.01 & -0.02 & 0.00 & -0.00 & 542,592 & 57.4\% \\
  MS & 0.53 & 0.02 & 0.00 & 0.01 & 0.01 & 0.01 & 347,507 & 54.0\% \\
  MT & 0.50 & 0.02 & -0.00 & -0.01 & 0.00 & -0.00 & 144,075 & 44.8\% \\
  NC & 0.48 & 0.02 & -0.01 & -0.01 & 0.00 & -0.00 & 941,681 & 63.5\% \\
  ND & 0.51 & 0.02 & -0.00 & -0.00 & 0.00 & 0.00 & 35,359 & 53.0\% \\
  NE & 0.46 & 0.02 & -0.01 & -0.02 & -0.00 & -0.00 & 82,868 & 56.9\% \\
  NH & 0.53 & 0.00 & 0.00 & 0.02 & 0.01 & 0.00 & 97,222 & 1.0\% \\
  NJ & 0.44 & 0.01 & -0.01 & -0.03 & -0.00 & 0.00 & 515,558 & 31.3\% \\
  NM & 0.48 & 0.02 & -0.01 & -0.01 & 0.00 & -0.01 & 180,588 & 43.7\% \\
  NV & 0.50 & 0.02 & -0.00 & -0.00 & -0.00 & -0.00 & 252,032 & 48.0\% \\
  NY & 0.48 & 0.01 & -0.01 & -0.02 & -0.00 & -0.00 & 1,809,793 & 44.2\% \\
  OH & 0.48 & 0.02 & -0.00 & -0.01 & -0.00 & 0.01 & 237,571 & 41.6\% \\
  OK & 0.45 & 0.02 & -0.01 & -0.03 & -0.01 & -0.01 & 115,848 & 45.5\% \\
  OR & 0.44 & 0.02 & -0.01 & -0.03 & 0.00 & -0.00 & 964,400 & 56.1\% \\
  PA & 0.46 & 0.02 & -0.01 & -0.02 & -0.00 & 0.00 & 503,189 & 53.6\% \\
  PR & 0.32 & 0.00 & -0.07 & -0.03 & 0.13 & 0.10 & 132,378 & 9.4\% \\
  RI & 0.44 & 0.02 & -0.01 & -0.03 & -0.01 & 0.00 & 32,518 & 55.8\% \\
  SC & 0.50 & 0.02 & -0.00 & -0.01 & -0.00 & -0.00 & 401,346 & 53.2\% \\
  SD & 0.52 & 0.02 & -0.00 & 0.00 & 0.00 & 0.00 & 63,093 & 54.8\% \\
  TN & 0.50 & 0.02 & -0.00 & -0.01 & 0.00 & -0.00 & 288,306 & 55.3\% \\
  TX & 0.49 & 0.02 & -0.01 & -0.01 & 0.00 & 0.00 & 2,885,499 & 51.4\% \\
  UT & 0.61 & 0.01 & 0.01 & 0.01 & 0.00 & -0.00 & 1,142,357 & 40.9\% \\
  VA & 0.50 & 0.02 & -0.00 & -0.01 & 0.00 & -0.00 & 235,347 & 58.9\% \\
  VT & 0.47 & 0.02 & -0.01 & -0.01 & 0.00 & -0.01 & 23,586 & 47.0\% \\
  WA & 0.44 & 0.00 & -0.01 & -0.03 & 0.00 & 0.00 & 571,452 & 11.1\% \\
  WI & 0.45 & 0.02 & -0.01 & -0.03 & -0.00 & -0.01 & 162,935 & 60.0\% \\
  WV & 0.52 & 0.01 & 0.00 & -0.00 & -0.00 & -0.00 & 211,501 & 33.6\% \\
  WY & 0.48 & 0.01 & -0.00 & -0.01 & -0.00 & -0.01 & 53,580 & 25.9\% \\
  \end{longtable}
}

\newcommand{\TableStateOccupationIpw}{%
  \begin{longtable}{>{\raggedright\arraybackslash}p{0.15\textwidth}rrrrrrrr}
  \caption{State (Occupation IPW)} \label{table:state_occupation_ipw} \\
  \toprule
   & $\text{Incidence}_g$ & $\text{Intensity}_g$ & $\overline{I_n}$ & $\overline{I^W_n}$ & $\overline{I^C_n}$ & $\overline{I^M_n}$ & Count & \% w/ Index \\
  \midrule
  \endfirsthead
  \caption[]{State (Occupation IPW)} \\
  \toprule
   & $\text{Incidence}_g$ & $\text{Intensity}_g$ & $\overline{I_n}$ & $\overline{I^W_n}$ & $\overline{I^C_n}$ & $\overline{I^M_n}$ & Count & \% w/ Index \\
  \midrule
  \endhead
  \midrule
  \multicolumn{9}{r}{Continued on next page} \\
  \midrule
  \endfoot
  \bottomrule
  \endlastfoot
  AK & 0.45 & 0.03 & -0.01 & -0.02 & 0.00 & 0.01 & 66,858 & 20.2\% \\
  AL & 0.67 & 0.10 & 0.07 & 0.15 & 0.01 & 0.00 & 409,383 & 0.1\% \\
  AR & 0.65 & 0.12 & 0.09 & 0.19 & 0.01 & 0.02 & 535,895 & 0.3\% \\
  AZ & 0.62 & 0.09 & 0.06 & 0.14 & -0.01 & 0.03 & 312,343 & 2.1\% \\
  CA & 0.67 & 0.14 & 0.11 & 0.23 & 0.02 & 0.02 & 1,528,270 & 1.2\% \\
  CO & 0.63 & 0.08 & 0.05 & 0.09 & 0.01 & -0.02 & 482,724 & 0.1\% \\
  CT & 0.60 & 0.10 & 0.06 & 0.15 & 0.03 & 0.02 & 242,953 & 1.0\% \\
  DC & 0.74 & 0.15 & 0.13 & 0.28 & 0.02 & 0.02 & 50,700 & 0.4\% \\
  DE & 0.59 & 0.11 & 0.06 & 0.20 & 0.05 & 0.10 & 75,127 & 0.6\% \\
  FL & 0.72 & 0.17 & 0.15 & 0.31 & 0.01 & 0.03 & 1,490,980 & 0.8\% \\
  GA & 0.65 & 0.12 & 0.09 & 0.26 & 0.08 & 0.09 & 1,053,080 & 0.1\% \\
  ID & 0.54 & 0.08 & 0.04 & 0.12 & -0.00 & 0.07 & 95,327 & 0.7\% \\
  IL & 0.58 & 0.08 & 0.05 & 0.12 & 0.01 & 0.03 & 271,329 & 4.1\% \\
  IN & 0.57 & 0.09 & 0.05 & 0.12 & 0.01 & 0.02 & 540,814 & 0.9\% \\
  KS & 0.56 & 0.10 & 0.06 & 0.17 & 0.04 & 0.06 & 144,026 & 0.7\% \\
  KY & 0.69 & 0.11 & 0.09 & 0.18 & 0.04 & -0.03 & 390,370 & 0.1\% \\
  LA & 0.61 & 0.11 & 0.07 & 0.21 & 0.06 & 0.09 & 300,414 & 0.2\% \\
  MA & 0.47 & 0.06 & 0.01 & 0.05 & 0.03 & 0.02 & 720,441 & 1.0\% \\
  MD & 0.51 & 0.08 & 0.04 & 0.13 & 0.04 & 0.06 & 288,332 & 0.3\% \\
  ME & 0.61 & 0.11 & 0.07 & 0.17 & 0.03 & 0.04 & 40,158 & 1.0\% \\
  MI & 0.57 & 0.07 & 0.04 & 0.10 & 0.01 & 0.05 & 1,125,769 & 1.7\% \\
  MO & 0.63 & 0.12 & 0.09 & 0.21 & 0.05 & 0.02 & 542,592 & 0.1\% \\
  MS & 0.63 & 0.08 & 0.06 & 0.13 & 0.01 & 0.01 & 347,507 & 24.0\% \\
  MT & 0.54 & 0.09 & 0.03 & 0.15 & 0.09 & 0.09 & 144,075 & 0.1\% \\
  NC & 0.71 & 0.14 & 0.11 & 0.23 & 0.03 & -0.01 & 941,681 & 0.1\% \\
  ND & 0.55 & 0.09 & 0.03 & 0.17 & 0.01 & 0.19 & 35,359 & 0.3\% \\
  NE & 0.61 & 0.09 & 0.06 & 0.12 & 0.01 & 0.00 & 82,868 & 0.3\% \\
  NH & 0.53 & 0.08 & 0.03 & 0.11 & 0.07 & 0.03 & 97,222 & 0.7\% \\
  NJ & 0.59 & 0.10 & 0.06 & 0.17 & 0.05 & 0.06 & 515,558 & 0.2\% \\
  NM & 0.68 & 0.17 & 0.15 & 0.31 & -0.01 & 0.03 & 180,588 & 0.1\% \\
  NV & 0.66 & 0.13 & 0.10 & 0.20 & 0.00 & -0.01 & 252,032 & 1.1\% \\
  NY & 0.46 & 0.04 & 0.02 & 0.09 & 0.09 & 0.01 & 1,809,793 & 0.0\% \\
  OH & 0.51 & 0.08 & 0.03 & 0.11 & 0.09 & 0.03 & 237,571 & 1.7\% \\
  OK & 0.59 & 0.09 & 0.05 & 0.15 & 0.01 & 0.10 & 115,848 & 1.6\% \\
  PA & 0.57 & 0.09 & 0.05 & 0.13 & 0.03 & 0.03 & 503,189 & 3.0\% \\
  PR & 0.38 & 0.01 & -0.02 & -0.03 & 0.00 & 0.00 & 132,378 & 2.4\% \\
  RI & 0.49 & 0.06 & 0.01 & 0.08 & 0.05 & 0.06 & 32,518 & 0.7\% \\
  SC & 0.66 & 0.11 & 0.08 & 0.17 & 0.01 & 0.01 & 401,346 & 0.6\% \\
  SD & 0.53 & 0.07 & 0.03 & 0.06 & 0.02 & -0.01 & 63,093 & 0.9\% \\
  TN & 0.65 & 0.11 & 0.08 & 0.19 & 0.01 & 0.03 & 288,306 & 1.3\% \\
  VA & 0.62 & 0.10 & 0.07 & 0.15 & -0.00 & 0.02 & 235,347 & 0.8\% \\
  VT & 0.59 & 0.07 & 0.04 & 0.09 & 0.03 & -0.02 & 23,586 & 0.5\% \\
  WI & 0.51 & 0.07 & 0.02 & 0.06 & 0.01 & 0.01 & 162,935 & 1.7\% \\
  \end{longtable}
}

\newcommand{\TableRegionalEconomicIndicatorsOccupationIpw}{%
  \begin{longtable}{>{\raggedright\arraybackslash}p{0.15\textwidth}rrrrrrrr}
  \caption{Regional Economic Indicators (Occupation IPW)} \label{table:regional_economic_indicators_occupation_ipw} \\
  \toprule
   & $\text{Incidence}_g$ & $\text{Intensity}_g$ & $\overline{I_n}$ & $\overline{I^W_n}$ & $\overline{I^C_n}$ & $\overline{I^M_n}$ & Count & \% w/ Index \\
  \midrule
  \endfirsthead
  \caption[]{Regional Economic Indicators (Occupation IPW)} \\
  \toprule
   & $\text{Incidence}_g$ & $\text{Intensity}_g$ & $\overline{I_n}$ & $\overline{I^W_n}$ & $\overline{I^C_n}$ & $\overline{I^M_n}$ & Count & \% w/ Index \\
  \midrule
  \endhead
  \midrule
  \multicolumn{9}{r}{Continued on next page} \\
  \midrule
  \endfoot
  \bottomrule
  \endlastfoot
  \multicolumn{9}{l}{\textit{WDB Unemployment Rate}} \\
  2–3\% & 0.56 & 0.08 & 0.05 & 0.11 & 0.02 & 0.01 & 1,189,955 & 0.4\% \\
  3–4\% & 0.52 & 0.06 & 0.02 & 0.06 & 0.02 & 0.02 & 5,582,159 & 0.6\% \\
  4–5\% & 0.56 & 0.07 & 0.03 & 0.08 & 0.01 & 0.02 & 4,656,425 & 1.0\% \\
  5–6\% & 0.63 & 0.09 & 0.07 & 0.15 & 0.01 & 0.01 & 3,085,903 & 1.6\% \\
  6–7\% & 0.60 & 0.08 & 0.05 & 0.11 & 0.01 & 0.01 & 1,256,781 & 1.6\% \\
  7–8\% & 0.61 & 0.08 & 0.06 & 0.12 & 0.00 & 0.01 & 684,247 & 3.4\% \\
  8–9\% & 0.59 & 0.08 & 0.06 & 0.11 & 0.00 & 0.00 & 650,020 & 1.0\% \\
  9–10\% & 0.55 & 0.07 & 0.04 & 0.09 & 0.01 & 0.00 & 257,986 & 2.6\% \\
  10\%+ & 0.66 & 0.10 & 0.07 & 0.16 & 0.01 & 0.01 & 404,906 & 1.0\% \\
  Not available & 0.54 & 0.06 & 0.03 & 0.07 & 0.01 & 0.03 & 5,984,057 & 0.6\% \\
  \midrule
  \multicolumn{9}{l}{\textit{WDB Household Debt to Income}} \\
  \textless{}1 & 0.55 & 0.09 & 0.05 & 0.14 & 0.07 & 0.01 & 698,911 & 0.7\% \\
  1–2 & 0.59 & 0.08 & 0.05 & 0.11 & 0.01 & 0.01 & 10,652,227 & 1.3\% \\
  2–3 & 0.61 & 0.09 & 0.06 & 0.13 & 0.01 & 0.01 & 5,944,404 & 0.8\% \\
  3–4 & 0.58 & 0.10 & 0.06 & 0.17 & 0.04 & 0.06 & 472,769 & 0.4\% \\
  Not available & 0.52 & 0.05 & 0.02 & 0.05 & 0.01 & 0.02 & 5,994,373 & 0.6\% \\
  \midrule
  \multicolumn{9}{l}{\textit{WDB Median Age}} \\
  30–34 & 0.55 & 0.09 & 0.05 & 0.11 & 0.02 & 0.01 & 1,497,012 & 0.6\% \\
  35–39 & 0.60 & 0.08 & 0.05 & 0.11 & 0.01 & 0.01 & 8,440,281 & 1.5\% \\
  40–44 & 0.54 & 0.07 & 0.04 & 0.10 & 0.02 & 0.02 & 6,743,004 & 0.7\% \\
  45–49 & 0.55 & 0.08 & 0.05 & 0.11 & 0.02 & 0.02 & 1,101,886 & 1.1\% \\
  50–54 & 0.61 & 0.12 & 0.09 & 0.23 & 0.08 & 0.03 & 97,858 & 0.8\% \\
  Not available & 0.54 & 0.06 & 0.03 & 0.07 & 0.01 & 0.03 & 5,864,899 & 0.6\% \\
  \midrule
  \multicolumn{9}{l}{\textit{WDB Population (per sq km)}} \\
  \textless{}10 & 0.44 & 0.03 & -0.01 & -0.01 & 0.00 & 0.01 & 847,519 & 1.2\% \\
  10–50 & 0.64 & 0.08 & 0.06 & 0.13 & 0.01 & 0.01 & 5,402,429 & 1.9\% \\
  50–200 & 0.60 & 0.10 & 0.06 & 0.14 & 0.02 & 0.03 & 5,267,410 & 0.5\% \\
  200–1000 & 0.53 & 0.08 & 0.04 & 0.11 & 0.02 & 0.02 & 4,400,228 & 0.8\% \\
  1000–4000 & 0.55 & 0.08 & 0.04 & 0.12 & 0.05 & 0.01 & 555,749 & 1.5\% \\
  4000+ & 0.59 & 0.10 & 0.06 & 0.14 & 0.03 & 0.00 & 815,750 & 0.3\% \\
  Not available & 0.53 & 0.06 & 0.03 & 0.07 & 0.01 & 0.02 & 6,473,599 & 0.8\% \\
  \midrule
  \multicolumn{9}{l}{\textit{WDB Diversity Index}} \\
  25–50 & 0.62 & 0.12 & 0.08 & 0.19 & 0.03 & 0.02 & 1,171,829 & 1.8\% \\
  50–75 & 0.50 & 0.06 & 0.02 & 0.06 & 0.02 & 0.02 & 4,673,886 & 0.9\% \\
  75–100 & 0.62 & 0.11 & 0.07 & 0.17 & 0.02 & 0.02 & 5,635,688 & 0.5\% \\
  100+ & 0.62 & 0.08 & 0.06 & 0.12 & 0.01 & 0.01 & 6,314,128 & 1.6\% \\
  Not available & 0.52 & 0.05 & 0.02 & 0.05 & 0.01 & 0.02 & 5,967,153 & 0.6\% \\
  \midrule
  \multicolumn{9}{l}{\textit{WDB Population}} \\
  \textless{}100K & 0.61 & 0.08 & 0.05 & 0.11 & 0.01 & 0.01 & 6,146,306 & 1.9\% \\
  100K–1M & 0.52 & 0.07 & 0.03 & 0.07 & 0.02 & 0.02 & 8,568,206 & 0.6\% \\
  +1M & 0.58 & 0.10 & 0.07 & 0.15 & 0.03 & 0.02 & 3,420,260 & 1.0\% \\
  Not available & 0.53 & 0.06 & 0.02 & 0.06 & 0.01 & 0.03 & 5,627,912 & 0.5\% \\
  \midrule
  \multicolumn{9}{l}{\textit{WDB Mean Commuting Time (Min)}} \\
  \textless{}15 & 0.43 & 0.03 & -0.01 & -0.02 & 0.00 & 0.01 & 23,069 & 14.4\% \\
  15–30 & 0.60 & 0.08 & 0.05 & 0.12 & 0.01 & 0.01 & 15,324,823 & 1.1\% \\
  30–45 & 0.56 & 0.09 & 0.05 & 0.12 & 0.02 & 0.02 & 2,566,869 & 0.7\% \\
  Not available & 0.54 & 0.06 & 0.03 & 0.07 & 0.01 & 0.03 & 5,847,923 & 0.6\% \\
  \midrule
  \multicolumn{9}{l}{\textit{WDB Median Income Level}} \\
  \$10–15K & 0.43 & 0.02 & -0.01 & -0.01 & 0.00 & -0.00 & 86,408 & 1.5\% \\
  \$15–20K & 0.62 & 0.08 & 0.06 & 0.13 & 0.01 & 0.00 & 418,287 & 3.6\% \\
  \$20–25K & 0.65 & 0.09 & 0.07 & 0.15 & 0.01 & 0.01 & 2,932,611 & 2.3\% \\
  \$25–30K & 0.58 & 0.07 & 0.04 & 0.10 & 0.02 & 0.01 & 5,488,976 & 0.9\% \\
  \$30–35K & 0.52 & 0.07 & 0.03 & 0.07 & 0.01 & 0.02 & 4,691,389 & 0.7\% \\
  \$35–40K & 0.50 & 0.05 & 0.01 & 0.04 & 0.01 & 0.02 & 2,354,731 & 0.7\% \\
  \$40–45K & 0.45 & 0.04 & 0.00 & 0.02 & 0.02 & 0.02 & 1,072,191 & 0.7\% \\
  \$45–50K & 0.54 & 0.09 & 0.05 & 0.13 & 0.02 & 0.05 & 520,082 & 0.4\% \\
  \$50–60K & 0.61 & 0.12 & 0.08 & 0.19 & 0.05 & 0.02 & 279,835 & 0.3\% \\
  Not available & 0.54 & 0.06 & 0.03 & 0.07 & 0.01 & 0.03 & 5,864,899 & 0.6\% \\
  \midrule
  \multicolumn{9}{l}{\textit{WDB Metro Status (USDA Rural-Urban Continuum Codes)}} \\
  \multicolumn{8}{l}{\textit{Metro:}} \\[0.2em]
  1M+ & 0.55 & 0.08 & 0.05 & 0.12 & 0.03 & 0.02 & 5,595,889 & 0.8\% \\
  250K–1M & 0.48 & 0.05 & 0.01 & 0.03 & 0.01 & 0.02 & 1,756,570 & 1.0\% \\
  \textless{}250K & 0.56 & 0.08 & 0.04 & 0.11 & 0.03 & 0.04 & 319,277 & 1.3\% \\
  \multicolumn{8}{l}{\textit{Metro Adjacent:}} \\[0.2em]
  20K+ & 0.62 & 0.10 & 0.07 & 0.15 & 0.02 & 0.00 & 387,726 & 0.3\% \\
  5K–20K & 0.64 & 0.09 & 0.06 & 0.16 & 0.03 & 0.05 & 78,840 & 1.9\% \\
  \multicolumn{8}{l}{\textit{Nonmetro Adjacent:}} \\[0.2em]
  20K+ & 0.65 & 0.12 & 0.09 & 0.17 & -0.00 & -0.00 & 209,212 & 0.2\% \\
  5K–20K & 0.53 & 0.10 & 0.06 & 0.16 & 0.03 & 0.06 & 159,004 & 0.2\% \\
  Not available & 0.59 & 0.07 & 0.05 & 0.11 & 0.01 & 0.01 & 15,255,884 & 1.0\% \\
  \end{longtable}
}

\newcommand{\TableLRPerformance}{%
\begin{table}[H]
\centering
\caption{Logistic Regression Classification Performance Metrics (Test Set)}
\label{tab:lr_performance}
\begin{tabular}{l c c c c}
\toprule
\textbf{Class} & \textbf{Precision} & \textbf{Recall} & \textbf{F1-Score} & \textbf{Support} \\
\midrule
$\text{I}_n \leq 0$ & 0.61 & 0.66 & 0.64 & 11{,}794 \\
$\text{I}_n > 0$    & 0.71 & 0.67 & 0.69 & 14{,}788 \\
\midrule
\textbf{Accuracy}     & & & 0.66 & 26{,}582 \\
\textbf{Macro Avg}    & 0.66 & 0.66 & 0.66 & 26{,}582 \\
\textbf{Weighted Avg} & 0.67 & 0.66 & 0.66 & 26{,}582 \\
\textbf{AUC-ROC}      & & & 0.724 & \\
\bottomrule
\end{tabular}
\end{table}
}

\newcommand{\TableLRCoefficients}{%
  \begin{longtable}{>{\raggedright\arraybackslash}p{0.45\textwidth}rrr}
  \caption{Logistic Regression Coefficients} \label{tab:lr_coefficients} \\
  \toprule
  Feature & Coefficient & Odds Ratio & $|\text{Coefficient}|$ \\
  \midrule
  \endfirsthead
  \caption[]{Logistic Regression Coefficients} \\
  \toprule
  Feature & Coefficient & Odds Ratio & $|\text{Coefficient}|$ \\
  \midrule
  \endhead
  \midrule
  \multicolumn{4}{r}{Continued on next page} \\
  \midrule
  \endfoot
  \bottomrule
  \endlastfoot
  Pre-Program Occupation & 4.3581 & 78.1118 & 4.3581 \\
  Pre-Program Industry Subsector & 1.7683 & 5.8607 & 1.7683 \\
  State & 0.9487 & 2.5825 & 0.9487 \\
  Local Workforce Board (WDB) & 0.6398 & 1.8960 & 0.6398 \\
  Highest Educational Level: Advanced Degree & 0.5501 & 1.7334 & 0.5501 \\
  Sex: Did not self-identify & 0.4925 & 1.6364 & 0.4925 \\
  Training Service Type: Registered Apprenticeship & 0.3861 & 1.4712 & 0.3861 \\
  Highest Educational Level: Bachelor's Degree & 0.3730 & 1.4521 & 0.3730 \\
  Training Service Type: Youth Occupational Skills Training & -0.2507 & 0.7783 & 0.2507 \\
  Training Service Type: Customized Training & 0.2446 & 1.2771 & 0.2446 \\
  Funding Stream: Dislocated Worker & -0.2250 & 0.7985 & 0.2250 \\
  Training Service Type: Occupational Skills Training & -0.1801 & 0.8352 & 0.1801 \\
  Training Service Type: On the Job Training & -0.1611 & 0.8512 & 0.1611 \\
  Race / Ethnicity: Black & -0.1553 & 0.8561 & 0.1553 \\
  Funding Stream: Wagner-Peyser & -0.1517 & 0.8593 & 0.1517 \\
  Training Service Type: ABE or ESL w/ Training & -0.1368 & 0.8722 & 0.1368 \\
  WBD Median Income Level: Low & 0.1288 & 1.1375 & 0.1288 \\
  Employment Status at Entry: Employed & 0.1223 & 1.1301 & 0.1223 \\
  Low Income Status: Yes & 0.1203 & 1.1278 & 0.1203 \\
  Highest Educational Level: Associate's Degree & 0.1185 & 1.1258 & 0.1185 \\
  \end{longtable}
}

\title{\textbf{Did US Worker Retraining Reduce Participant Automation Exposure?}\thanks{We thank Ben Ansell, David Autor, Stephanie Chan, Sebastien Krier, Fabien Curto Millet, Iason Gabriel, Alex Imas, Bouke Klein Teeselink, and David Rueda for their comments and feedback on this paper. The views expressed in this work do not necessarily reflect the views of the University of Oxford, Google DeepMind, or the Forecasting Research Institute. }}

\author[1,2]{Julian Jacobs }
\author[3]{Jordan Canedy }

\affil[1]{ University of Oxford}
\affil[2]{ Google DeepMind}
\affil[3]{ Forecasting Research Institute}

\date{\today}

\begin{document}

\maketitle

\begin{abstract}
\noindent This paper evaluates whether the U.S. Workforce Innovation and Opportunity Act (WIOA) supported American worker resilience to technological automation. Analyzing over 23 million WIOA participation records (2017-2023), we introduce the “Retrainability Index,” which measures program outcomes through post-intervention wage recovery and shifts in Routine Task Intensity (RTI).  We show WIOA rarely shifts workers into less automation-exposed work, with a significant portion of participants simply returning to their prior field. Successful outcomes driven mostly by wage gains, possibly due to “catch-up” mean reversion, rather than changes in occupation. Outcomes are moderated by a person's prior occupational skill set and area of work, as well as their local economy. We find evidence that employer led programs—notably apprenticeships—are associated with the highest incidence of success. This suggests the United States'  existing public active labor market programming can support baseline wage recovery for vulnerable populations, but is not well-equipped to support the large-scale, cross-industry labor transitions. 
\end{abstract}
\thispagestyle{empty} 
\pagestyle{fancy}
\fancyhead{} 
\fancyhead[R]{Jacobs \& Canedy} 
\fancyfoot{} 
\fancyfoot[C]{\thepage} 

\section{Introduction}

This paper evaluates whether the United States' primary active labor market program (ALMP) successfully insulates workers from technology-enabled labor automation. By analyzing over 23 million participation records from the U.S. Workforce Innovation and Opportunity Act (WIOA) between 2017 and 2023, we introduce the ``Retrainability Index,'' a novel composite metric assessing program success through post-intervention wage recovery and shifts in Routine Task Intensity (RTI). In doing so, we hope to better understand the extent to which ALMPs---especially worker retraining programs---can support worker upskilling and resilience in the face of technological change. We feel this question is of particular importance as artificial intelligence (AI) systems improve and diffuse throughout economies. 

The mismatch between public exhortations for retraining and the poor evidence of its efficacy remains striking. While active labor market programming is ubiquitous in policy proposals for managing technological change, the literature on government-funded ALMPs is generally patchy, marked by disappointing randomized control trials and difficult-to-interpret observational data \citep{card2018works, heckman1999economics}. As scholarship by \citet{barnow2015employment} suggests, the US may have a particularly poor record of ALMP and retraining among rich countries. This paper aims to present a more nuanced picture of these programs by examining the outcomes of participants in WIOA, the primary public funder of retraining across the US \citep{andersson2024does}. We distinguish between WIOA's light-touch job search assistance and heavier-touch formal retraining to understand how well these interventions insulate workers from automation, and how these outcomes are shaped by local labor market dynamics.

There are three central findings from this paper. The first is that ALMP success in the US appears moderated by a person's prior occupation, as well as their local economy. In other words, programmatic success appears unsurprisingly related to a person existing area of work and the presence of new employment opportunities in their locality. We also identify a potential age penalty, with rates of success declining as participants age. Second, we find convincing evidence that heavier touch employer led programs are most successful, especially apprenticeships. This is in line with our expectations and prior literature, which has consistently suggested reskilling efforts that actively engage employers generate the best outcomes. Despite this, only about $5\%$ of WIOA participants receive retraining of any kind, and this sample is highly self-selecting.   

Crucially, we show that WIOA rarely supports worker resilience to automation, with $45\%$ of all WIOA participants returning to their prior industry of work, and 27\% staying in the same occupation. As a result, positive outcomes are rare and driven chiefly by higher wages, as opposed to lower RTI scores due to shifted occupations. Moreover, wage gains may be the consequence of a structural “catch-up” dynamic, whereby the most vulnerable program participants see a reversion to the mean.  As a result, we argue that WIOA acts primarily as a “second chance” program, offering the most vulnerable participants a potential wage floor. We do not find evidence, however, that WIOA was able to support widespread labor upskilling and resilience in the face of technological disruption and automation. 

\section{A Brief History of US Active Labor Market Programs}

Over the last 80 years, US public active labor market programs (ALMPs) have steadily evolved from centralized New Deal-style interventions—beginning with the 1933 Wagner-Peyser Act \citep{barnow2015employment, oleary2008reemployment}—to decentralized efforts targeting local private sector employers. While earlier programs heavily targeted basic skills and poverty reduction \citep{barnow2015employment}, the 1998 Workforce Investment Act (WIA) significantly widened participation and introduced ``Individual Training Accounts'' (ITAs) so individuals could choose the skills and sectors to invest their time in \citep{barnow2015employment, heinrich2008workforce}. In 2014, the Workforce Innovation and Opportunity Act (WIOA) streamlined this system, providing most US federally-funded active labor market programming today \citep{barnow2015employment, hyman2025retrainable}. 

Every year, approximately 500,000 participants take part in the WIOA's `Adult' and `Dislocated Worker' streams. Most `Adult' participants are near the poverty line \citep{socialpolicy2022wioa}, while `Dislocated Worker' participants are typically older workers who have lost stable employment. Of these participants, around 200,000 receive training vouchers at a cost of \$500 million---a relatively low figure compared to the \$25 billion spent annually on Pell Grants \citep{deming2023investing}. Since local providers bid for federal funding based on regional employer needs and political considerations, the skills imparted and methods of instruction vary considerably across US states, resulting in a programmatic patchwork \citep{barnow2015employment}. For instance, according to data from 2023-24, less than 10\% of WIOA training involved \textit{paid} on-the-job training, and just 2\% involved apprenticeships \citep{doleta2024wioa}.

Evaluating ALMPs is methodologically challenging. Researchers tracking post-exit employment and earnings \citep{barnow2015employment} generally struggle to show statistically significant benefits. A primary challenge is non-random selection; program participants are usually not representative of the wider displaced population, making it difficult to isolate the impact of program participation from a willingness to take part \citep{heckman1999economics, lalonde1986evaluating}. While quasi-experiments attempt to construct counterfactuals \citep{card2018works, rothstein2022caal}, unobservable characteristics and the huge variance in state programming make reliable causal claims difficult \citep{barnow2015employment}. The existing evidence provides cause for skepticism. Historical randomized controlled trials found no statistically significant improvements in employment or earnings, and recent 10-year evaluations suggest WIOA's retraining streams did not improve outcomes \citep{barnow2015employment}. While 70\% of recent WIOA participants were employed within a year of exit \citep{doleta2024wioa}, these outcomes lack control groups. Furthermore, \citet{deming2023investing} suggests approximately 40\% of participants are driven into low-wage, high-demand support roles (e.g., nursing assistants) that offer little career growth and remain highly vulnerable to automation.

There are at least three good reasons why ALMPs may be poorly suited to support labor transitions during economic shocks, whether they are the consequence of technology, outsourcing, or some exogenous event. The first is \textit{participant barriers}, which include prohibitive costs, lack of childcare, and short-run opportunity costs that may limit participation \citep{hendra2016encouraging}. Older workers, in particular, may be reluctant to retrain \citep{maestas2016does, neumark2013mechanisms}, and participants are often bewildered by an overly-complex voucher system offering over 75,000 programs without comparable performance data \citep{deming2023investing}. Second, local provider quality varies widely, and public programs struggle to supply easily transferable skills \citep{deming2023investing}. These challenges are particularly pronounced given the obvious difficulties in anticipating future labor market demands \citep{acemoglu2011skills}. Finally,  there simply may not be enough secure and well paid work to retrain into. Literature has previously suggested technology shocks can induce a temporary mismatch between the availability of desirable work, relative to the number of people that want that work. \citep{autor2015polyanyi, acemoglu2011skills, acemoglu2020robots}. It's not clear how any retraining program could overcome this sort of structural labor market challenge. 

\section{Dataset and Index Construction}\label{section:dataset_and_index_construction}

To construct the \textit{Retrainability Index}, we utilize WIOA Participant Individual Record Layout (PIRL) files submitted by state workforce agencies to the Employment and Training Administration (ETA) of the U.S. Department of Labor \citep{doleta2024wioa}. The PIRL serves as the primary administrative reporting system for WIOA Titles I and III, providing granular data on demographic characteristics, program participation, services received, and post-exit labor market outcomes. As noted by \citet{barnow2015employment}, administrative databases like the PIRL offer substantial advantages over survey data for evaluating training interventions, particularly regarding the accuracy of service receipt timing and earnings histories.

We compile all available PIRL files from Program Year (PY) 2017, Quarter 4 through PY 2024, Quarter 3, treating each distinct period of participation as a unique observational unit. To ensure the analysis focuses on substantive interventions, we restrict the sample by removing all observations classified as \textit{Reportable Individuals} (who receive only self-services or information-only services). Additionally, we exclude records lacking an observed program exit date, yielding a final dataset of approximately 23 million unique and completed participation periods.

While requiring an observed program end date and post-program earnings is essential for calculating the index, this restriction inherently conditions on a successful labor market re-attachment. It drops individuals who abandon the program, remain persistently enrolled, or exit into unemployment—plausibly the worst outcomes. Consequently, this selection mechanism almost certainly biases our index results upward, meaning the aggregate incidence of success detailed in this paper likely inflates the true share of WIOA participants experiencing positive outcomes. Furthermore, this survivorship bias interacts directly with our core findings: the robust ``catch-up'' effect among low earners may partially reflect selection and mean reversion. Beyond this, the significant ``age penalty" we observe, whereby older workers experience lower incidence of success, may very well be understated. 

Each participation period includes occupation codes capturing employment in the three quarters prior to and four quarters following participation. To address data inconsistencies, we prioritize the occupation code recorded temporally closest to the participation window. To enrich this data, we merge the 2018 Standard Occupational Classification (SOC) structure to assign detailed occupational attributes.\footnote{Specifically, we use the 2018 SOC Structure file, available for download at \url{https://www.bls.gov/soc/2018/}.} Furthermore, we extract the nearest North American Industry Classification System (NAICS) code for pre- and post-program employment, collapsing these to the 3-digit subsector level to harmonize industrial classifications.\footnote{We use the Bureau of Labor Statistics \textit{National Employment Matrix} to identify occupation–industry employment by SOC and NAICS codes, available at \url{https://www.bls.gov/emp/data/occupational-data.htm}.}

To capture local labor market heterogeneity, each participation period is assigned to a \textit{Local Workforce Development Board} (WDB), the primary sub-state governance unit coordinating WIOA service delivery \citep{barnow2015employment}. After matching annual WDB designations (PY 2016--2023) and prioritizing primary codes where multiple exist, we attach jurisdiction-level demographic and economic indicators---including population, median household income, median age, and population density---aggregated to the board level while preserving program-year identifiers.

Our index is then constructed as a composite of changes in wages and routine task intensity (RTI) prior to and following program participation. We use quarterly wages reported prior to program entry and following program exit, which are converted to constant 2010 dollars using the annual Consumer Price Index (CPI) to ensure comparability across program years \citep{barnow2015employment} (adopting 2010 as the base year for comparability with \citet{hyman2025retrainable}). We compute mean quarterly wages pre- and post-program, ignoring missing wage quarters, and construct the inverse hyperbolic sine (IHS) difference in these means. We then merge occupation-level RTI measures \citep{acemoglu2011skills} to characterize the routine task environment of participants' pre- and post-program employment.\footnote{The task measures are available in the replication files provided by the authors and can be downloaded directly as \texttt{onet-soc.dta} from \href{https://economics.mit.edu/people/faculty/daron-acemoglu/data-archive}{https://economics.mit.edu/people/faculty/daron-acemoglu/data-archive} \citep{acemoglu2011skills}.} We define these subcomponents in additional detail in Appendix \ref{appendix:subcomponent-definitions}.  

While our reliance on a four-quarter post-exit window captures immediate re-employment dynamics, it inherently misses longer-term wage trajectories. As noted by \citet{jacobson1993earnings}, displaced workers entering new industries often face initial wage penalties before experiencing gradual, long-term earnings growth, meaning the short-term window we are able to study may underestimate the ultimate value of certain retraining transitions. For instance, while the retraining effects of Trade Adjustment Assistance (TAA)—a WIOA program often regarded as a policy failure—can take years to materialize, recent evidence shows its wage insurance provisions may successfully increase immediate short-run employment probabilities and long-term cumulative earnings \citep{hyman2024wageinsurance}.

To capture resilience to technological change, we argue that RTI is the correct proxy for the particular form of automation that American workers were exposed to in the period we study (2017 to 2023) \citep{acemoglu2011skills, frey2017future}. While RTI effectively captures computerization-driven automation, it does not capture exposure to AI. A worker transitioning into less routine work might score as a ``success'' in our Index while simultaneously increasing their exposure to AI-enabled automation. In this regard, our methods differ from concurrent research by \citet{hyman2025retrainable},\footnote{This paper was developed independently and contemporaneously alongside our own work. We thank the authors for discussions regarding the methodological distinctions between our approaches.} who find wage penalties for WIOA participants explicitly targeting AI-intensive roles.

Despite this, AI-exposure—perhaps the dominant empirical proxy for AI labor market impacts—remains an inappropriate tool for our analysis on three grounds: (1) AI exposure does not tell us whether a worker is experiencing \textit{complementarity} or \textit{displacement}, and in many cases seems to correspond most to wage-boosting complementing effects; (2) AI exposure is concentrated in highly paid sectors where WIOA rarely places vulnerable workers; and (3) the AI-labor shock was either nascent or nonexistent prior to 2023. By contrast, RTI provides an empirically grounded proxy for the real automation exposure workers may have previously experienced. In this regard, our work is backwards looking, attempting to discern how WIOA helped workers adapt to the prior technological shocks.

Our index is constructed such that a participant experiences ``no change'' in their outcomes at $0$. So a participant exiting into an occupation with identical RTI and wages receives exactly $0$, while departures move proportionally toward $1$ (best outcome) or $-1$ (worst outcome). Concretely, $\text{I}_n > 0$ indicates labor market improvement, $\text{I}_n = 0$ indicates stasis, and $\text{I}_n < 0$ indicates deterioration. 

The composite score uses a weighted linear combination. Wage change receives 50\% weight, while routine cognitive and routine manual task shifts receive 25\% each. Since the index is defined at the individual participation period, it can be flexibly aggregated to any geographic or demographic unit without loss of consistency. The index is computed strictly for the subset of participants with observed pre- and post-participation wage and occupation data (unless otherwise specified) which we report as `\% w/ Index' in our results. We provide a complete formal specification in Appendix \ref{appendix:detailed-definition}.

\section{Data Summary and Descriptive Analysis}

We begin by analyzing the distribution of the occupation-level index (as described in Equation \ref{eq:index_subcomponents_occupation}) across a random subset ($10\%$) of the full sample participation periods. Table \ref{tab:summary_stats_occupation} presents the summary statistics for the index and its primary components.

\begin{table}[H]
\centering
\caption{Summary Statistics of the (Occupation-Level) Index and its Components}
\label{tab:summary_stats_occupation}
\resizebox{\textwidth}{!}{%
\begin{tabular}{lcccccccc}
\toprule
\textbf{Component} & \textbf{Count} & \textbf{Mean} & \textbf{Std Dev} & \textbf{Min} & \textbf{25\%} & \textbf{50\% (Median)} & \textbf{75\%} & \textbf{Max} \\
\midrule
$I_n$ (Index) & 22,972 & 0.05 & 0.16 & -0.54 & -0.03 & 0.01 & 0.08 & 0.87 \\
$I^W_n$ (Wage Change) & 22,972 & 0.11 & 0.27 & -0.89 & -0.02 & 0.02 & 0.10 & 1.00 \\
$I^C_n$ (Routine Cognitive Task Shift) & 22,972 & 0.01 & 0.22 & -1.00 & 0.00 & 0.00 & 0.00 & 1.00 \\
$I^M_n$ (Routine Manual Task Shift) & 22,972 & 0.01 & 0.20 & -0.98 & 0.00 & 0.00 & 0.03 & 0.98 \\
\bottomrule
\end{tabular}
}
\end{table}

\begin{figure}[H]
    \centering
    \includegraphics[width=1\linewidth]{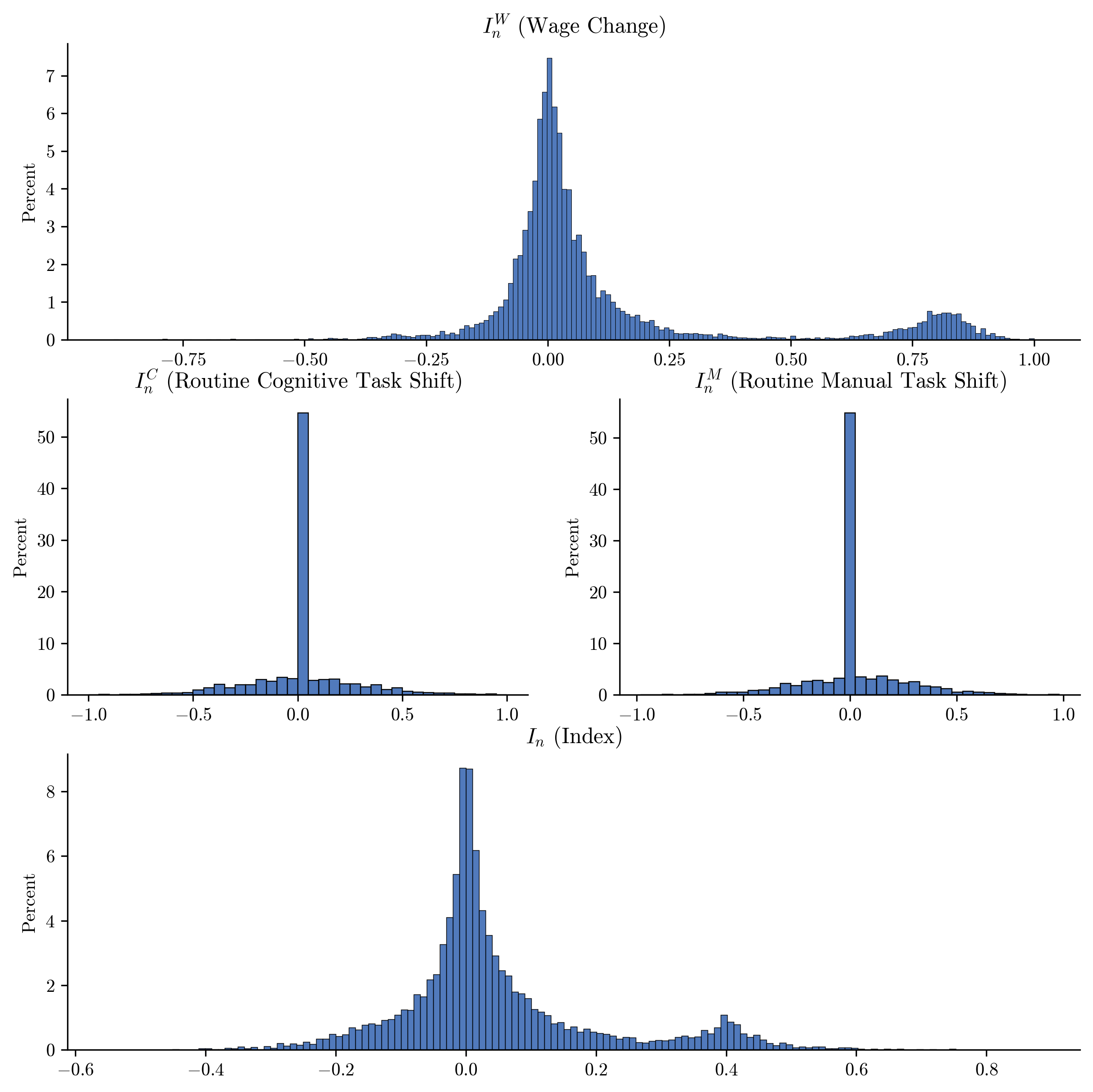}
    \caption{Component and composite index distributions illustrating how wage gains and routine task intensity shift measures combine into the overall index. Note the higher variance in wage outcomes relative to the peaked distribution of RTI changes.}
    \label{fig:image-2}
\end{figure}
As shown in Figure~\ref{fig:image-2}, the composite index $I_n$ exhibits a sharply 
peaked distribution centered near zero (Median $= 0.01$, Mean $= 0.05$, 
SD $= 0.16$). The composite variance is primarily driven by wage changes $I_n^W$, which show a wider dispersion ($\text{SD} = 0.27$) and have a secondary cluster near $0.8$, pointing to a distinct subgroup of participants who transitioned into substantially higher-wage occupations. The routine cognitive and manual task shift components ($I_n^C$, $I_n^M$) are heavily concentrated at zero which indicates that most participants transition into occupations with similar routine task content. In fact, for participation periods with adequate occupation and industry data, approximately $27\%$ of participants remained in the same occupation after program participation, and $45\%$ return to the same industry.
\begin{figure}[H]
    \centering
    \includegraphics[width=1\linewidth]{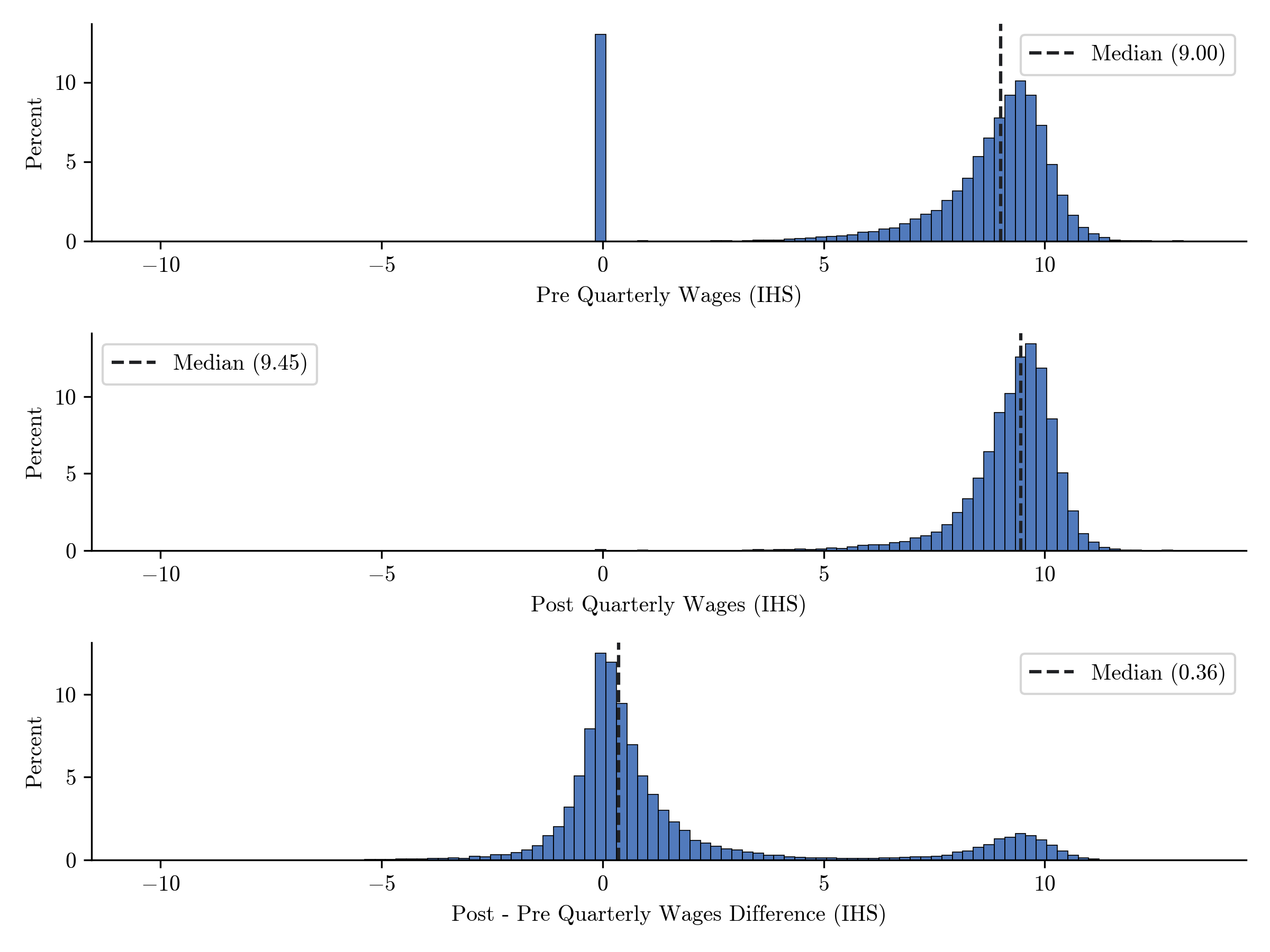}
    \caption{Distribution of IHS-Transformed Mean Quarterly Wages, Pre- and Post-Program Participation}
    \label{fig:image-IHS Quarterly Wages}
\end{figure}

When aggregated to program and participant demographic characteristics, the mean index score ($I_g)$ falls between $0.05$ and $0.10$ as shown in Table \ref{tab:summary_stats}. Across 52 state-level entities (including territories), the index ranges from $0.01$ to $0.36$ with a standard deviation of $0.07$. Local Workforce Development Boards ($N=438$) show higher variation (Range: $0$--$0.58$), reflecting local economic heterogeneity, yet the interquartile range remains narrow. The 14 distinct training service types exhibit the highest average incidence of positive outcome, with a mean of $0.10$. Substantively, this suggests that participants in formal retraining interventions tend to experience better outcomes when compared with the baseline, and non-training WIOA programs. 

\begin{table}[H]
\centering
\caption{Summary Statistics of the Aggregated Retrainability Index ($I_g$)}
\label{tab:summary_stats}
\resizebox{\textwidth}{!}{%
\begin{tabular}{lrrrrrrrr}
\toprule
\textbf{Aggregation} & \textbf{Count} & \textbf{Mean} & \textbf{Std Dev} & \textbf{Min} & \textbf{25\%} & \textbf{50\% (Median)} & \textbf{75\%} & \textbf{Max} \\
\midrule
Local Workforce Board (WDB) & 438 & 0.09 & 0.09 & 0.00 & 0.03 & 0.08 & 0.13 & 0.58 \\
Employment Status at Entry & 4 & 0.05 & 0.02 & 0.03 & 0.03 & 0.06 & 0.06 & 0.08 \\
Program Year & 8 & 0.07 & 0.03 & 0.03 & 0.07 & 0.09 & 0.09 & 0.10 \\
Pre-Program Occupation & 582 & 0.08 & 0.09 & 0.00 & 0.02 & 0.06 & 0.11 & 0.67 \\
Received Training Indicator & 2 & 0.08 & 0.01 & 0.07 & 0.07 & 0.09 & 0.09 & 0.09 \\
Funding Stream & 5 & 0.09 & 0.04 & 0.06 & 0.06 & 0.07 & 0.12 & 0.15 \\
Training Service Type & 14 & 0.10 & 0.04 & 0.06 & 0.07 & 0.09 & 0.11 & 0.21 \\
Participant Age & 71 & 0.08 & 0.07 & 0.00 & 0.06 & 0.07 & 0.08 & 0.41 \\
State & 50 & 0.10 & 0.07 & 0.01 & 0.06 & 0.09 & 0.13 & 0.36 \\
Race / Ethnicity & 8 & 0.07 & 0.02 & 0.04 & 0.06 & 0.08 & 0.08 & 0.08 \\
Low Income Status & 2 & 0.08 & 0.02 & 0.07 & 0.07 & 0.10 & 0.10 & 0.10 \\
Highest Educational Level & 9 & 0.07 & 0.01 & 0.05 & 0.06 & 0.07 & 0.08 & 0.09 \\
Routine Cognitive Intensity (Pre-Program) & 537 & 0.08 & 0.09 & 0.00 & 0.02 & 0.06 & 0.11 & 0.67 \\
Routine Manual Intensity (Pre-Program) & 537 & 0.08 & 0.09 & 0.00 & 0.02 & 0.06 & 0.11 & 0.67 \\
Sex & 3 & 0.09 & 0.02 & 0.07 & 0.08 & 0.08 & 0.11 & 0.11 \\
\bottomrule
\end{tabular}
}
\end{table}

Across all WIOA participation in our sample, the median age is 39. Participants are racially and ethnically diverse: $43.3\%$ White, $24.2\%$ Black, $18.5\%$ Hispanic, 2.4\% Asian and 10.4\% other races/ethnicities. Gender representation is nearly balanced ($51.2\%$ Female, $47.3\%$ Male, $1.4\%$ Did Not Identify). Notably, the majority ($83.9\%$) were unemployed at the time of program entry.  The proportion of unemployed participants generally increases with age. While roughly $79\%$ of participants aged 20 are unemployed, this figure rises to nearly $89\%$ for those approaching retirement age. This suggests that older workers engaging with the system are doing so from a position of significantly deeper labor market detachment compared to their younger counterparts, who are more likely to be employed or not in the labor force at entry. This also helps clarify WIOA's practical function in American economic policy: it serves chiefly as a kind of 'second-chance' intervention aimed at re-introducing participants to stable employment, rather than a means of supporting upskilling. 

\begin{figure}[H]
    \centering
    \includegraphics[width=1\linewidth]{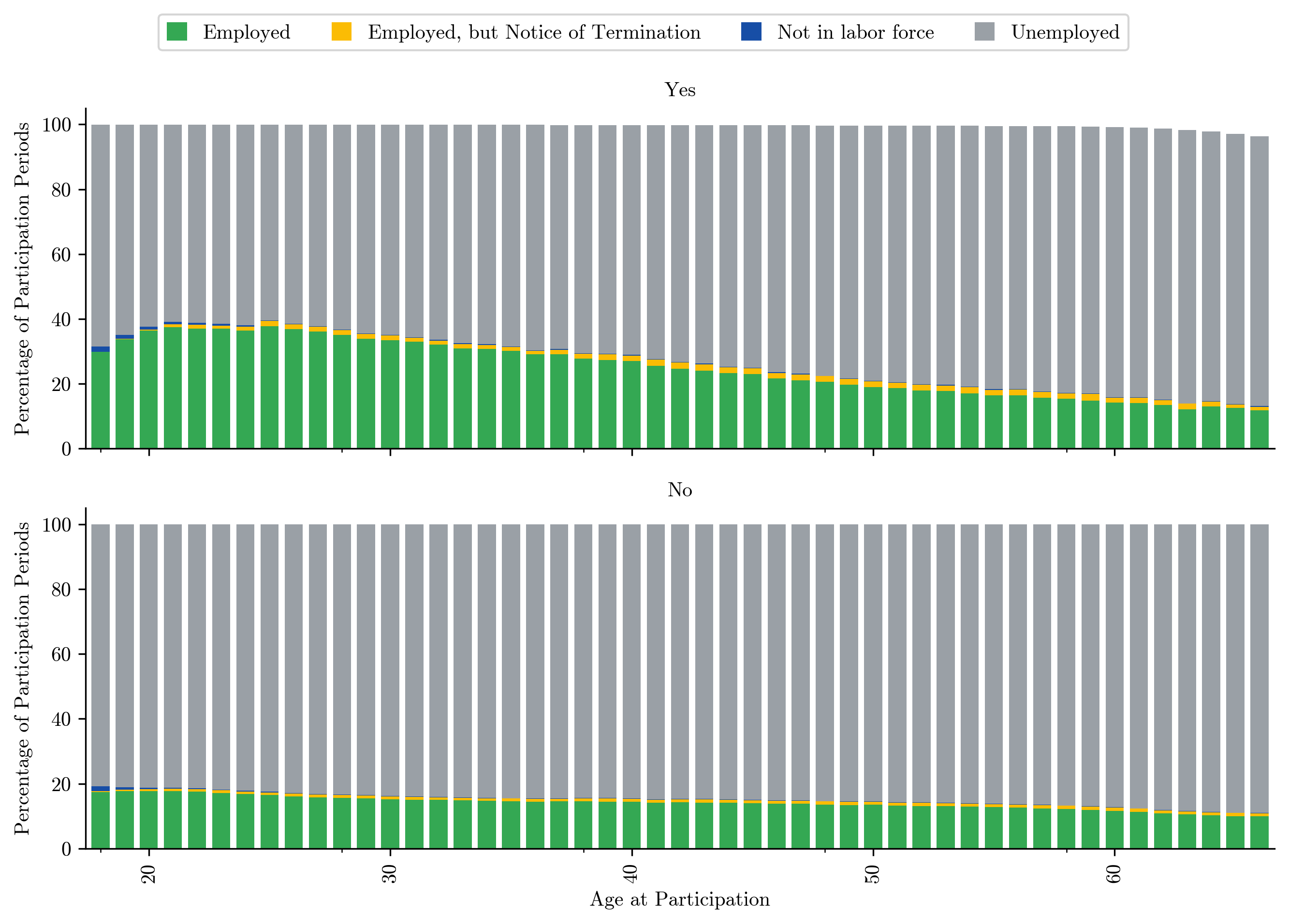}
    \caption{Percentage of participant periods by employment status and age.}
    \label{fig:image-10}
\end{figure}

Figure \ref{fig:image-10} further contextualizes \textit{Employment Status at Entry} and its strong relationship to age, particularly among participants who received formal training. Among these participants, we observe a distinct negative gradient between age and labor market attachment; while approximately 30\% of participants aged 18--24 enter the program while maintaining some form of employment, this proportion contracts steadily over time. By age 60, the cohort is characterized by high unemployment ($>80\%$). This distribution suggests that the \textit{Employment Status} variable effectively absorbs the variance associated with age-related structural exclusion. It implies that for older workers, the primary barrier to ALMP success may not be just age itself, but also the higher levels of labor market detachment that can accompany displacement later in one's career \citep{jacobson1993earnings}.

\begin{figure}[H]
    \centering
    \includegraphics[width=1\linewidth]{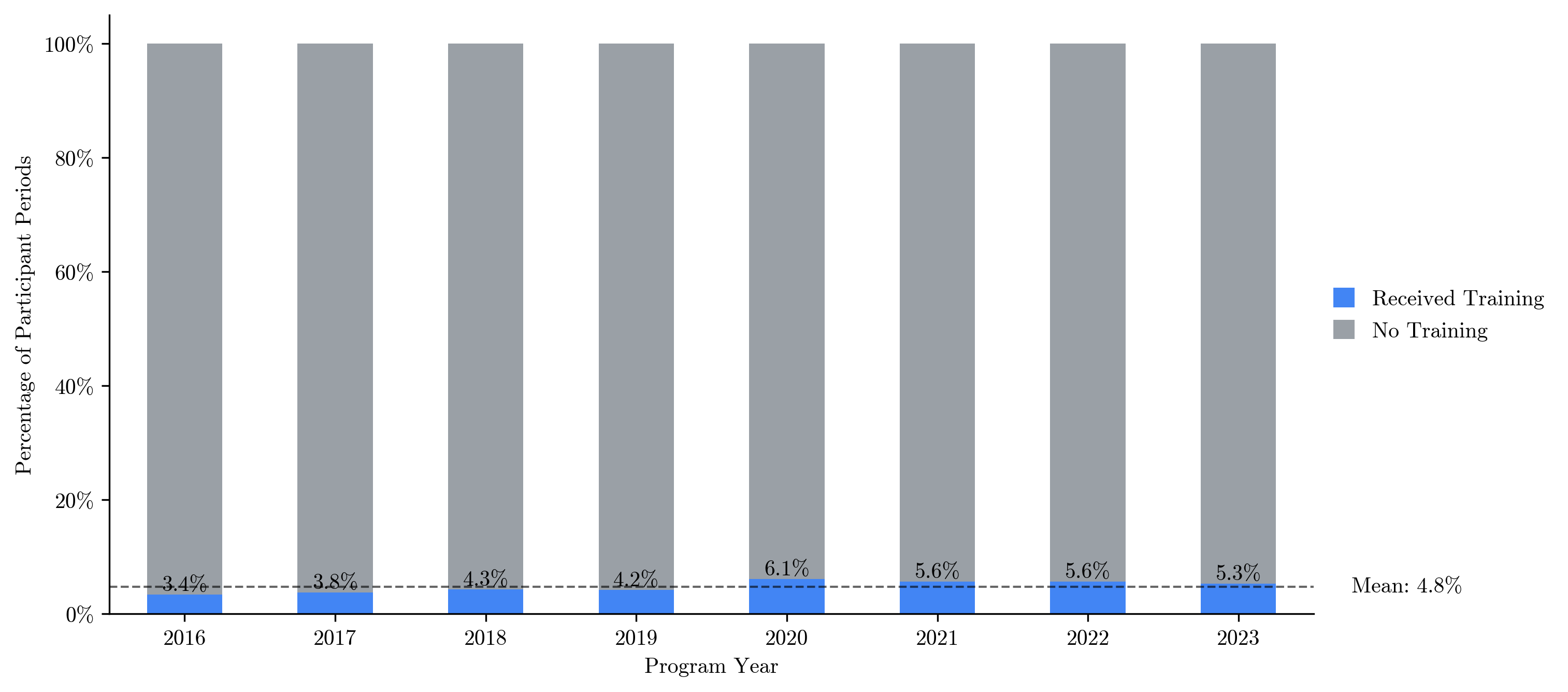} 
    \caption{Percentage of WIOA participants receiving training services by program year (2016--2023). The consistently low training rate ($<6.1\%$) highlights that deep upskilling interventions are concentrated on a small minority of the WIOA population.}
    \label{fig:training_rate}
\end{figure}

To better understand the \textit{Retrainability Index }results, we now look at the descriptive characteristics of the population group that received formal training services, which are primarily funded through the WIOA Adult program ($52.6\%$) and Dislocated Worker program ($22.8\%$). This cohort consists of approximately $1.08 $ million unique participation periods, representing approximately $4.5\%$ of the total sample in our analysis. As illustrated in Figure \ref{fig:training_rate}, the rate of training received has remained consistently low across program years, fluctuating between $3.4\%$ and $6.1\%$ with a yearly mean of $4.8\%$. 

In other words, the vast majority of people that participate in the US WIOA system do not receive formal training services. The training that is provided is heavily skewed toward hard skills acquisition, with 'Occupational Skills Training' appearing as the dominant training service type, accounting for $56.3\%$ of participation periods that received training, followed by 'Skill Upgrading' ($12.4\%$), 'Youth Occupational Skills Training' ($12.2\%$) and 'On-the-Job Training' ($9.3\%$). In reality, however, the majority of WIOA participants receive lighter touch active-labor market support, such as job matching and career counseling. 

As we previously argued, Americans that receive formal retraining are \textit{self-selecting}. Individuals who enter WIOA-funded training programs generally differ considerably from the general pool of non-training WIOA participants in both observable and unobservable ways (e.g., motivation, career intent, or eligibility for specific funding streams). As a consequence, we are reticent to make causal determinations about formal training impacts on program outcomes using this observational dataset.

\begin{figure}[H]
    \centering
    \includegraphics[width=1\linewidth]{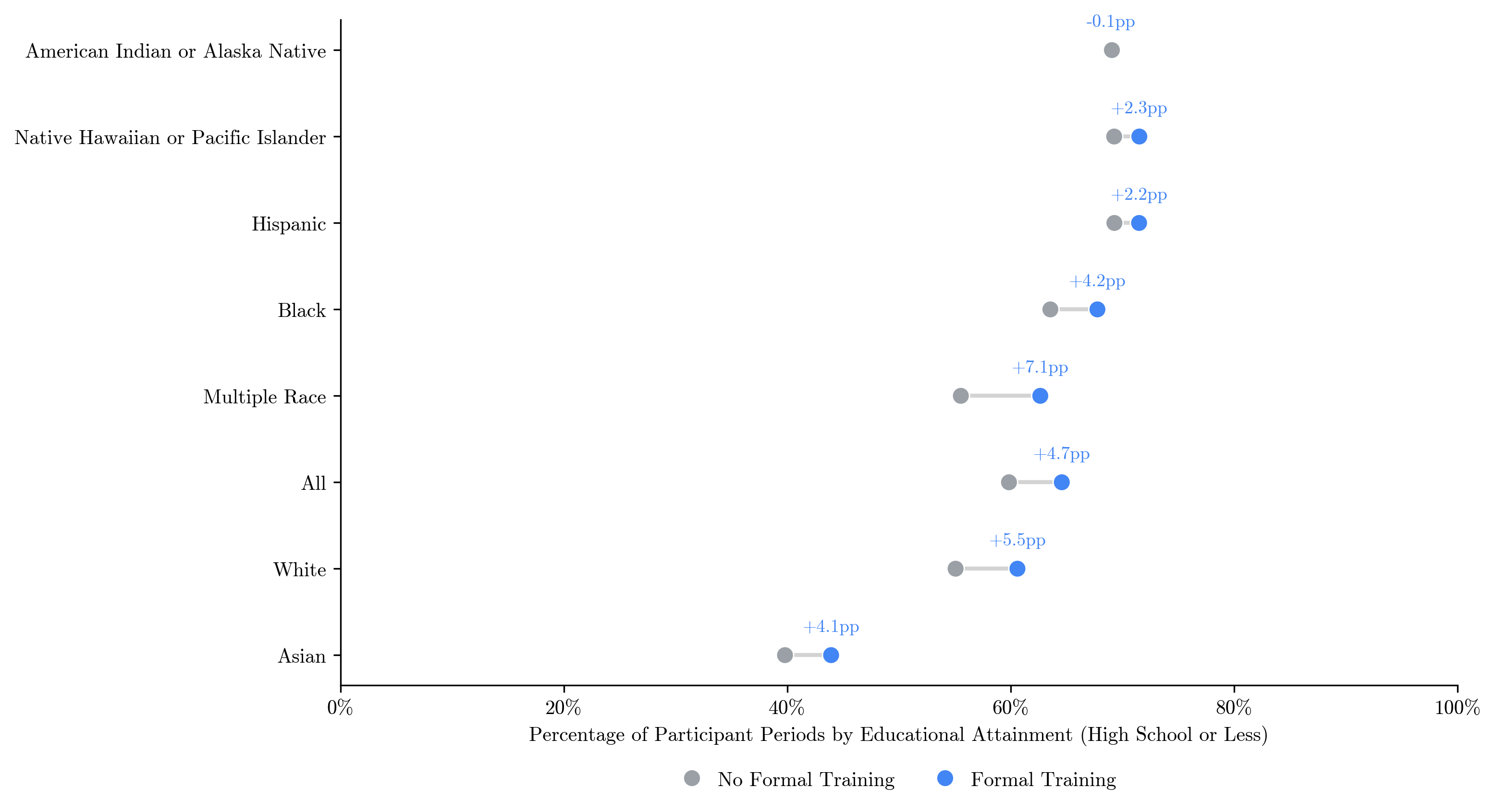} 
    \caption{Comparison of educational attainment between training and non-training cohorts. Training recipients are more likely to have a high school education or less, indicating successful targeting of lower-skilled workers.}
    \label{fig:employment}
\end{figure}

People who participate in formal training have lower levels of educational attainment than those that do not receive formal training. This finding could be explained under the logic that people with less education have more to gain from additional formal training, including classroom time. However, we equally might have assumed that prior-education at least partly predicts a person's likelihood of benefiting from further formal training. We show that formal trainers are more racially diverse ($55.9\%$ vs.\ $48.1\%$ non-White), economically vulnerable ($62.5\%$ vs.\ $20.2\%$ low income), younger (median age $30$ vs.\ $39$), and have lower educational baselines when compared with non-formal trainers. In Figure \ref{fig:employment},  we show that a larger proportion of WIOA participants who received training have a high school education or less compared to those who did not receive training. The rate of \textit{lower} educational attainment is $4.7$ percentage points higher among the training cohort than the non-training cohort. This trend holds across almost all demographic groups, with the widest gaps observed for White ($+5.5$pp) and Multiple Race ($+7.1$pp) participants. 

Among training recipients, we see evidence of potential upskilling. Prior to training, the most common origin subsectors were \textit{Administrative and Support Services} ($16.9\%$) and \textit{Food Services and Drinking Places} ($8.9\%$)---sectors often characterized by highly routine task profiles. Post-training, the share of employment in Food Services drops to $5.7\%$. Conversely, we observe inflows into the healthcare and logistics sectors: \textit{Ambulatory healthcare services} ($7.7\%$ [$+2.8\%$]), \textit{Hospitals} ($7.5\%$ [$+3.7\%$]), \textit{Nursing and Residential Care} ($7.4\%$ [$+2.0\%$]), and \textit{Truck Transportation} ($6.2\%$ [$+5.2\%$]) emerge as top destinations. 

This potential upskilling dynamic is further mirrored in occupational data. The most common pre-program occupations---\textit{Customer Service Representatives} ($4.2\%$), \textit{Cashiers} ($3.5\%$), and \textit{Laborers (Freight, Stock, and Material Movers)} ($2.6\%$)---are largely generalist roles with higher exposure to automation. Following program exit, the distribution shifts toward licensed and credentialed technical roles. \textit{Heavy and Tractor-Trailer Truck Drivers} becomes the single largest occupational category ($16.1\%$), followed by a cluster of specialized healthcare roles: \textit{Registered Nurses} ($5.2\%$), \textit{Licensed Practical/Vocational Nurses} ($4.8\%$),  and \textit{Medical Assistants} ($3.1\%$) with \textit{Customer Service Representatives} falling to $3.2\%$.

\section{Characteristics of Program Success}\label{section:Characteristics of Success}

We now examine the individual and environmental factors that correspond with successful WIOA outcomes, as measured by our index ($I_n$). Our analysis indicates that success in the U.S. workforce development system is strongly associated with structural program factors and local labor market conditions, rather than participant demographics. We note the possibility of a ceiling effect, whereby WIOA participants in more affluent regions or industries benefit \textit{less} than participants in more vulnerable positions. 

To assess the combined predictive power of observable characteristics, we employ a gradient-boosted decision tree model (XGBoost) to predict whether participation resulted in a positive outcome (defined as $\text{I}_n > 0$). We prefer gradient boosting over standard logistic regression because tree-based models naturally capture nonlinear relationships and higher-order interactions between features without requiring manual specification of interaction terms.\footnote{A logistic regression model is included in Appendix \ref{appendix:logistic-regression-results} as a baseline for comparison.} The model uses program participation and structure variables, regional economic indicators, and participant demographics and individual characteristics as features.\footnote{We exclude post-program attributes (e.g., post-training occupation, post-exit wages) from the feature set to prevent data leakage, as these components are used to calculate the target Index itself.} The model is tuned via randomized hyperparameter search and evaluated on a held-out test set. We emphasize again that this exercise is predictive rather than causal and do not claim that any feature identified as important by the model necessarily has a causal effect on retraining outcomes. Rather, our aim is to understand which observable characteristics carry the most information about whether a participant experiences a positive transition, recognizing that many of these features are likely proxies for deeper, unobserved factors.

Table \ref{tab:model_performance} summarizes the classification metrics. The model achieves an overall accuracy of $72\%$ and an ROC-AUC of 0.791. These results indicate that positive retrainability outcomes are difficult to predict based solely on the administrative features we have included in our model, which is in line with our expectations. This suggests that success in retraining—defined as wage gains and RTI shifts—is driven in part by unobserved heterogeneity such as individual motivation, specific provider quality, or idiosyncratic employer matches rather than broad demographic or regional profiles. 

\begin{table}[H]
\centering
\caption{XGBoost Classification Performance Metrics (Test Set)}
\label{tab:model_performance}
\begin{tabular}{l c c c c}
\toprule
\textbf{Class} & \textbf{Precision} & \textbf{Recall} & \textbf{F1-Score} & \textbf{Support} \\
\toprule
$\text{I}_n \leq 0$ & 0.66 & 0.58 & 0.62 & 18{,}279 \\
$\text{I}_n > 0$    & 0.74 & 0.81 & 0.77 & 27{,}715 \\
\toprule
\textbf{Accuracy} & & & 0.72 & 45{,}994 \\
\textbf{Macro Avg} & 0.70 & 0.69 & 0.70 & 45{,}994 \\
\textbf{Weighted Avg} & 0.71 & 0.72 & 0.71 & 45{,}994 \\
\toprule
\end{tabular}
\end{table}

\begin{table}[H]
\centering
\caption{Feature Importance for Predicting Positive Index Outcomes}
\label{tab:feature_importance}
\resizebox{\textwidth}{!}{%
\begin{tabular}{llllr}
\toprule
Feature & Model (Gain) Importance & Permutation Importance & SHAP Importance & Avg Rank \\
\midrule
\multicolumn{2}{l}{\textit{Program Participation \& Structure}} \\
Local Workforce Board (WDB) & 0.11 (3) & 0.01 (3) & 0.15 (3) & 3 \\
Funding Stream & 0.06 (5) & 0.0 (4) & 0.09 (4) & 4 \\
State & 0.07 (4) & 0.0 (5) & 0.07 (5) & 5 \\
Training Service Type & 0.03 (6) & 0.0 (6) & 0.04 (8) & 6 \\
Program Year & 0.02 (9) & 0.0 (8) & 0.02 (11) & 9 \\
Employment Status at Entry & 0.03 (7) & 0.0 (11) & 0.02 (12) & 10 \\
\multicolumn{4}{r}{\textit{Category Avg Rank:}} & \textbf{6.2} \\
\midrule
\multicolumn{2}{l}{\textit{Demographics \& Individual Characteristics}} \\
Pre-Program Occupation & 0.22 (1) & 0.13 (1) & 0.72 (1) & 1 \\
Pre-Program Industry Subsector & 0.15 (2) & 0.06 (2) & 0.4 (2) & 2 \\
Participant Age & 0.02 (10) & 0.0 (7) & 0.06 (6) & 7 \\
Low Income Status & 0.03 (8) & 0.0 (9) & 0.05 (7) & 8 \\
Highest Educational Level & 0.02 (16) & 0.0 (12) & 0.03 (10) & 12 \\
Race / Ethnicity & 0.02 (21) & 0.0 (14) & 0.01 (20) & 21 \\
Sex & 0.02 (22) & -0.0 (15) & 0.01 (18) & 21 \\
\multicolumn{4}{r}{\textit{Category Avg Rank:}} & \textbf{10.3} \\
\midrule
\multicolumn{2}{l}{\textit{Regional Economic Indicators}} \\
WDB Unemployment Rate & 0.02 (12) & 0.0 (10) & 0.03 (9) & 11 \\
WDB Diversity Index & 0.02 (11) & -0.0 (18) & 0.01 (17) & 14 \\
WDB Population (per sq km) & 0.02 (14) & -0.0 (19) & 0.01 (13) & 14 \\
WDB Median Income Level & 0.02 (13) & -0.0 (17) & 0.0 (21) & 16 \\
WDB High Household Debt-to-Income Ratio & 0.02 (18) & -0.0 (21) & 0.01 (15) & 18 \\
WDB Median Age & 0.02 (20) & -0.0 (20) & 0.01 (14) & 18 \\
WDB Population & 0.02 (17) & -0.0 (22) & 0.01 (16) & 21 \\
WDB Metro Status & 0.02 (19) & -0.0 (16) & 0.0 (22) & 22 \\
\multicolumn{4}{r}{\textit{Category Avg Rank:}} & \textbf{16.8} \\
\bottomrule
\end{tabular}
}
\end{table}

Table \ref{tab:feature_importance} presents the hierarchy of feature importance measures, isolating the observable characteristics that are associated with positive index outcomes. We report \textit{Model Feature Importance} (measured by Gain, which quantifies the improvement in split quality when using each feature), \textit{Permutation Feature Importance} (which measures the decrease in model performance when feature values are randomly shuffled), and \textit{SHAP Feature Importance} (which quantifies each feature's average contribution to predictions across all samples based on Shapley values from cooperative game theory) \citep{lundberg2017unified}. We rank each importance metric independently and report the average of these ranks. The analysis indicates that structural program factors possess slightly higher predictive power than demographic characteristics, with average ranks of 6.2 and 10.3 respectively.

Substantively, the model indicates that success within the U.S. retraining system may be primarily dictated by a worker's occupational starting point and their local administrative geography. A participant's prior occupation and industry serve as the strongest predictors of their trajectory, largely because these factors define their baseline capacity for wage recovery; workers displaced from low-wage and routine roles naturally possess a higher mathematical ceiling for marginal improvement compared to high-earners facing diminishing returns. 

Geography and programming quality appear to act as additional significant predictors of success. The specific local workforce board (WDB) and U.S. state managing the intervention are strongly associated with programmatic success outweighing broader macroeconomic conditions like regional unemployment or median income. A worker's ability to successfully transition appears deeply path-dependent and bounded by highly localized factors---such as the quality of administrative leadership, the specific funding stream utilized, and regional employer networks \citep{barnow2015employment}. In other words, successful WIOA outcomes appear partly contingent on the existence of local employment opportunities to retrain into, as well as the quality of the program offered in that locality.

These dynamics are corroborated by our  logistic regression model detailed in Appendix C. Consistent with the gradient-boosted tree findings, the regression identifies a participant's structural starting point (Pre-Program Occupation and Industry Subsector) and administrative geography (State and Local WDB) as most associated with positive outcomes. 

We note that this predictive hierarchy should be carefully contextualized within the mechanical behavior of gradient-boosted decision trees, which aggressively push aside correlated features. The consistently low importance of demographic variables (\textit{Race/Ethnicity}, \textit{Sex}, \textit{Participant Age}) and \textit{Highest Educational Level} does not imply an equitable or credentials-blind labor market. Rather, because the U.S. economy is characterized by deep occupational segregation and geographic stratification, these structural inequalities are effectively absorbed by variables like \textit{Pre-Program Occupation} and \textit{Local Workforce Board}. Once the algorithm splits on a participant's job title and jurisdiction, explicit demographic tags or degree statuses yield minimal additional information gain. Similarly, the geographic identifiers function as spatial fixed effects that swallow the variance of granular economic statistics. Consequently, the model surfaces the most proximate indicators of human capital and administrative environment, systematically masking the deeper confounding variables that initially sort workers into those specific roles and regions.

To improve the granularity of the primary model, we decompose the feature space into targeted sub-models, isolating specific categories of variables to observe their relative importance (detailed results are provided in Appendix \ref{app:feature_tables}). 

Focusing exclusively on \textit{Program Participation \& Structure}, we find that the \textit{Funding Stream} (Avg Rank 1) and local administrative boundaries (\textit{Local Workforce Board (WDB)}, Avg Rank 2) act as the primary features corresponding to program success. The predictive dominance of the funding stream indicates that the targeted structural design and resources of a specific intervention (e.g., Youth vs. Dislocated Worker) may dictate outcomes more heavily than generalized training types or immediate labor market attachment. Meanwhile, the persistent importance of the WDB reinforces the proxy effect of unobserved local administrative quality and employer networks.

In our isolated analysis of \textit{Regional Economic Indicators}, we highlight the importance of urban density and labor market tightness. \textit{WDB Population (per sq km)} (Avg Rank 1) and the \textit{WDB Unemployment Rate} (Avg Rank 2) emerge as the strongest regional predictors, higher than the \textit{WDB Diversity Index} (Avg Rank 3). This suggests again that successful labor market transitions are heavily reliant on the sheer volume of alternative employment opportunities and aggregate labor demand within a given commuting zone. Finally, evaluating \textit{Demographics and Individual Characteristics} in isolation suggests that \textit{Participant Age} and \textit{Low Income Status} are the most informative. This provides evidence for both an ``age penalty''---where older workers face steeper time-horizon barriers to recouping human capital investments \citep{jacobson1993earnings}---and the wage catch-up dynamic inherent to economically vulnerable populations.

A question, however, is what a `successful' WIOA experience means in practice. Are positive outcomes driven by meaningful shifts in wages, reduced exposure to automation, or both? We argue that for the formal training and non-formal training cohorts, wage recovery seems to play a disproportionate role in shaping positive index scores.

\begin{figure}[H]
    \centering
    \includegraphics[width=1\linewidth]{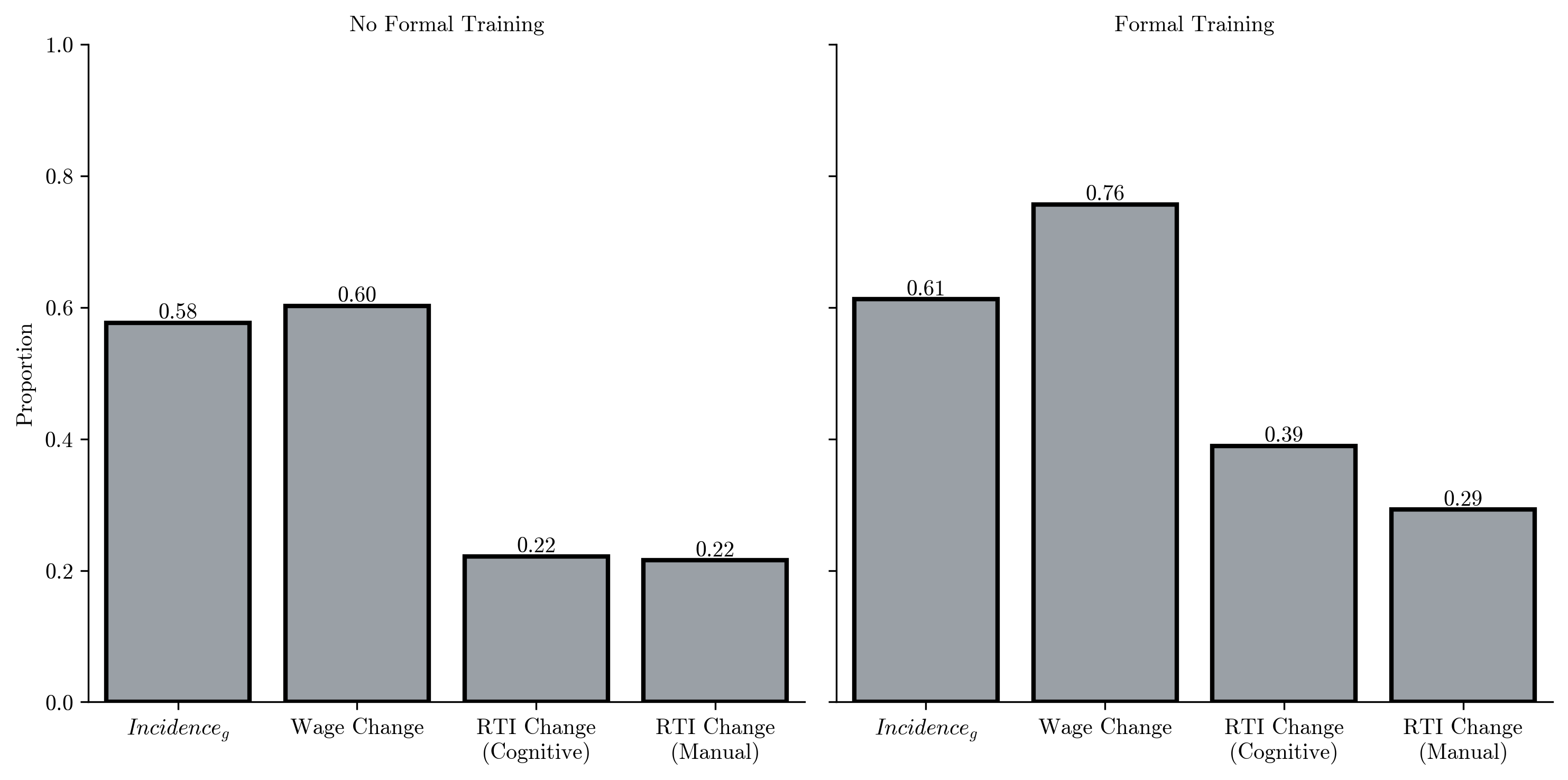}
    \caption{Proportion of positive outcomes by index subcomponent for non-formal and formal training cohorts.}
    \label{fig:wage_rti_index_outcomes}
\end{figure}

Figure \ref{fig:wage_rti_index_outcomes} decomposes the frequency of positive contributions by subcomponent, indicating a contrast in the underlying drivers of participant outcomes. While the subset of participants with complete occupational data exhibits notably stronger overall wage growth than the broader WIOA population, this economic recovery rarely coincides with improved resilience to technological automation. For the non-formal-training cohort (those receiving only basic career services), a majority record positive wage trajectories upon re-entry. However, their corresponding RTI change remains minimal. Since only a small fraction of non-trainers transition into roles with lower routine task intensity, the data suggests that while basic WIOA services correspond to successful labor market re-attachment and wage catch-up, they largely churn workers back into occupations with identical or worse automation vulnerabilities.

Participants who received formal retraining were more likely to have positive outcomes. However, this is primarily driven by positive wage changes at 0.76, indicating that over 75\% of trainees achieved real earnings growth. Even with formal training, the share of participants who experienced a reduction in the routine nature of their work increased only slightly from 0.22 to 0.39 and 0.29 for routine cognitive and routine manual tasks respectively. This suggests that many formal training participants are simply entering better-paid but equally automation-susceptible roles. 

The divergence in outcomes should also be contextualized by the baseline disparities between the two cohorts. As illustrated in the pre-participation wage trends, participants who received formal training consistently entered the program with significantly lower earnings histories than their non-training counterparts. In 2023, for instance, the gap in pre-program quarterly wages was approximately \$2,592, with trainees earning roughly half of what non-trainees earned prior to entry. 

We argue the strong performance of trainees on the Index is partly driven by a structural ``catch-up'' dynamic. Since these participants originate from the lower tail of the wage distribution, they possess greater ``headroom'' for marginal improvement. This pattern aligns with a classic Ashenfelter dip: workers often enter training programs following a transitory earnings decline, implying that a significant portion of their subsequent wage recovery is likely mean reversion rather than a direct treatment effect of the training itself---a dynamic we formally isolate and confirm via matched estimates below \citep{ashenfelter1978estimating, card1988estimating}. For a worker entering with near-poverty wages, even a transition into a modest entry-level role registers as a substantial positive gain ($\Delta w$). Conversely, the non-training cohort enters with higher prior earnings, making them statistically more susceptible to ``wage scars''---the well-documented phenomenon where displaced high-earners re-enter the workforce at lower pay levels.

Consequently, the high Retrainability Index scores for the training group should be interpreted with caution. While they indicate successful re-attachment to the labor market, they do not necessarily prove that training generates a unique skill premium that outperforms the broader market. Rather, they reflect that WIOA training services effectively function as a floor for the most vulnerable, facilitating a reversion to mean earnings, while the non-training population faces the steeper challenge of replicating prior incomes.

\begin{figure}[H]
    \centering
    \includegraphics[width=1\linewidth]{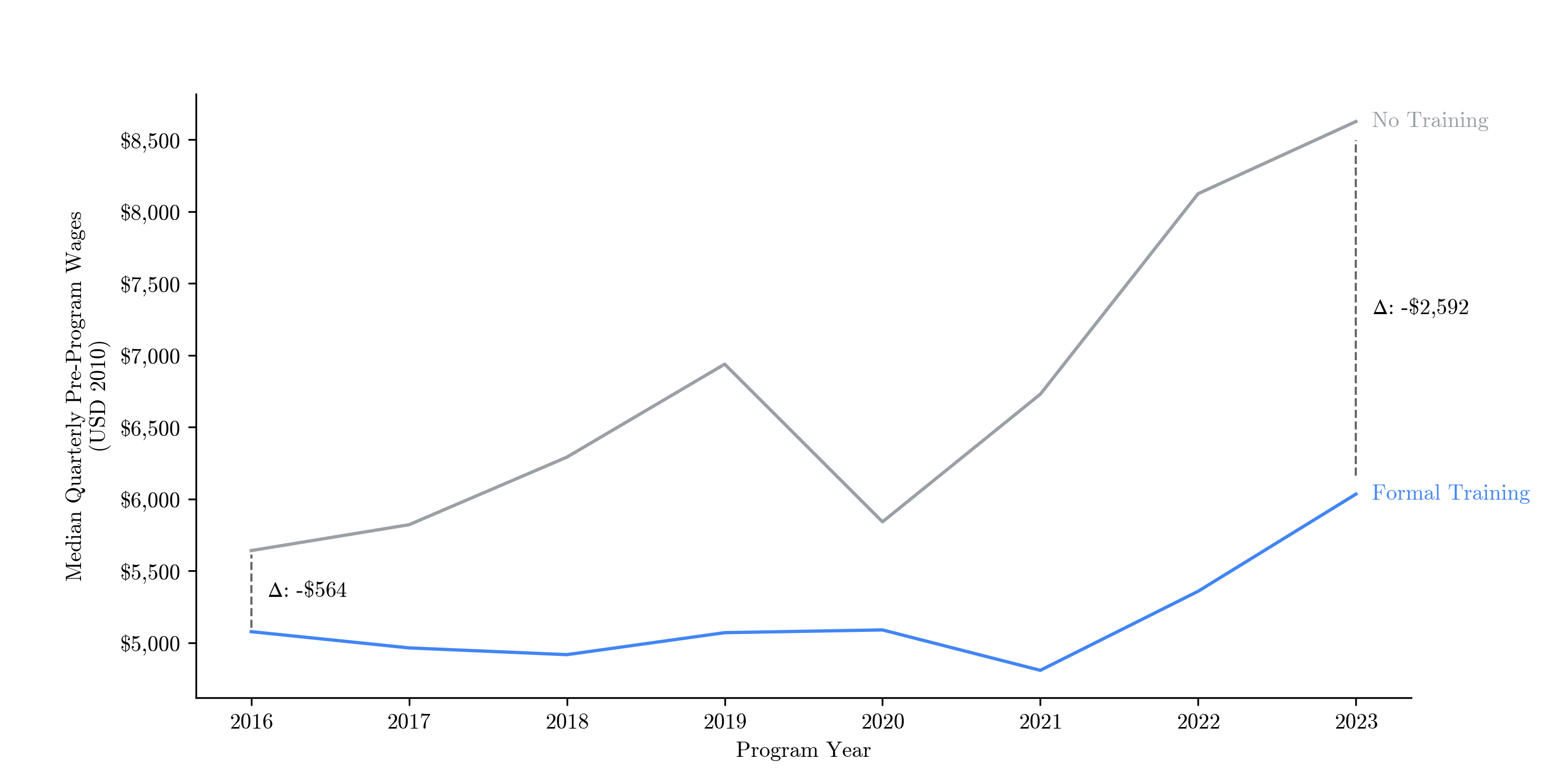}
    \caption{The median (mean) quarterly wages pre-program participation by program year for participation periods with $I_n$ (Subsector).}
    \label{fig:wages by year (subsector)}
\end{figure}

\begin{figure}[H]
    \centering
    \includegraphics[width=1\linewidth]{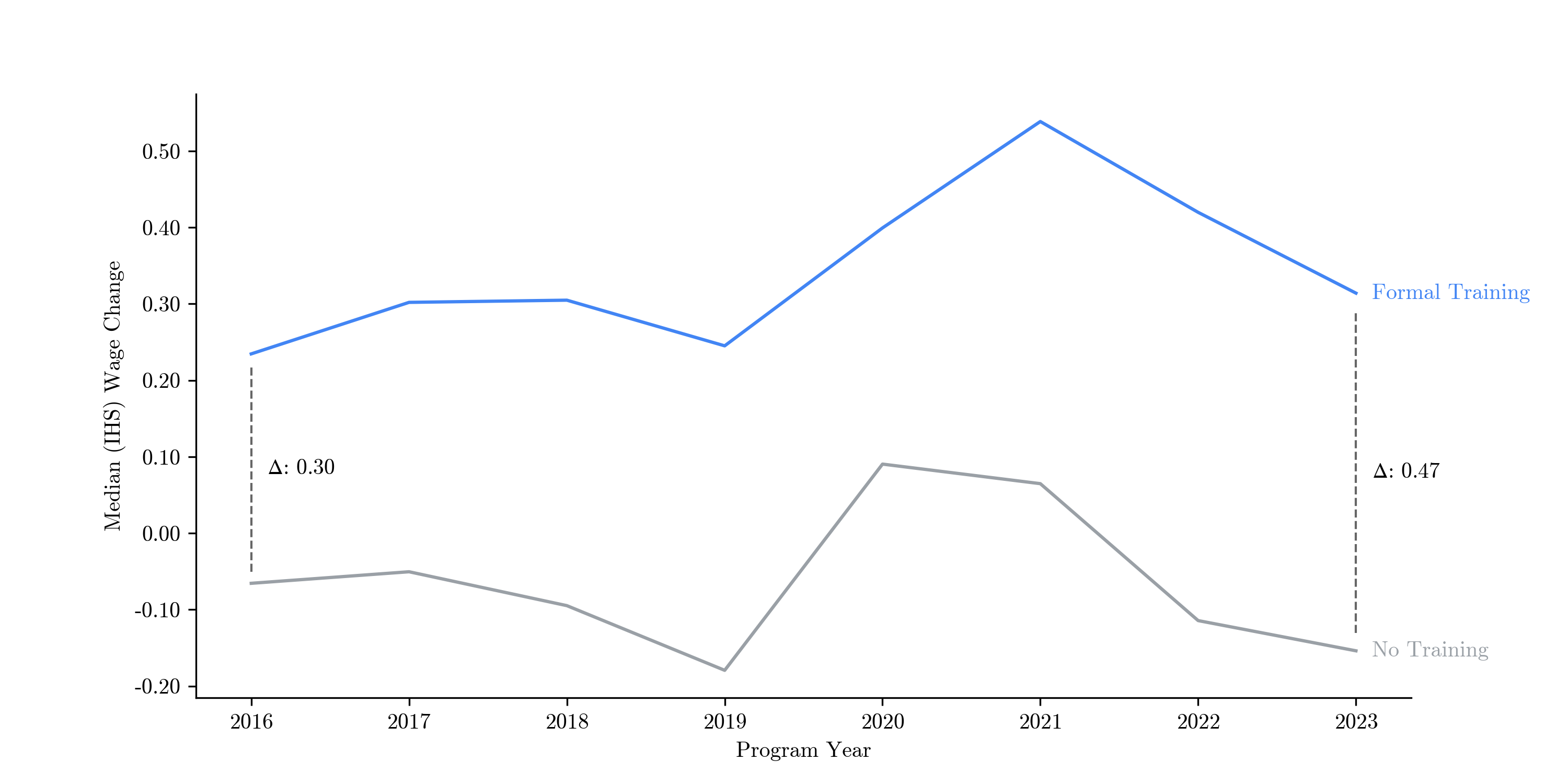}
    \caption{Median difference in inverse hyperbolic sine (IHS)-transformed average quarterly wages between post- and pre-WIOA program participation (see Appendix \ref{appendix:subcomponent-definitions} for details), by program year for participation periods with $I_n$ (Subsector).}
    \label{fig:SINE_subsector}
\end{figure}

This finding is unsurprising when viewed in conjunction with the predictive hierarchy previously noted. As suggested earlier, \textit{Pre-Program Occupation} serves as a potential determinant of index outcomes specifically because it sets the structural ``ceiling'' for potential gains. Indeed, Figures \ref{fig:wages by year (subsector)} and \ref{fig:SINE_subsector} explicitly illustrate this dynamic. Read together, they show trainees entering with earnings roughly half those of non-trainees and then experiencing large post-program absolute dollar gains. For the non-training cohort, who held higher pre-participation wages, the lack of positive index movement likely reflects the difficulty of lateral or upward mobility without new credentialing. For the training cohort, the intervention offers a potential `catch-up' effect, moving workers from a low-level equilibrium (unemployment or low-wage work) to a higher one. 

\begin{figure}[H]
    \centering
    \includegraphics[width=1\linewidth]{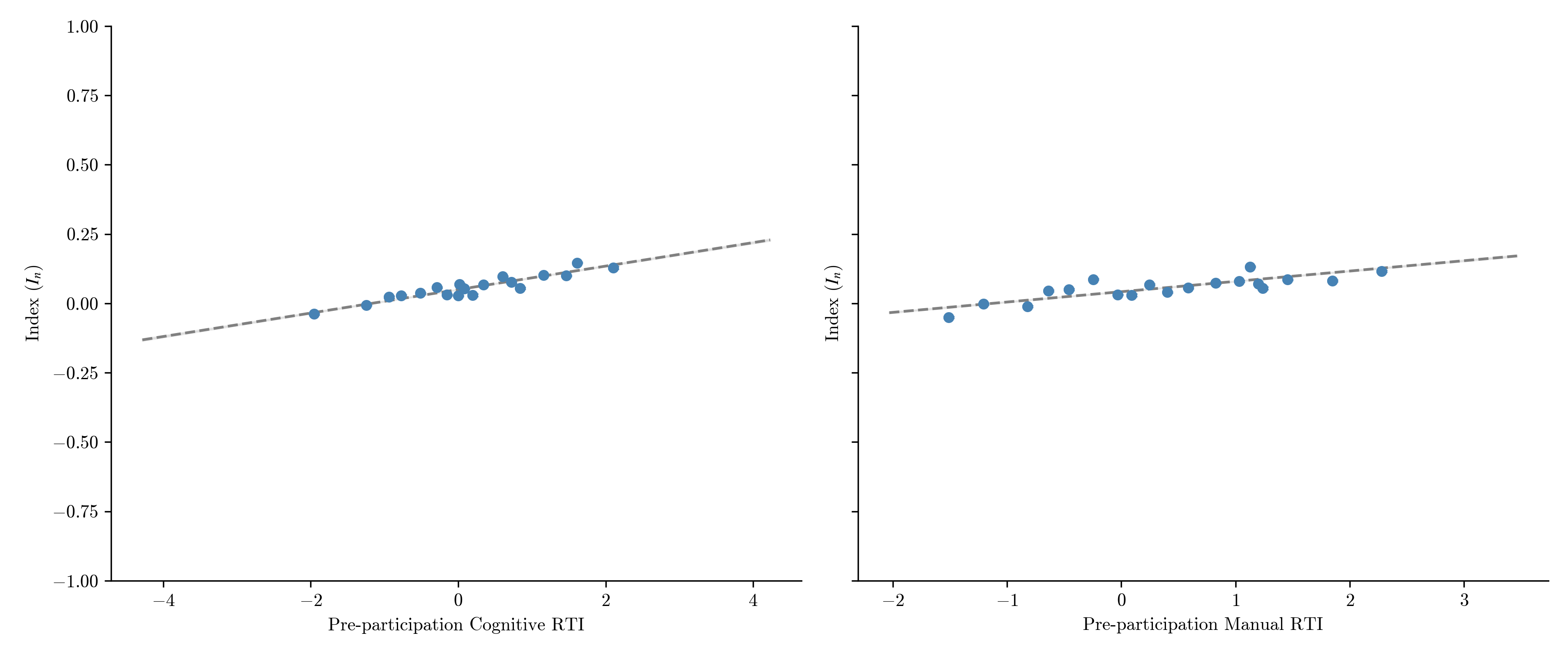}
    \caption{Binscatter of pre-participation Routine Cognitive and Manual Task Intensity against the Index $I_n$. Each point represents the mean index value within an equal-frequency bin. The dashed line shows linear regression (OLS) estimates fitted to the underlying (non-binned) data. For cognitive RTI: $\hat{\beta} = 0.042$, $R^2 = 0.060$. For manual RTI: $\hat{\beta} = 0.037$, $R^2 = 0.050$.}
    \label{fig:RTI v Index}
\end{figure}

\begin{figure}[H]
    \centering
    \includegraphics[width=1\linewidth]{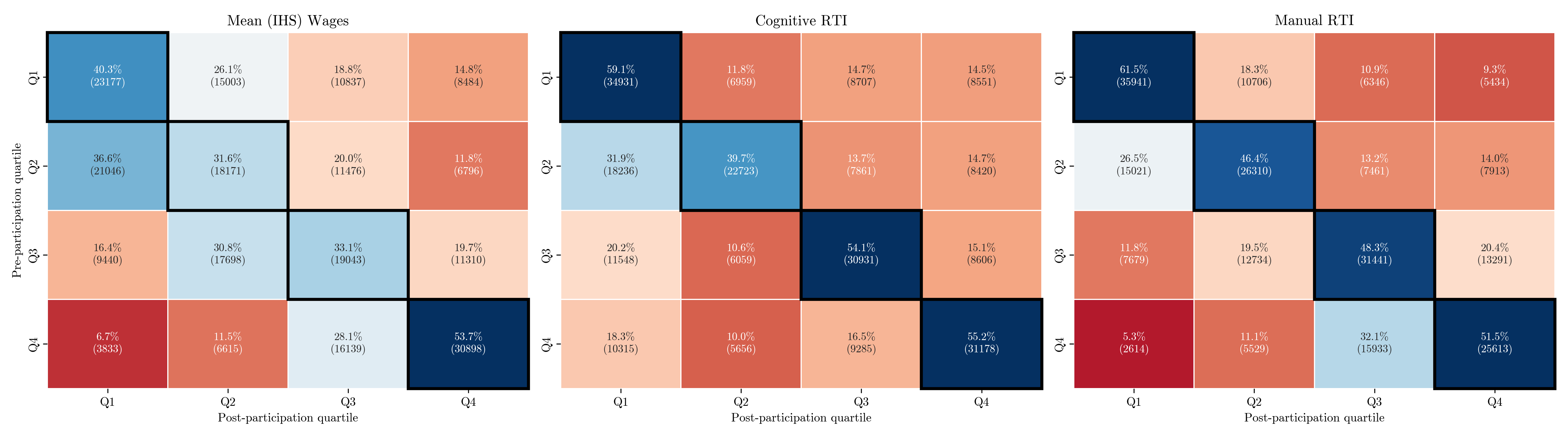}
    \caption{Quartile transitions for Mean (IHS) Wages, Routine Cognitive Task Intensity (RTI), and Routine Manual Task Intensity (RTI). Each cell shows the percentage of participation periods that started in pre-participation quartile QX and ended in post-participation quartile QY, where rows sum to 100\%. Shading reflects deviation from the 25\% benchmark expected under independence: red indicates below 25\% and blue indicates above 25\%.}
    \label{fig:RTI Pre-Post}
\end{figure}

If we find evidence of potential ``catch-up'' gains and a reversion to the mean in wages, do we see the same thing in RTI scores? In Figure  \ref{fig:RTI v Index}, we look at the relationship between a person's pre-WIOA RTI scores and Index outcomes, showing that individuals who start with higher RTI scores (i.e. their work is more routine and susceptible to automation) experience better Index outcomes. This is again likely because they possess greater headroom for RTI score reductions in post-WIOA employment outcomes. 
A key question that emerges from this, however, is how pre-program RTI scores condition post-program RTI in an \textit{absolute} sense, as a opposed to relative to a participant's previous score.  In Figure \ref{fig:RTI Pre-Post}, we can see clearly that a person's previous RTI score and wages is strongly associated with their post-WIOA RTI scores, in line with expectations and our observation that a significant portion of participants simply return to their original occupation or industry. This provides important context for interpreting the Index. Participants who enter the program with low RTI scores (i.e., less automation-susceptible roles) inherently face a ceiling effect. Consequently, they may register lower or negative Index scores, even if their absolute post-program wages and RTI remain vastly superior to those of the broader participant pool.

As an additional test,  we employed Propensity Score Matching (PSM), inspired by \citet{hyman2025retrainable}  in their concurrent research, to estimate the underlying impact of these interventions (this approach is detailed fully in Appendix \ref{appendix:psm_estimates}). This model finds potential evidence of statistically significant positive impacts of  \textit{Registered Apprenticeships} on Index outcomes. However, we argue this approach is ultimately  unable to account for unobservables, which limits our ability to make causal claims.  

\section{Subgroup Analysis}

Next we present the aggregated Retrainability Index results across key programmatic and demographic subgroups. To interpret these findings, we report six summary statistics for each grouping:

\begin{itemize}
    \item \textbf{Incidence} ($Incidence_g$): The share of participation periods within the group that achieved a positive outcome (defined as an Index score $I_n > 0$). This functions as a ``headcount ratio'' for successful transitions.
    \item \textbf{Intensity} ($Intensity_g$): The average index score among only participation periods that achieved a positive outcome. This measures the magnitude or ``depth'' of the success for positive outcomes.
    \item \textbf{Individual Index} ($\overline{I_n}$): The mean Index score for an individual participation period.
    \item \textbf{Subcomponents} ($\overline{I_{n}^{W}}$, $\overline{I_{n}^{C}}$, $\overline{I_{n}^{M}}$): The normalized mean scores for Wage Change ($W$), Routine Cognitive Task shift ($C$), and Routine Manual Task shift ($M$). Values above $0$ indicate average growth/improvement, while values below $0$ indicate regression.
\end{itemize}

\TableReceivedTrainingOccupationIpw

The primary bifurcation in the dataset exists between the training and non-training cohorts. As detailed in Table \ref{table:received_training_occupation_ipw}, participants who received substantive training interventions achieved an incidence score of 0.61, representing a modest improvement over the 0.58 baseline observed in the non-training population. Decomposing this result reveals the mechanism of impact. While both cohorts experience positive wage momentum, the ``training premium'' remains distinctly associated with a stronger wage trajectory ($\overline{I_{n}^{W}}$ rises from 0.10 for non-trainees to 0.17 for trainees). In contrast, shifts in RTI show more subdued improvements ($\overline{I_{n}^{C}}$ rises marginally from 0.01 to 0.02, and $\overline{I_{n}^{M}}$ from 0.01 to 0.06). This suggests that while formal training is associated with accelerated earnings restoration, it is not broadly associated with a sectoral pivot away from routine work.

When comparing these broader trends against the occupation-level estimates, we should contextualize the administrative selection bias inherent to the narrower sample. It is highly likely that the 1.2\% of records with complete pre- and post-program occupational codes are concentrated within specific geographic districts with superior reporting infrastructure. Consequently, the local workforce boards or states that possess the institutional capacity and resources to accurately track granular occupational data are also more likely to oversee higher-performing interventions or operate within stronger regional labor markets, pulling the average incidence of success upward relative to the broader subsector population.

\TableEmploymentStatusOccupationIpw

Table \ref{table:employment_status_occupation_ipw} highlights the role of labor market attachment. Participants who were \textit{Employed} at the time of entry achieved the highest incidence of positive outcomes (0.61), compared to 0.57 for those who were \textit{Unemployed}. One explanation for this is that already-employed workers likely possess unobserved soft skills, recent on-the-job experience, and professional networks that facilitate smoother occupational transitions. Conversely, participants classified as \textit{Not in Labor Force}\footnote{WIOA defines \textit{Not in the labor force} as a participant who is unemployed and not actively looking for work at the time they entered the program.} experienced the lowest incidence of positive outcomes (0.52). Since all participants in our index calculation must have recorded wages in the three quarters prior to participation, this lower success rate is not a symptom of long-term unemployment, but rather likely reflects the steep challenges of overcoming sudden, acute labor market detachment or other structural barriers that cause individuals to stop actively searching for work.

\TableProgramYearOccupationIpw

Table \ref{table:program_year_occupation_ipw} hints at the sensitivity of retrainability to the macroeconomic cycle. As revealed by the occupation-level data, the trajectory of program efficacy is characterized less by a steady recovery and more by a temporary pandemic-era spike followed by a decline. Incidence rates peaked at a high of 0.65 in Program Year 2020 before reverting back to roughly their pre-pandemic levels. This volatility may indicate that the index is highly elastic to aggregate labor demand. In exceptionally tight or volatile labor markets (e.g., the immediate post-pandemic reopening), employer barriers to entry fall, significantly boosting transition scores independent of training quality. The subsequent decline as the macroeconomic environment shifted warns against interpreting year-over-year index changes as purely a function of programmatic improvement.

\TableTrainingServiceOccupationIpw

Table \ref{table:training_service_occupation_ipw} shows that employer-linked models are associated with some of the most positive outcomes in the dataset. We again find evidence that \textit{Registered Apprenticeships} and \textit{Customized Training} record the highest incidence rates (0.71 for both) alongside strong wage intensity ($\overline{I_{n}^{W}}=0.10$ and $0.17$, respectively). This aligns with prior literature suggesting that training models which bypass search friction by binding skill acquisition directly to employment---such as manufacturing credentials and employer-linked models---often correspond with better outcomes than traditional classroom-based approaches \citep{brown2024impact}.  Meanwhile, lighter-touch interventions such as \textit{Job Readiness Training} correspond to a surprisingly high incidence (0.73) coinciding with large wage intensity ($\overline{I_{n}^{W}}=0.31$). However, this group also exhibits a positive shift in cognitive routine task intensity ($\overline{I_{n}^{C}}=0.08$), indicating a transition into more routine work. We argue that, while lighter-touch services coincide with rapid labor market re-entry and wage catch-up, this pattern typically reflects workers entering highly routine roles rather than an association with genuine upskilling against automation.

\begin{figure}[H]
    \centering
    \includegraphics[width=1\linewidth]{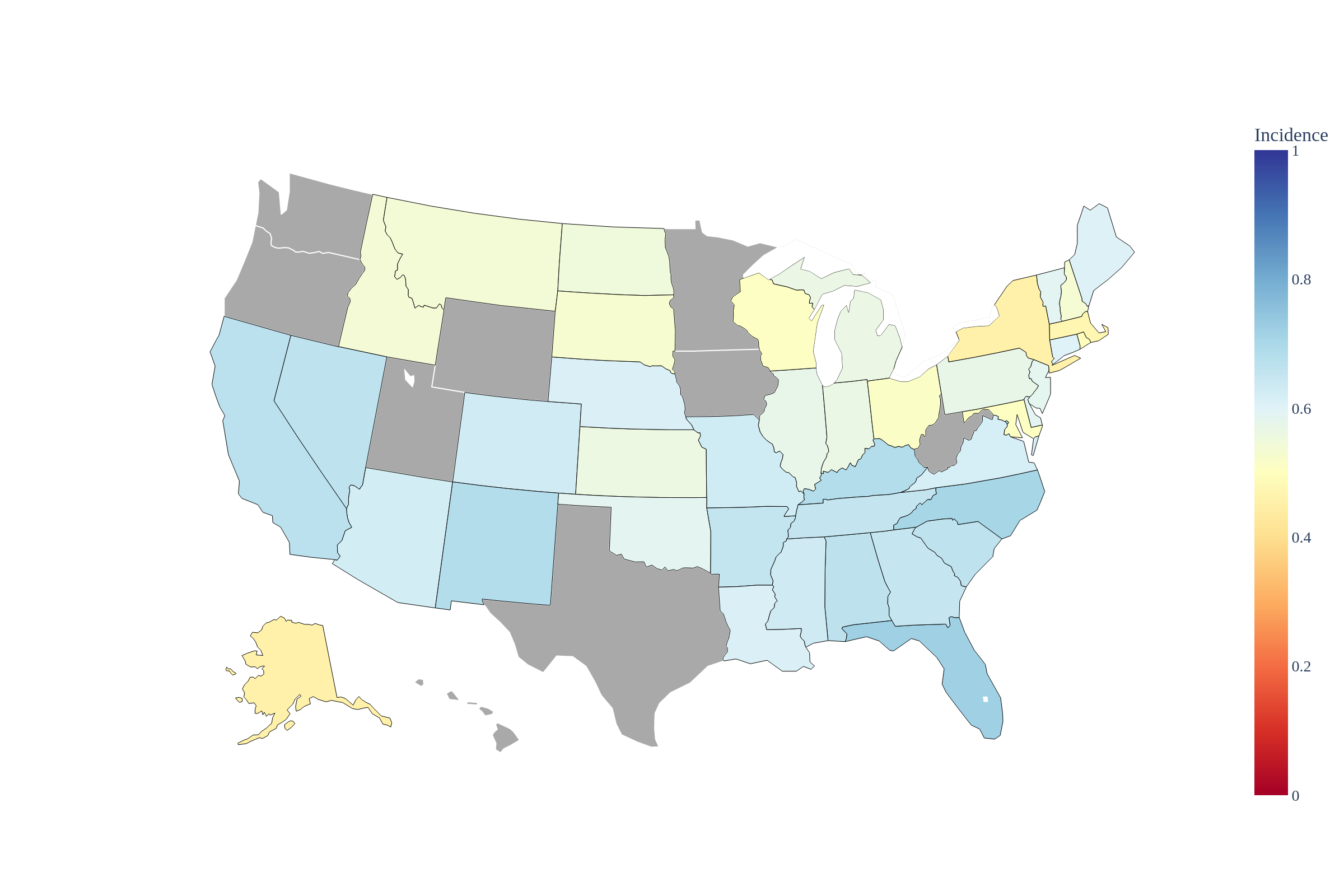}
    \caption{$\text{Incidence}_{g}$ (Occupation IPW) by State. $\text{Incidence}_g$ is not calculated for states in gray with less than 100 participation periods with $I_n$.}
    \label{fig:image-16}
\end{figure}

Our analysis of state-level outcomes reveals significant geographic heterogeneity. We observe that states with lower baseline costs of living and historically lower wages---such as Alabama (0.67), Mississippi (0.63), and Arkansas (0.65)---often achieve higher incidence rates than wealthy, high-skill economies---such as New York (0.46) and Massachusetts (0.47). This inverse relationship appears to support the ``catch-up'' hypothesis identified earlier in the participant-level analysis. In lower-wage labor markets, the threshold for a positive index outcome is structurally lower; a transition from unemployment to a standard entry-level role represents a significant relative wage gain ($\Delta w$). Conversely, in high-cost, high-wage states like New York, displaced workers face a potential ``ceiling effect.'' Replicating a previous high-status salary requires a far more complex professional transition, making the marginal gains captured by the index more difficult to achieve. The low wage intensity scores in high-skill states (e.g., NY $\overline{I_{n}^{W}} =  0.09$ vs. FL $\overline{I_{n}^{W}} = 0.31$) suggest that displacement in elite labor markets often results in steeper relative wage scarring.

\begin{figure}[H]
    \centering
    \includegraphics[width=1\linewidth]{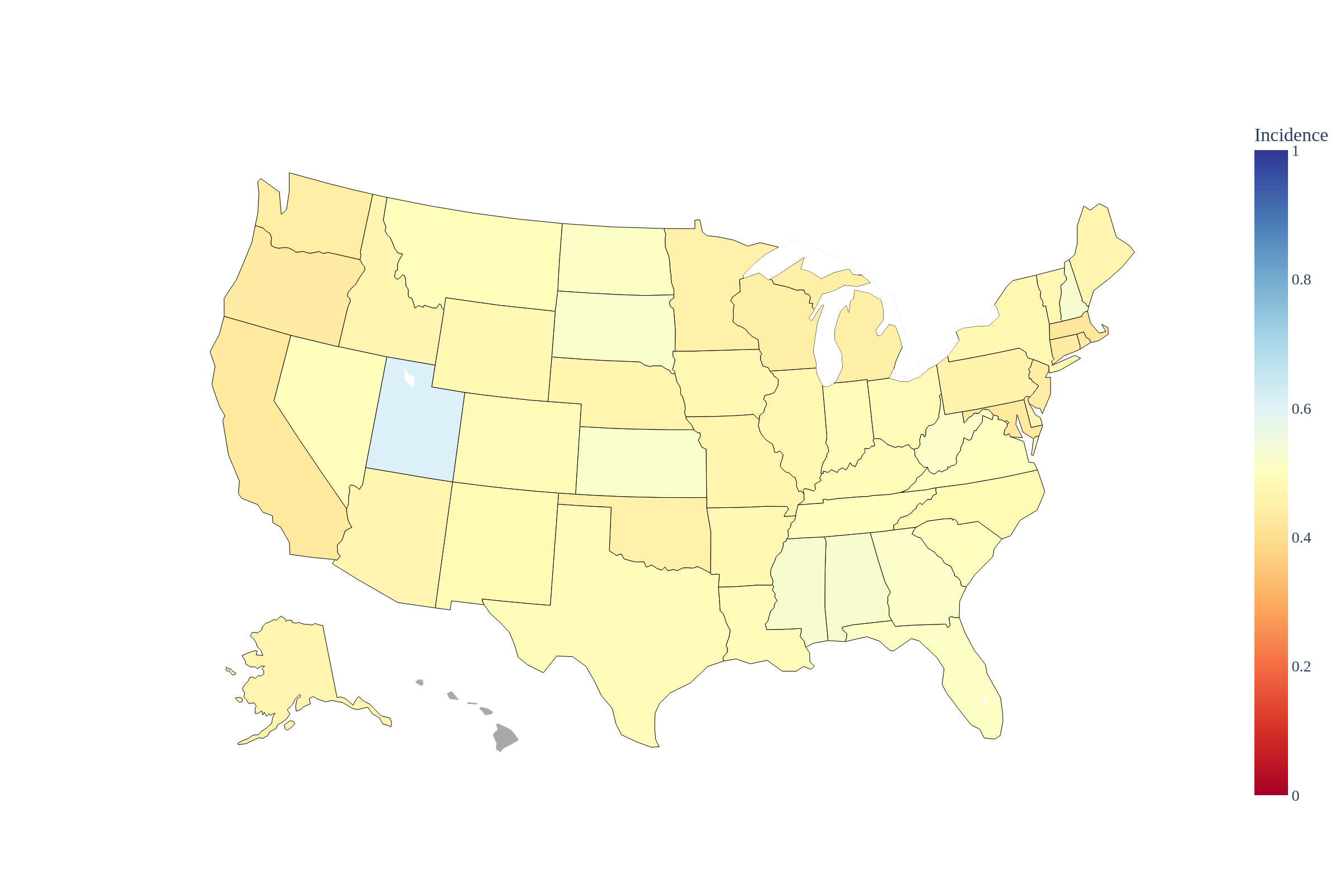}
    \caption{$\text{Incidence}_{g}$ (Subsector) by State. $\text{Incidence}_g$ is not calculated for states in gray with less than 100 participation periods with $I_n$.}
    \label{fig:image-16-subsector}
\end{figure}

Comparing these primary occupation-level estimates against the broader subsector-level data (Figure \ref{fig:image-16-subsector}) reinforces the administrative selection bias discussed previously. The occupation-level map displays significant data sparsity—with several states lacking the granular reporting infrastructure required to meet the 100-participant threshold—yet exhibits higher overall success rates (heavily concentrated in the 0.60 to 0.74 range). In contrast, the subsector map provides nearly complete national coverage but reflects more subdued, conservative incidence rates (largely clustering between 0.40 and 0.55). Despite this level shift, the relative geographic trend holds: within the broader subsector population, lower-wage states like Mississippi (0.53) and Alabama (0.52) continue to report higher transition success than wealthy states like New York (0.48) and Massachusetts (0.42), showing the "ceiling effect" is a robust structural phenomenon.

\TableAgeOccupationIpw

Meanwhile, demographic decomposition of the index reinforces our theory of diminishing marginal returns based on baseline advantage. Table \ref{table:age_occupation_ipw} illustrates a clear negative gradient between age and retrainability. Incidence rates peak among early-career participants—reaching 0.75 for 15-19-year-olds and stabilizing around 0.65 for those in their early 20s—before declining steadily to 0.50 by age 60. This age penalty is driven by a divergence in wage growth potential; while younger workers benefit from rapid initial human capital accumulation, older workers barriers to replicating their earnings and wage recovery. 

\TableDemographicsOccupationIpw
As shown in Table \ref{table:demographics_occupation_ipw}, while incidence rates exhibit some variation across \textit{Race/Ethnicity}---with Black participants recording an incidence of 0.62 compared to 0.56 for both Hispanic and White participants---the magnitude of these positive outcomes remains notably stable. Specifically, the wage intensity ($\overline{I_{n}^{W}}$) for all three groups clusters tightly between 0.10 and 0.11. In contrast, the distinction between \textit{Low Income Status} and non-low-income participants is pronounced across both metrics. Low-income participants record a higher incidence rate (0.63 vs. 0.56) and double the wage intensity ($\overline{I_{n}^{W}} = 0.16$ vs. $0.08$) of their non-low-income counterparts. This divergence suggests that the WIOA system's association with "catch-up" wage growth is most acutely observed along lines of baseline economic vulnerability rather than demographic categories like race or sex. In Appendix \ref{appendix:additional-index-results}, Table \ref{table:highest_educational_level_occupation_ipw} decomposes outcomes by educational attainment, providing further evidence for the ``ceiling effect'' hypothesis. Participants with the lowest formal credentials—those with \textit{No Educational Level}—achieve an incidence rate of $0.60$, identical to those with a secondary school diploma ($0.60$). However, as educational attainment rises, the incidence of positive outcomes declines.

\section{Implications for the AI Transition}

In this paper, we have introduced the \textit{Retrainability Index} as a means of evaluating the effectiveness of the American WIOA program in supporting worker resilience to technological automation. We found that success is moderated by a person's occupation, local labor market, and age, among other characteristics. We show that employer-led programming---notably apprenticeships---are most associated with successful outcomes. Yet we find that WIOA rarely supports worker upskilling, with a significant portion of participants returning to their prior field, and successful outcomes are driven mostly by wage gains, possibly due to catch up mean reversion. This suggests WIOA has been largely unable to support worker resilience in the face of technological automation. 

We insist that these findings describe WIOA as it currently operates, not ALMP or retraining programs in general. Indeed, a very small share of this paper's sample receive formal retraining, and many scholars have previously suggested WIOA is underfunded and overly-complex. More research is needed to understand the potential for future ALMP programming. The evidence collected in this paper does, however, indicate that existing American public ALMP infrastructure may be poorly suited for managing technological transitions. 

We pursued this research paper with an eye toward the AI transition, amid growing expectations that AI---particularly future advanced AI systems---could kindle labor displacement. Our goal was to better understand the potential role retraining might play in supporting future workforce transitions. Although our findings are endogenous only to the computerization shock and America's existing public infrastructure, we believe there are several implications of our research for the AI transition. 

In the first instance, our research suggests governments should support private-sector retraining partnerships and apprenticeships, which have consistently demonstrated the highest incidence of success in both our work and in prior literature. Beyond this, retraining providers should invest in experiments that can help us better understand the causal impact of programming on outcomes. Only through such efforts can we truly disentangle rates of success from self-selecting biases, which cloud our ability to make sense of WIOA outcomes. Additionally, given the complexity of existing program offerings, there may be a role for AI itself in helping WIOA participants navigate the system and identify best-fit interventions. 

Moreover, as we have shown in this paper, successful programmatic outcomes appear at least partially conditioned by a person's prior occupation, their local labor market, and their age, among other factors. Policymakers and researchers should work to better understand these dimensions to determine either a.) how to adapt program offerings to a person's specific context or b.) whether a person would benefit more from a non-ALMP intervention, such as direct income transfers, or programs that do not require the existence of continued work. 

\quad

\bibliographystyle{plainnat}

\begin{thebibliography}{99}

\bibitem[Acemoglu and Autor(2011)]{acemoglu2011skills}
Acemoglu, D., and Autor, D. (2011).
\newblock Skills, tasks and technologies: Implications for employment and earnings.
\newblock \emph{Handbook of Labor Economics}, 4, 1043--1171.

\bibitem[Acemoglu and Restrepo(2019)]{acemoglu2020robots}
Acemoglu, D., and Restrepo, P. (2020).
\newblock Robots and jobs: Evidence from US labor markets.
\newblock \emph{Journal of Political Economy}, 128(6), 2188--2244.

\bibitem[Alkire and Foster(2011)]{alkire2011counting}
Alkire, S., and Foster, J. (2011).
\newblock Counting and multidimensional poverty measurement.
\newblock \emph{Journal of Public Economics}, 95(7-8), 476--487.

\bibitem[Andersson et al.(2024)]{andersson2024does}
Andersson, F., Holzer, H. J., Lane, J. I., Rosenblum, D., and Smith, J. (2024).
\newblock Does Federally Funded Job Training Work?: Nonexperimental Estimates of WIA Training Impacts Using Longitudinal Data on Workers and Firms.
\newblock \emph{Journal of Human Resources}, 59(4), 1244--1283.

\bibitem[Ashenfelter(1978)]{ashenfelter1978estimating}
Ashenfelter, O. (1978).
\newblock Estimating the Effect of Training Programs on Earnings.
\newblock \emph{The Review of Economics and Statistics}, 60(1), 47--57.

\bibitem[Autor(2010)]{autor2010polarization}
Autor, D. H. (2010).
\newblock The polarization of job opportunities in the US labor market: Implications for employment and earnings.
\newblock \emph{Center for American Progress and The Hamilton Project}.

\bibitem[Autor(2015)]{autor2015polyanyi}
Autor, D. H. (2015).
\newblock Why are there still so many jobs? The history and future of workplace automation.
\newblock \emph{Journal of Economic Perspectives}, 29(3), 3--30.

\bibitem[Barnow and Smith(2015)]{barnow2015employment}
Barnow, B. S., and Smith, J. (2015).
\newblock Employment and training programs.
\newblock In \emph{Economics of Means-Tested Transfer Programs in the United States, Volume 2} (pp. 127--234). University of Chicago Press.

\bibitem[Brown et al.(2024)]{brown2024impact}
Brown, C., et al. (2024).
\newblock The Impact of Manufacturing Credentials on Earnings and the Probability of Employment.
\newblock \emph{ILR Review}.

\bibitem[Brynjolfsson et al.(2018)]{brynjolfsson2018machine}
Brynjolfsson, E., Mitchell, T., and Rock, D. (2018).
\newblock What can machines learn and what does it mean for occupations and the economy?
\newblock \emph{AEA Papers and Proceedings}, 108, 43--47.

\bibitem[Card and Sullivan(1988)]{card1988estimating}
Card, D., and Sullivan, D. G. (1988).
\newblock Measuring the Effect of Subsidized Training Programs on Movements In and Out of Employment.
\newblock \emph{Econometrica}, 56(3), 497--530.

\bibitem[Card et al.(2018)]{card2018works}
Card, D., Kluve, J., and Weber, A. (2018).
\newblock What works? A meta-analysis of recent active labor market program evaluations.
\newblock \emph{Journal of the European Economic Association}, 16(3), 894--931.

\bibitem[{CareerSource Florida}(2023)]{careersource_florida_2023}
{CareerSource Florida} (2023).
\newblock 2022-2023 WIOA Annual Statewide Performance Report.
\newblock \emph{Florida Department of Commerce}. \url{https://careersourceflorida.com/wp-content/uploads/2023/12/2022-23-WIOA-Annual-Performance-Report.pdf}.

\bibitem[Deming(2023)]{deming2023investing}
Deming, D. J. (2023).
\newblock Investing in the Workforce: The Impact of WIOA Training on Employment and Earnings.
\newblock \emph{Harvard Kennedy School Working Paper}.

\bibitem[ETA(2024)]{doleta2024wioa}
Employment and Training Administration. (2024).
\newblock WIOA Performance Results: Program Year 2023.
\newblock U.S. Department of Labor.

\bibitem[{Florida Office of Economic and Demographic Research}(2024)]{florida_edr_2024}
{Florida Office of Economic and Demographic Research} (2024).
\newblock An Economic Overview of Florida.
\newblock \emph{The Florida Legislature}. \url{https://edr.state.fl.us/content/presentations/economic/FlEconomicOverview_1-22-24.pdf}.

\bibitem[Frey and Osborne(2017)]{frey2017future}
Frey, C. B., and Osborne, M. A. (2017).
\newblock The future of employment: How susceptible are jobs to computerisation?
\newblock \emph{Technological Forecasting and Social Change}, 114, 254--280.

\bibitem[Goldin and Katz(2008)]{goldin2008race}
Goldin, C., and Katz, L. F. (2008).
\newblock \emph{The Race Between Education and Technology}.
\newblock Harvard University Press.

\bibitem[Heckman et al.(1999)]{heckman1999economics}
Heckman, J. J., LaLonde, R. J., and Smith, J. A. (1999).
\newblock The economics and econometrics of active labor market programs.
\newblock \emph{Handbook of Labor Economics}, 3, 1865--2097.

\bibitem[Heinrich et al.(2008)]{heinrich2008workforce}
Heinrich, C. J., Mueser, P. R., and Troske, K. R. (2008).
\newblock Workforce Investment Act non-experimental net impact evaluation.
\newblock \emph{IMPAQ International}.

\bibitem[Hendra et al.(2016)]{hendra2016encouraging}
Hendra, R., Greenberg, D. H., Hamilton, G., Oppenheim, A., Pennington, A., Schaberg, K., and Tessler, B. (2016).
\newblock Encouraging Evidence on a Sector-Focused Advancement Strategy.
\newblock \emph{MDRC}.

\bibitem[Hernán and Robins(2020)]{hernan2020causal}
Hernán, M. A., and Robins, J. M. (2020).
\newblock \emph{Causal Inference: What If}.
\newblock Boca Raton: Chapman \& Hall/CRC.

\bibitem[Hyman et al.(2025)]{hyman2025retrainable}
Hyman, B. G., Lahey, B., Ni, K., and Pilossoph, L. (2025).
\newblock How Retrainable are AI-Exposed Workers?
\newblock \emph{NBER Working Paper No. 34174}.

\bibitem[Hyman et al.(2024)]{hyman2024wageinsurance}
Hyman, B. G., Kovak, B. K., and Leive, A. (2024).
\newblock Wage Insurance for Displaced Workers.
\newblock \emph{NBER Working Paper No. 32464}.

\bibitem[Jacobson et al.(1993)]{jacobson1993earnings}
Jacobson, L. S., LaLonde, R. J., and Sullivan, D. G. (1993).
\newblock Earnings losses of displaced workers.
\newblock \emph{The American Economic Review}, 685--709.

\bibitem[Kish(1965)]{kish1965survey}
Kish, L. (1965).
\newblock \emph{Survey Sampling}.
\newblock New York: John Wiley \& Sons.

\bibitem[LaLonde(1986)]{lalonde1986evaluating}
LaLonde, R. J. (1986).
\newblock Evaluating the econometric evaluations of training programs with experimental data.
\newblock \emph{The American Economic Review}, 604--620.

\bibitem[Lundberg and Lee(2017)]{lundberg2017unified}
Lundberg, S. M., and Lee, S. I. (2017).
\newblock A unified approach to interpreting model predictions.
\newblock \emph{Advances in Neural Information Processing Systems}, 30.

\bibitem[Maestas et al.(2016)]{maestas2016does}
Maestas, N., Mullen, K. J., and Powell, D. (2016).
\newblock The effect of population aging on economic growth, the labor force, and productivity.
\newblock \emph{American Economic Journal: Macroeconomics}, 8(3), 1--28.

\bibitem[Neumark and Song(2013)]{neumark2013mechanisms}
Neumark, D., and Song, J. (2013).
\newblock Do stronger age discrimination laws make it harder for older workers to find jobs?
\newblock \emph{Journal of Public Economics}, 92, 58--75.

\bibitem[OECD(2008)]{oecd2008handbook}
OECD. (2008).
\newblock \emph{Handbook on Constructing Composite Indicators: Methodology and User Guide}.
\newblock OECD Publishing.

\bibitem[O'Leary and Eberts(2008)]{oleary2008reemployment}
O'Leary, C. J., and Eberts, R. W. (2008).
\newblock The Wagner-Peyser Act and US Employment Service: Seventy-five years of matching job seekers and employers.
\newblock \emph{W.E. Upjohn Institute for Employment Research}.

\bibitem[Rothstein et al.(2022)]{rothstein2022caal}
Rothstein, J., Santillano, R., von Wachter, T., Khan, W., and Yang, M. (2022).
\newblock CAAL-Skills: Study of Workforce Training Programs in California.
\newblock \emph{California Policy Lab}.

\bibitem[Social Policy Research Associates(2022)]{socialpolicy2022wioa}
Social Policy Research Associates. (2022).
\newblock PY 2020 WIOA Adult Program Performance.
\newblock \emph{Prepared for the U.S. Department of Labor}.

\bibitem[Stuart(2010)]{stuart2010matching}
Stuart, E. A. (2010).
\newblock Matching methods for causal inference: A review and a look forward.
\newblock \emph{Statistical Science}, 25(1), 1--21.

\bibitem[Susskind(2020)]{susskind2020world}
Susskind, D. (2020).
\newblock \emph{A World Without Work: Technology, Automation, and How We Should Respond}.
\newblock Metropolitan Books.

\end{thebibliography}

\clearpage

\appendix
\part*{Appendix}

\section{Index Construction Methodology}

\subsection{Subcomponent Details}\label{appendix:subcomponent-definitions}

Wage difference refers to the gap between an individual's average quarterly wages in the pre-participation window and the post-participation window. This provides a direct measure of earnings progression associated with the participation in job retraining or other workforce development programs. The change in routine cognitive and routine manual task intensity associated with the occupation or industry subsector in which the individual is employed before and after program participation proxies the technological vulnerability of the work performed, allowing us to assess whether participants transition into occupations or industries that rely less on routine, automatable tasks.

Formally, let \( w_{n,t} \) denote the quarterly wage of individual \( n \) in quarter \( t \). We index quarters relative to the first quarter of WIOA participation, with \( \tau = 0 \) representing the participation quarter. We define the pre- and post-period sets as \( \mathcal{P} = \{-3,-2,-1\} \) and \( \mathcal{Q} = \{1,2,3,4\} \), respectively. The average pre- and post-participation wages for individual \( n \) are:
\begin{align}
    \overline{w}_{n}^{\text{pre}}  &= \sinh^{-1} \left( \frac{1}{|\mathcal{P}|} \sum_{\tau \in \mathcal{P}} w_{n,\tau} \right), \\
     \overline{w}_{n}^{\text{post}} &= \sinh^{-1} \left( \frac{1}{|\mathcal{Q}|} \sum_{\tau \in \mathcal{Q}} w_{n,\tau} \right),
\end{align}
where we utilize inverse hyperbolic sine (IHS) to capture the percentage gain in wage changes and mitigate level-effect distortions. Subsequently, we apply min-max normalization across the joint set of average pre- and post-participation wages\footnote{Prior to normalization, wages are winsorized at the 1st and 99th percentiles to reduce the influence of extreme tails.} (i.e., $\mathcal{W} = \left\{ \bar{w}_{n}^{s} : n = 1, \dots, N, \; s \in \{\text{pre}, \text{post}\} \right\}$) so that $1$ represents the highest wage and $-1$ represents the lowest wage:
\begin{equation}\label{eq:wage_normalization}
    \overline{w}_{n}^{\text{pre\_norm}} = \frac{\overline{w}_{n}^{\text{pre}}- \min(\mathcal{W})}{\max(\mathcal{W}) - \min(\mathcal{W})} \qquad \overline{w}_{n}^{\text{post\_norm}} = \frac{\overline{w}_{n}^{\text{post}}- \min(\mathcal{W})}{\max(\mathcal{W}) - \min(\mathcal{W})}.
\end{equation}

We then define the change in wages for participation period $n$ as:
\begin{equation}
    \Delta w_n = \overline w_{n}^{\text{post\_norm}} - \overline w_{n}^{\text{pre\_norm}}.
\end{equation}

To quantify the extent to which participants transition into more or less routine forms of work, we measure changes in routine task intensities along two dimensions: routine \emph{cognitive} tasks and routine \emph{manual} tasks. Let \( p_{ij} \) denote the employment share of occupation \( i \) within industry subsector \( j \), and let \( N \) denote the total number of occupations. Occupation-level task intensities are fixed parameters taken from O*NET and related sources \citep{acemoglu2011skills}.
\begin{align*}
    r^{C_i} &= \text{routine cognitive task intensity of occupation } i, \\
    r^{M_i} &= \text{routine manual task intensity of occupation } i.
\end{align*}

For each participation period $n$, we define the change in routine task intensity as the difference in occupation-level intensities associated with a participant's pre- and post-participation employment:
\begin{align}\label{eq:rti_difference}
    \Delta r^{C}_n & = r^{C_y}_n - r^{C_x}_n & \Delta r^{M}_n & = r^{M_y}_n - r^{M_x}_n
\end{align}
where \( r_{n}^{C_x} \) and \( r_{n}^{C_y} \) denotes the routine cognitive task intensities of the occupations of pre-program and post-program employment (with routine manual task intensities defined analogously).\footnote{We apply the same winsorization and min-max normalization strategy to the joint set of pre- and post program routine cognitive/manual task intensities as described in Equation \ref{eq:wage_normalization}.}

Additionally, we define industry subsector-level routine task intensities as employment-share-weighted averages of the task intensities of the occupations that comprise each subsector:
\begin{align}\label{eq:subsector_rti}
    R^{C_j} &= \sum_{i=1}^{N} p_{ij} \, r^{C_i} & R^{M_j} &= \sum_{i=1}^{N} p_{ij} \, r^{M_i}.
\end{align}

Then for each period of participation \( n \), the change in industry subsector-level routine task intensity is defined (analogously to Equation \ref{eq:rti_difference}) as:
\begin{align}\label{eq:subsector_rti_change}
    \Delta R_{n}^{C} &= R_{n}^{C_y} - R_{n}^{C_x} & \Delta R_{n}^{M} &= R_{n}^{M_y} - R_{n}^{M_x}.
\end{align}

\subsubsection{Interpreting Industry Subsector Routine Task Intensity}

A methodological limitation is that subsector-level routine task intensity is only weakly correlated with the occupation-level task intensities of the specific jobs held by WIOA participants. To quantify this discrepancy between occupation and industry subsector routine task intensity, we exploit the subset of participation periods for which both pre- and post-program occupation codes are observed (approximately 1.2\% of participation periods), comparing changes in actual occupation-level routine task intensities ($\Delta r$) with changes computed using our subsector-level proxies ($\Delta R$).

The empirical relationship is indeed weak. The fitted relationships
\begin{align}
    \Delta r^{C}_n &\approx 0.25 \, \Delta R^{C}_n + 0.043 \quad (R^2 = 0.013), \\
    \Delta r^{M}_n &\approx 0.25 \, \Delta R^{M}_n + 0.098 \quad (R^2 = 0.027),
\end{align}
indicate that subsector-level changes explain only 1--3\% of the variation in occupation-level task changes for the subset with complete data. These patterns imply substantial within-subsector occupational heterogeneity.

\begin{figure}[H]
    \centering
    \includegraphics[width=0.75\linewidth]{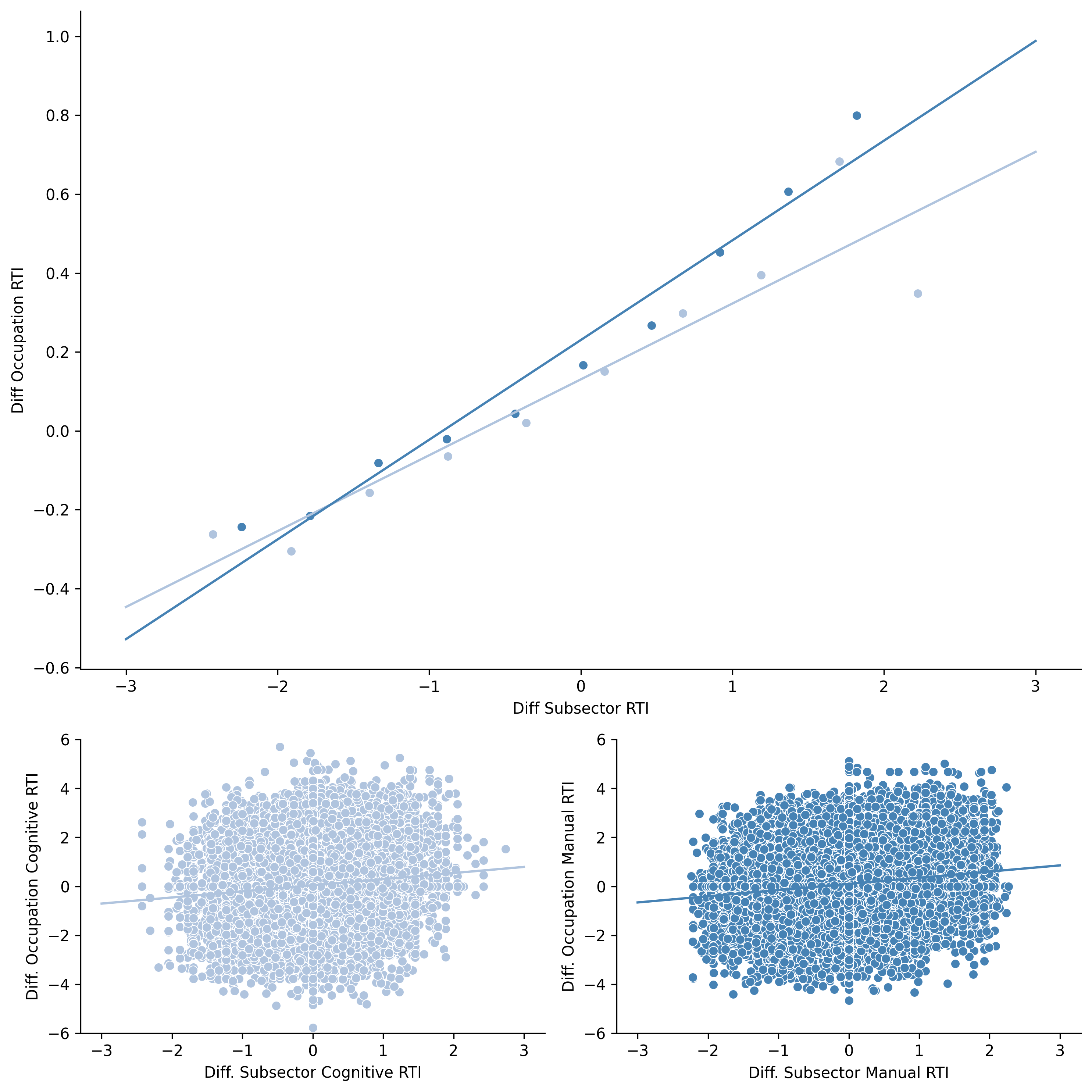}
    \caption{Comparison of occupation-level and subsector-level changes in routine cognitive (bottom left) and routine manual (bottom right) task intensity with scatter plot binned by subsector-level changes (top).}
    \label{fig:image-6}
\end{figure}

This arises because industries contain heterogeneous mixtures of occupations with vastly different routine task intensity compositions. Despite the low correlation with individual job tasks, we retain the subsector-level index as a measure of sectoral exposure. As noted in Equation \ref{eq:subsector_rti}, $R^{C_j}$ and $R^{M_j}$ quantify the technological environment that prevails in the subsector as a whole. In the context of workforce development, placing a participant into a sector with low Routine Task Intensity (RTI) is a valuable outcome regardless of their specific entry-level role, as it implies a transition into an ecosystem with greater long-term resilience against automation \citep{acemoglu2020robots}. Therefore, $\Delta R_n^M$ and $\Delta R_n^C$ should be interpreted as the change in the \textit{technological vulnerability of the labor market segment} the participant occupies, rather than a precise measure of their daily ergonomic or cognitive functions.

To address the sparsity of occupation codes in future iterations of this index, we propose estimating the expected change in occupation-level task intensity conditional on observable characteristics. Throughout this analysis, subsector-level routine task intensities are treated as \emph{industry characteristics} that proxy for the routine intensity of the sectoral environment into which participants are placed, rather than as approximations of individual job-level tasks. This distinction allows the industry subsector-level index to function as a macro-indicator of workforce mobility toward resilient industries, even in the absence of granular occupational coding.

\subsection{Weighting \& Aggregation Details}\label{appendix:detailed-definition}
Throughout this paper, we offer two formulations of the index: one using occupation-level routine task intensities and the other using industry subsector-routine task intensities. The occupation-level formulation uses the following subcomponents:
\begin{equation}\label{eq:index_subcomponents_occupation}
    I_{n}^{W} \equiv \Delta w_n, \qquad
    I_{n}^{C, occ} \equiv \Delta r_{n}^{C}, \qquad
    I_{n}^{M, occ} \equiv \Delta r_{n}^{M}
\end{equation}
and the industry subsector-level formulation uses:
\begin{equation}\label{eq:index_subcomponents_subsector}
    I_{n}^{W} \equiv \Delta w_n, \qquad
    I_{n}^{C, sub} \equiv \Delta R_{n}^{C}, \qquad
    I_{n}^{M, sub} \equiv \Delta R_{n}^{M}
\end{equation}

For convenience, we refer to the subcomponents of the index as $I_{n}^{W}$, $I_{n}^{C}$, and $I_{n}^{M}$, specifying in context whether we are referring to the occupation-level formulation (Equation \ref{eq:index_subcomponents_occupation}) or the industry subsector-level formulation (Equation \ref{eq:index_subcomponents_subsector}).

\subsubsection{Subcomponent Weighting}

The final individual index is a weighted linear combination of the three components. By construction, $I_{n}^{W}$, $I_{n}^{C}$, and $I_{n}^{M}$ all fall between $-1$ and $1$, and therefore the index will also fall within this range. We assign a 50\% weight to wage growth (reflecting immediate economic benefits) and split the remaining 50\% equally between the two resilience metrics (cognitive and manual task exposure). However, we illustrate in Appendix \ref{appendix:index-sensitivity} that the index shows low sensitivity to the particular choice of weights.

\begin{align}\label{eq:individual_level_index}
I_{n}
=
0.5 \cdot I_{n}^{W}
-
0.25 \cdot I_{n}^{C}
-
0.25 \cdot I_{n}^{M}.
\end{align}

\subsubsection{Inverse Probability Weighting}\label{appendix:inverse_probability_weighting}

Not all participant records yield a calculable index score. In particular, many participation periods lack sufficient wage or occupation data to construct $I_n$ which could vary systematically with participant characteristics. To correct for this potential bias, we apply inverse probability weighting (IPW) to reweight the calculable subsample so that it resembles the full sample population as defined in Section \ref{section:dataset_and_index_construction}.

\textit{Propensity score estimation.} We define an indicator $C_n \in \{0,1\}$ equal to one if participant period $n$ yields a calculable index value. We then estimate the propensity score $\hat{p}_n \equiv \hat{P}(C_n = 1 \mid X_n)$ via logistic regression of $C_n$ on a vector of pre-outcome covariates $X_n$, comprising: sex, race, highest educational level, employment status at program entry, funding stream, program year, age, low-income status indicator, and received-training indicator. Categorical variables are one-hot encoded after mode imputation; the continuous age variable is median imputed and standardized; binary indicators are mode imputed.

\textit{Weight construction.} We construct stabilized IPW weights for each calculable participant period:
\begin{align}
    \hat{w}_n = \frac{\hat{P}(C_n = 1)}{\hat{p}_n},
\end{align}
where $\hat{P}(C_n = 1)$ is the marginal calculability rate in the full sample. We use stabilized rather than simple weights ($1/\hat{p}_n$), as stabilized weights reduce variance while preserving the reweighting property \citep{hernan2020causal}.

\textit{Diagnostics.} We assess balance using the standardized mean difference (SMD) for each covariate before and after weighting. Table~\ref{tab:ipw_balance} reports the SMD for each covariate before and after reweighting. Prior to weighting, 16 of 39 covariate levels exhibited meaningful imbalance ($|\text{SMD}| \geq 0.1$), with the largest imbalances observed for Wagner-Peyser funding stream
($\text{SMD} = -0.96$), received training ($\text{SMD} = 0.85$), Adult funding stream ($\text{SMD} = 0.66$), and low-income status ($\text{SMD} = 0.53$). After applying stabilized IPW, the number of imbalanced covariates falls to 6 of 39, and no covariate exceeds $|\text{SMD}| = 0.19$, indicating substantially improved comparability between the calculable subsample and the
full participant population.

\begin{table}[H]
\centering
\caption{Covariate Balance Before and After IPW (for $|\text{SMD Before}| \geq 0.10$)}
\label{tab:ipw_balance}
\small
\begin{tabular}{lcc}
\toprule
\textbf{Covariate} & \textbf{SMD Before} & \textbf{SMD After} \\
\midrule
Funding Stream: Wagner-Peyser       & -0.960 & -0.128 \\
Received Training                   &  0.846 &  0.030 \\
Funding Stream: Adult               &  0.661 &  0.058 \\
Low-Income Status                   &  0.526 &  0.102 \\
Funding Stream: Dislocated Worker   &  0.458 &  0.023 \\
Race: Black                         &  0.352 &  0.189 \\
Employment Status: Unemployed               & -0.293 & -0.075 \\
Age                                 & -0.287 & -0.118 \\
Employment Status: Employed              &  0.284 &  0.077 \\
Education Level: Secondary School Diploma               &  0.222 &  0.048 \\
Education Level: No Educational Level Completed              & -0.220 &  0.009 \\
Race: White                          & -0.207 & -0.072 \\
Race: Hispanic                          & -0.159 & -0.153 \\
Education Level: Bachelor's Degree               & -0.127 & -0.044 \\
Education Level: Advanced Degree               & -0.119 & -0.047 \\
Sex: Did not self-identify                             & -0.105 & -0.091 \\
Funding Stream: Youth & 0.051 & 0.133 \\
\bottomrule
\end{tabular}
\begin{minipage}{\linewidth}
\end{minipage}
\end{table}

\begin{figure}[H]
    \centering
    \includegraphics[width=1\linewidth]{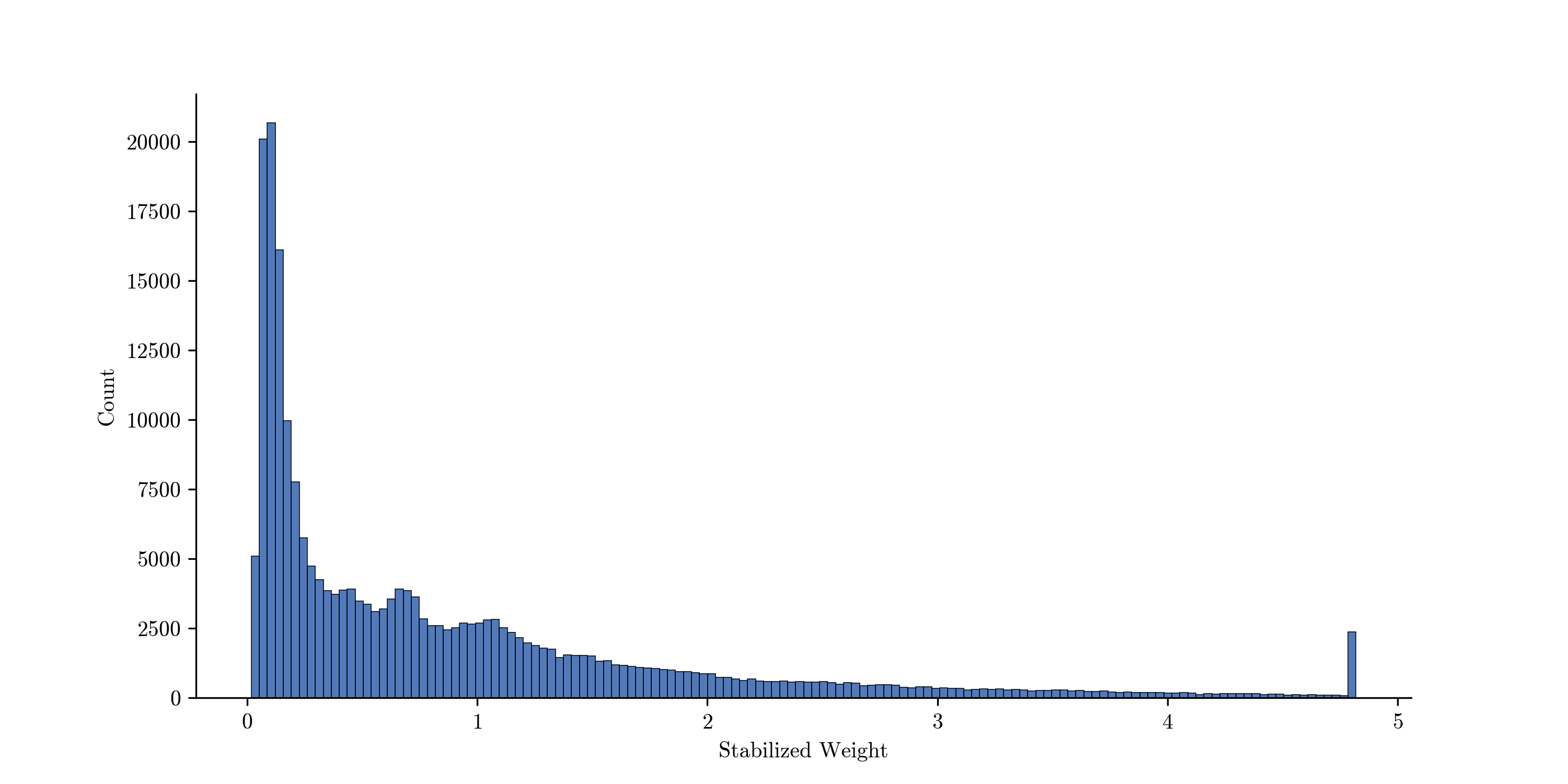}
    \caption{The distribution of stabilized weights for calculable sample.}
    \label{fig:image-18}
\end{figure}

To assess weight stability, we examine the distribution of stabilized weights among the calculable subsample ($n = 229{,}966$) and compute the effective sample size (ESS) using the formula, $\text{ESS} = (\sum w_i)^2 / \sum w_i^2$ \citep{kish1965survey}. As shown in Figure \ref{fig:image-18}, the stabilized weights are concentrated between 0 and 1, consistent with the expectation that stabilized weights are centered near unity, though a right skew is evident. The ESS is approximately 100,200, representing a 44\% retention of the calculable sample ($n = 229{,}966$).

\subsubsection{Aggregation}

Because the index is defined at the level of the participation period, it can be flexibly aggregated to any administrative or geographic unit without loss of consistency. Let $g$ be a group from a grouping of interest (i.e., State, a Local Workforce Development Board (WDB), Industry Subsector). Furthermore, let $g^{+} \subseteq g$ be the set of participation periods such that $I_n > 0$. Let $\omega_n$ denote the inverse probability weight for participation period $n$, and define the effective weighted counts $W_{g} = \sum_{n \in g} \omega_n$ and $W_{g^{+}} = \sum_{n \in g^{+}} \omega_n$. We then define the share of participation periods that experience a positive outcome and the intensity of these positive outcomes as
\begin{align}
    \text{Incidence}_g & = \frac{W_{g^{+}}}{W_{g}} \\
    \text{Intensity}_g & = \frac{\sum_{n \in g^{+}} \omega_n I_n}{W_{g^{+}}}
\end{align}
Subsequently, we define the group-level index as their product:\footnote{We construct our index similarly to the United Nations Development Programme's Global Multidimensional Poverty Index (MPI) which is the product of the incidence of multidimensional poverty and its intensity. Further technical details about the MPI can be found in \cite{alkire2011counting}.}
\begin{align}
I_{g} & = \text{Incidence}_g \cdot \text{Intensity}_g = \frac{\sum_{n \in g^{+}} \omega_n I_n}{W_{g}}.
\end{align}

\subsubsection{Assessing Subcomponent Weight Sensitivity} \label{appendix:index-sensitivity}

To assess the robustness of the index to alternative value judgments on indicator weighting, we conducted a Monte Carlo sensitivity analysis on a 10\% sample of the dataset. For each of 500 simulations, a random weight vector was drawn from a Dirichlet distribution, and the index was recomputed. Results were compared to the baseline using Spearman rank correlations ($\rho$) and absolute changes in percentile rank. We report results in Table \ref{tab:mc_sensitivity} for both the occupation-level formulation\footnote{Group-level aggregations for the occupation-level fomulation of the index use unweighted means. Since IPW weights are fixed across iterations, they are unlikely to materially affect rank stability conclusions, though groups with highly uneven weighting may see some reordering relative to the IPW-weighted index.} (Equation \ref{eq:index_subcomponents_occupation}) and the industry subsector-level formulation (Equation \ref{eq:index_subcomponents_subsector}), where primary values correspond to the occupation-level formulation and parenthetical entries report the difference between the subsector-level and occupation-level results. The number of groups denotes the number of distinct aggregated index values within each grouping.

\begin{table}[H]
\centering
\caption{Sensitivity of Index Rankings to Weight Perturbations (500 Monte Carlo Simulations)}
\label{tab:mc_sensitivity}
\resizebox{\textwidth}{!}{%
\begin{tabular}{lrrrrr}
\toprule
\textbf{Grouping} & \textbf{\# Groups} & \textbf{Mean $\rho$} & \textbf{Min $\rho$} & \textbf{Median Pct. Shift} & \textbf{Max Pct. Shift} \\
\midrule
        Participation Period ($\text{I}_n$) & 22,720 ($+$1,114,949) & 0.76 ($-$0.02) & 0.58 ($-$0.05) & 0.06 ($-$0.02) & 0.84 ($+$0.09) \\
        Local Workforce Board (WDB) & 436 ($+$205) & 0.68 ($+$0.19) & 0.50 ($+$0.31) & 0.07 ($-$0.03) & 0.91 ($+$0.09) \\
        Employment Status at Entry & 4 ($+$0) & 0.81 ($+$0.18) & 0.80 ($+$0.00) & 0.12 ($-$0.11) & 0.24 ($-$0.22) \\
        Program Year & 8 ($+$0) & 0.96 ($+$0.02) & 0.86 ($+$0.07) & 0.04 ($-$0.04) & 0.13 ($-$0.04) \\
        Pre-Program Occupation & 586 ($+$658) & 0.68 ($+$0.04) & 0.38 ($+$0.21) & 0.05 ($+$0.01) & 0.87 ($+$0.08) \\
        Received Training Indicator & 2 ($+$0) & $-$0.42 ($+$1.42) & $-$1.00 ($+$2.00) & 0.35 ($-$0.35) & 0.35 ($-$0.35) \\
        Funding Stream & 5 ($+$0) & 0.93 ($+$0.03) & 0.80 ($+$0.10) & 0.06 ($-$0.06) & 0.08 ($-$0.01) \\
        Training Service Type & 14 ($+$0) & 0.49 ($+$0.36) & 0.19 ($+$0.52) & 0.17 ($-$0.09) & 0.57 ($-$0.22) \\
        Participant Age & 71 ($+$17) & 0.84 ($+$0.12) & 0.72 ($+$0.12) & 0.05 ($-$0.03) & 0.77 ($-$0.08) \\
        State & 49 ($+$3) & 0.68 ($+$0.28) & 0.40 ($+$0.52) & 0.09 ($-$0.06) & 0.76 ($-$0.47) \\
        Race / Ethnicity & 8 ($+$0) & 0.80 ($+$0.18) & 0.29 ($+$0.67) & 0.07 ($-$0.07) & 0.29 ($-$0.19) \\
        Low Income Status & 2 ($+$0) & 1.00 ($+$0.00) & 1.00 ($+$0.00) & 0.00 ($+$0.00) & 0.00 ($+$0.00) \\
        Highest Educational Level & 9 ($+$1) & 0.95 ($-$0.19) & 0.38 ($+$0.29) & 0.06 ($+$0.08) & 0.12 ($+$0.25) \\
        Sex & 3 ($+$0) & 0.50 ($+$0.50) & 0.50 ($+$0.50) & 0.33 ($-$0.33) & 0.33 ($-$0.33) \\
\bottomrule
\end{tabular}
}
\end{table}

At the individual participant period level, the index demonstrates \emph{moderate robustness}: the mean Spearman correlation is approximately $\rho \approx 0.76$, and the median percentile rank shift is approximately $6\%$. Strong stability is found for groupings with moderate-to-large $N$, including \textit{Program Year} ($\rho = 0.96$), \textit{Funding Stream} ($\rho = 0.93$), \textit{Highest Educational Level} ($\rho = 0.95$), and \textit{Participant Age} ($\rho = 0.84$). Apparent instability in groupings such as \textit{Received Training Indicator}, \textit{Sex}, and \textit{Training Service Type} should be interpreted cautiously, as rank correlation statistics are inherently volatile when the number of groups is small, where a single rank swap has an outsized effect on $\rho$. The subsector-level formulation generally performs comparably or better across most groupings, with the largest improvements in \textit{Training Service Type} and \textit{Employment Status at Entry}. Overall, the analysis confirms that the index is structurally stable and not driven by specific weighting choices, particularly at the aggregation levels.

\begin{figure}[H]
    \centering
    \includegraphics[width=1\linewidth]{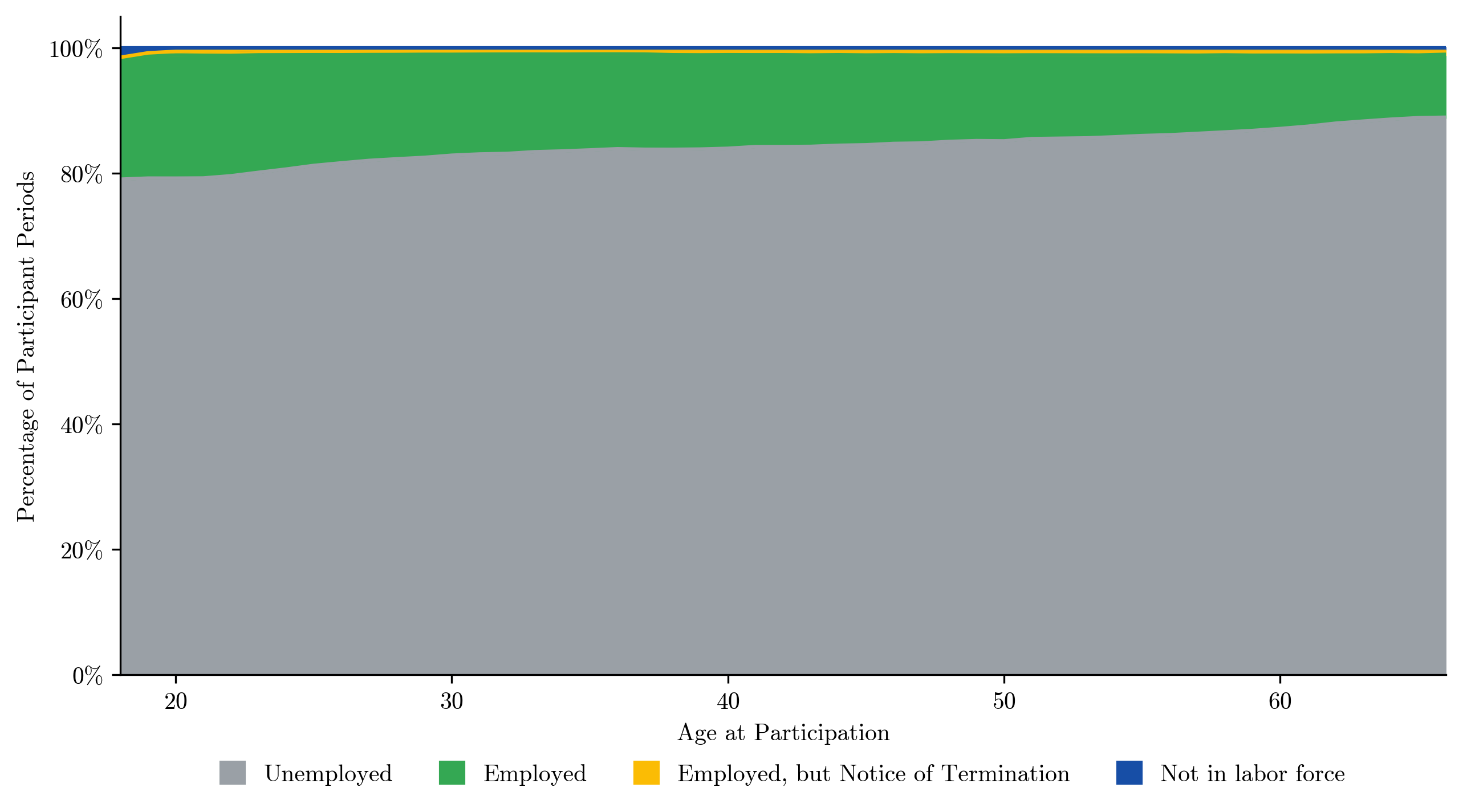} 
    \caption{ Employment status at participation by age for participants who received formal training. Unemployment (gray area) dominates across all age groups but expands significantly as age increases, indicating deeper labor market detachment among older trainees.}
    \label{fig:education}
\end{figure}

\begin{figure}[H]
    \centering
    \includegraphics[width=1\linewidth]{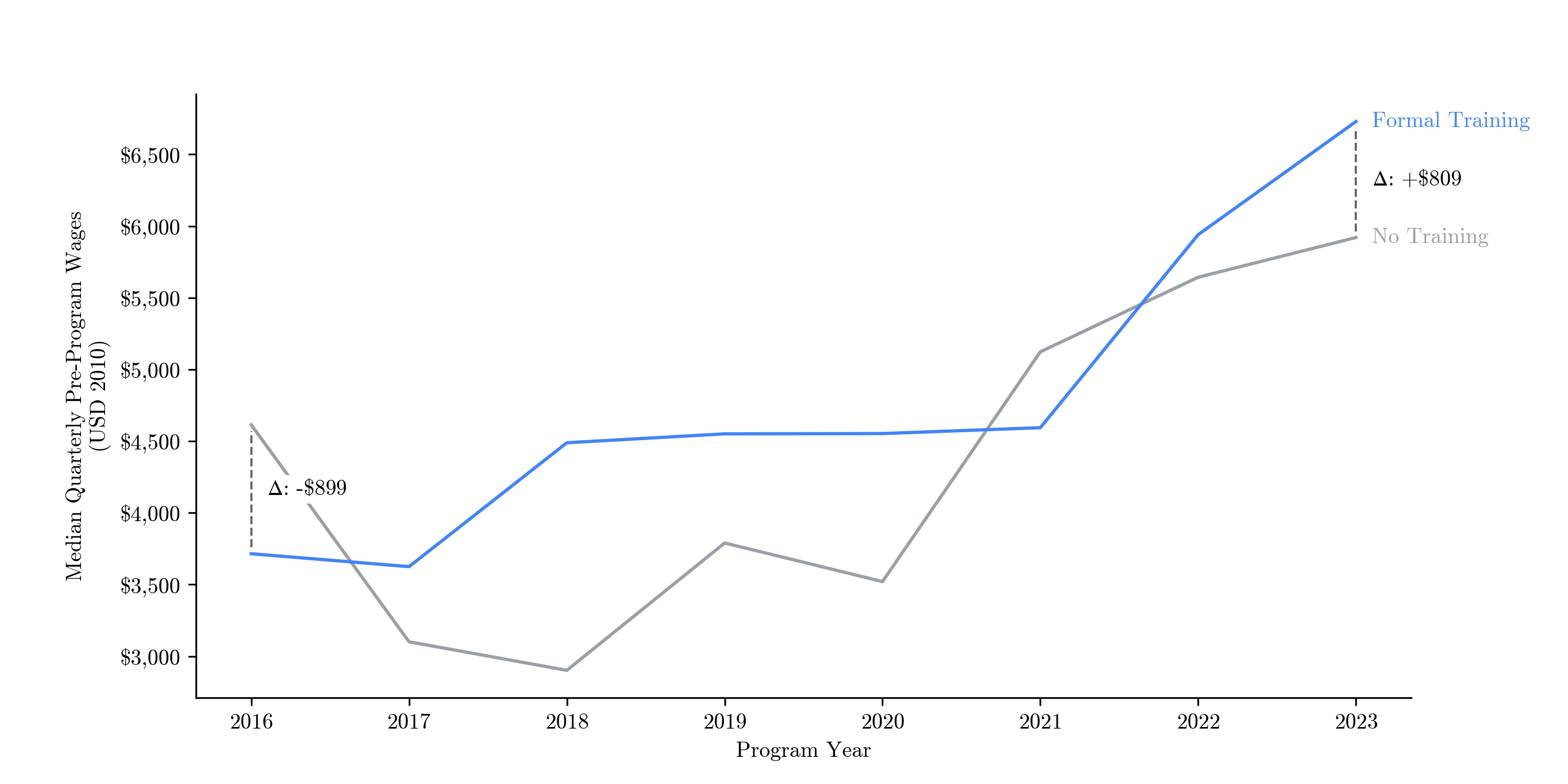}
    \caption{The median (mean) quarterly wages pre-program participation by program year for participation periods with $I_n$ (Occupation).}
    \label{fig:wages by year}
\end{figure}

\begin{figure}[H]
    \centering
    \includegraphics[width=1\linewidth]{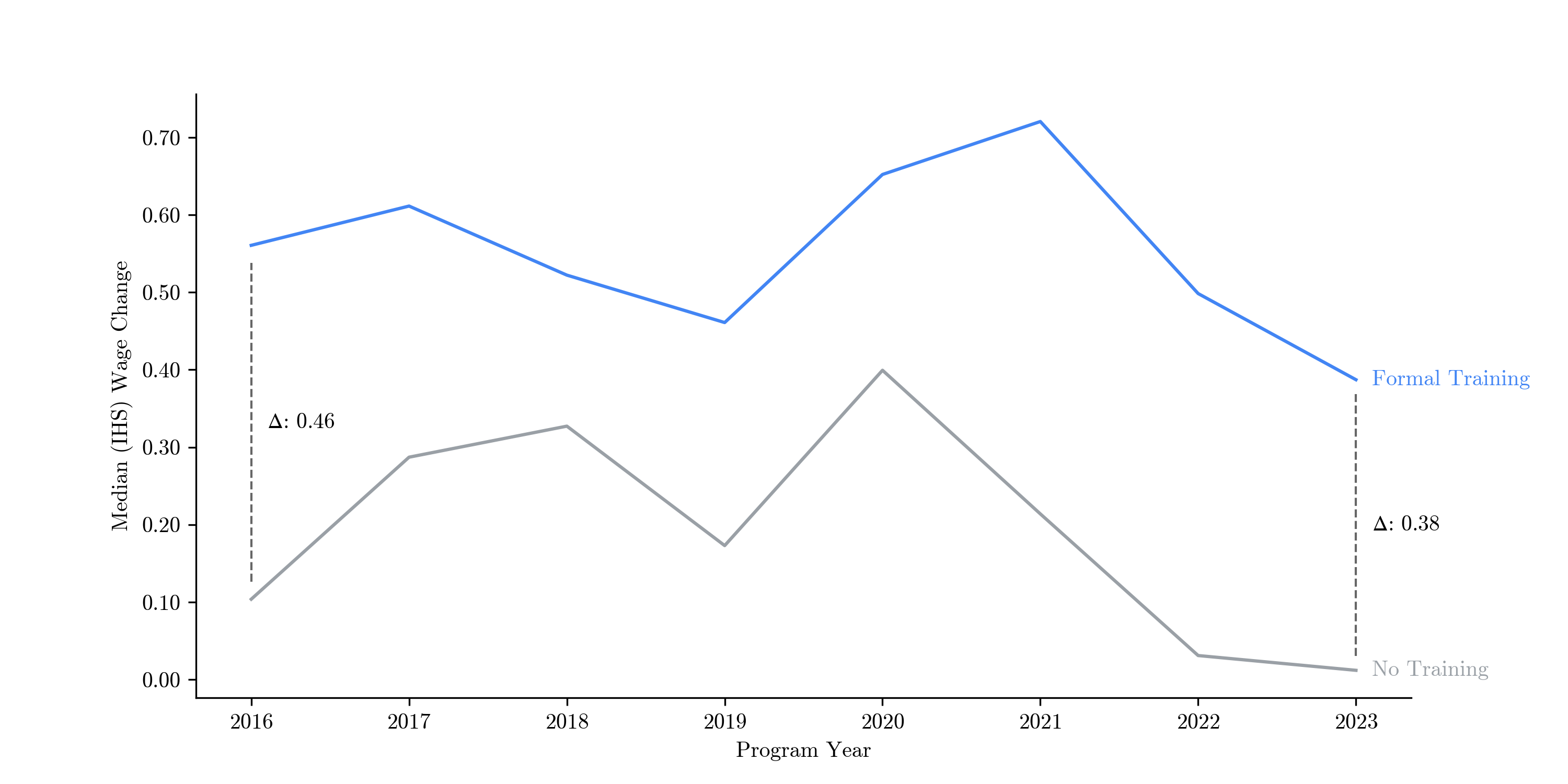}
    \caption{Median difference in inverse hyperbolic sine (IHS)-transformed average quarterly wages between post- and pre-WIOA program participation (see Appendix \ref{appendix:subcomponent-definitions} for details), by program year for participation periods with $I_n$ (Occupation).}
    \label{fig:SINE}
\end{figure}

\section{Feature Importance Sub-Models}\label{app:feature_tables}

The following tables present the feature importance metrics (Model Gain, Permutation, and SHAP) for the targeted sub-models, isolating specific categories of variables to observe their relative predictive power for positive index outcomes.

\begin{table}[H]
\centering
\caption{Feature Importance for Predicting Positive Index Outcomes: Program Participation \& Structure}
\label{tab:feature_importance_2}
\resizebox{\textwidth}{!}{%
\begin{tabular}{llllr}
\toprule
Feature & Model (Gain) Importance & Permutation Importance & SHAP Importance & Avg Rank \\
\midrule
Funding Stream & 0.3 (1) & 0.03 (1) & 0.19 (1) & 1 \\
Local Workforce Board (WDB) & 0.25 (2) & 0.01 (2) & 0.14 (2) & 2 \\
Program Year & 0.11 (4) & 0.01 (3) & 0.1 (3) & 3 \\
State & 0.13 (3) & 0.01 (4) & 0.08 (4) & 4 \\
Employment Status at Entry & 0.11 (5) & 0.0 (6) & 0.04 (5) & 5 \\
Training Service Type & 0.1 (6) & 0.0 (5) & 0.04 (6) & 6 \\
\bottomrule
\end{tabular}
}
\end{table}

\begin{table}[H]
\centering
\caption{Feature Importance for Predicting Positive Index Outcomes: Regional Economic Indicators}
\label{tab:feature_importance_3}
\resizebox{\textwidth}{!}{%
\begin{tabular}{llllr}
\toprule
Feature & Model (Gain) Importance & Permutation Importance & SHAP Importance & Avg Rank \\
\midrule
WDB Population (per sq km) & 0.12 (4) & 0.0 (1) & 0.04 (3) & 1 \\
WDB Unemployment Rate & 0.09 (7) & 0.0 (2) & 0.07 (1) & 2 \\
WDB Diversity Index & 0.13 (3) & 0.0 (3) & 0.04 (5) & 3 \\
WDB Median Income Level & 0.18 (1) & 0.0 (4) & 0.01 (8) & 5 \\
WDB Population & 0.09 (6) & 0.0 (5) & 0.04 (2) & 5 \\
WDB Median Age & 0.07 (9) & 0.0 (6) & 0.04 (4) & 6 \\
WDB Metro Status & 0.16 (2) & 0.0 (9) & 0.01 (9) & 8 \\
WDB Mean Commuting Time (Min) & 0.1 (5) & 0.0 (8) & 0.01 (7) & 8 \\
WDB High Household Debt-to-Income Ratio & 0.08 (8) & 0.0 (7) & 0.03 (6) & 9 \\
\bottomrule
\end{tabular}
}
\end{table}

\begin{table}[H]
\centering
\caption{Feature Importance for Predicting Positive Index Outcomes: Demographics}
\label{tab:feature_importance_4}
\resizebox{\textwidth}{!}{%
\begin{tabular}{llllr}
\toprule
Feature & Model (Gain) Importance & Permutation Importance & SHAP Importance & Avg Rank \\
\midrule
Pre-Program Occupation & 0.44 (1) & 0.13 (1) & 0.67 (1) & 1 \\
Pre-Program Industry Subsector & 0.29 (2) & 0.07 (2) & 0.48 (2) & 2 \\
Participant Age & 0.06 (3) & 0.01 (3) & 0.1 (3) & 3 \\
Low Income Status & 0.06 (4) & 0.0 (6) & 0.07 (4) & 4 \\
Race / Ethnicity & 0.06 (5) & 0.0 (5) & 0.04 (6) & 5 \\
Highest Educational Level & 0.05 (6) & 0.0 (7) & 0.05 (5) & 7 \\
Sex & 0.04 (7) & 0.0 (4) & 0.03 (7) & 7 \\
\bottomrule
\end{tabular}
}
\end{table}

\section{Logistic Regression Results}\label{appendix:logistic-regression-results}

We provide the following results for a logistic regression model trained to predict whether participation resulted in a positive outcome (defined as $I_n > 0$).

\TableLRPerformance

\TableLRCoefficients

\section{Causal Estimates of Training and Apprenticeship Effects}\label{appendix:psm_estimates}
As a complement to the observational analysis in Section \ref{section:Characteristics of Success}, we re-purpose the matching methodology used by \citet{hyman2025retrainable} in their concurrent and related research. We do this to  estimate the effect of two interventions on the individual-level Retrainability Index $I_n$ (as defined in Equation \ref{eq:individual_level_index}): receiving formal training and participating in a registered apprenticeship.\footnote{We note that these estimates should be interpreted against the individual-level mean of $I_n$ reported in the tables of Section~\ref{section:Characteristics of Success}.} Using nearest-neighbor propensity score matching within exact strata defined by program year, we estimate propensity scores using logistic regression on various participant demographics including pre-program occupation and industry subsector, local workforce board assignment, state, age, and pre-program wages \citep{stuart2010matching}. This approach is similar to the one taken by\citet{hyman2025retrainable}.  Treated units are then matched to their nearest neighbor within a caliper of $0.1$ on the propensity score. We compute treatment effects using weights (as defined in Section \ref{appendix:inverse_probability_weighting}) and bootstrap confidence intervals for both the average treatment effect on the treated (ATT) and the average treatment effect (ATE). We caution that the matching diagnostics for both analyses are mixed---while balance improves substantially after matching, residual imbalances persist for several covariates and propensity score overlap is limited---and these estimates should therefore be interpreted cautiously and in conjunction with the broader observational evidence presented in Section \ref{section:Characteristics of Success}. We discuss these diagnostic limitations in detail in the subsection below.

\begin{table}[H]
\centering
\caption{Propensity Score Matching: Treatment Effect Estimates}
\label{tab:psm_effects}
\begin{tabular}{lccccc}
\toprule
Intervention & N Pairs & ATT (95\% CI) & ATE (95\% CI) \\
\midrule
Received Formal Training  & 64,527 & $-0.009\ (-0.011,\ -0.008)$ & $-0.007\ (-0.008,\ -0.005)$ \\
Registered Apprenticeship &  1,483 & $0.038\ (0.032,\ 0.045)$ & $0.044\ (0.033,\ 0.054)$\\
\bottomrule
\end{tabular}
\end{table}

As shown in Table~\ref{tab:psm_effects}, the estimates of average treatment effect on the treated (ATT) and the average treatment effect (ATE) for receiving formal training are negative, but small in magnitude suggesting that after conditioning on observable characteristics, receiving formal training is associated with a \textit{slightly} lower index score relative to not receiving formal training. In contrast for participating in registered apprenticeship, the estimates of the ATT and ATE are positive, suggesting that after conditioning on observable characteristics, participating in a registered apprenticeship is associated with a meaningfully higher index score relative to not participating. An ATT of $0.038$ is a modest, but notably larger effect in magnitude than the near-zero effects observed for general training receipt.

\subsection{Diagnostics \& Limitations}

\begin{table}[htbp]
\centering
\caption{Propensity Score Matching: Diagnostics}
\label{tab:psm_diagnostics}
\begin{tabular}{lccccc}
\toprule
Intervention & Treated & Matched & Match Rate & Duplicate Controls & AUC \\
\midrule
Received Formal Training  & 64,527 & 64,527 & 100.0\% & 40,008 & 0.83 \\
Registered Apprenticeship &  1,701 &  1,483 &  87.2\% &    639 & 0.89 \\
\bottomrule
\end{tabular}
\end{table}

\begin{figure}[H]
    \centering
    \includegraphics[width=0.85\textwidth]{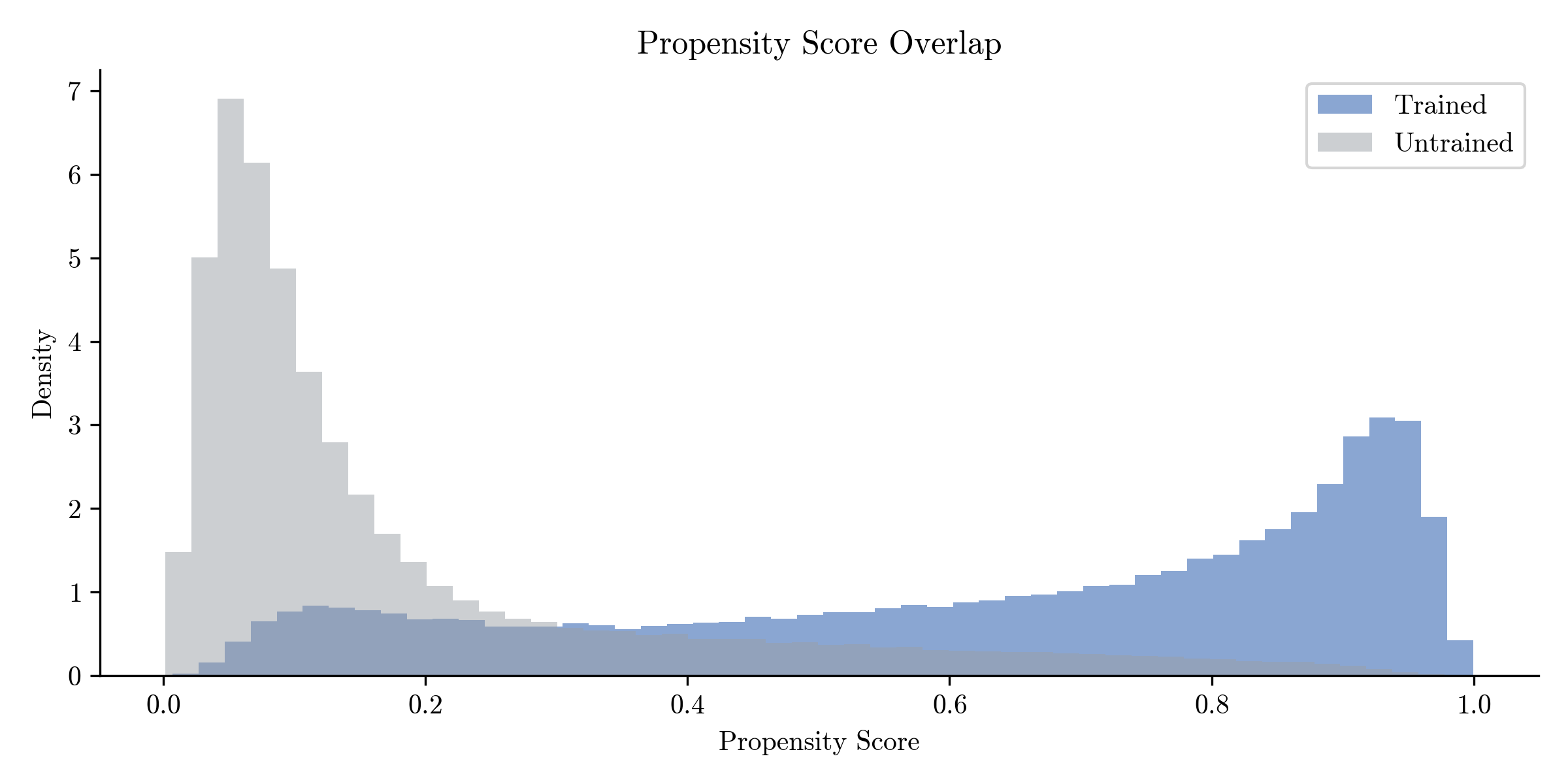}
    \caption{Propensity score overlap for the received formal training analysis.}
    \label{fig:pscore_training}
\end{figure}

\begin{figure}[H]
    \centering
    \includegraphics[width=0.85\textwidth]{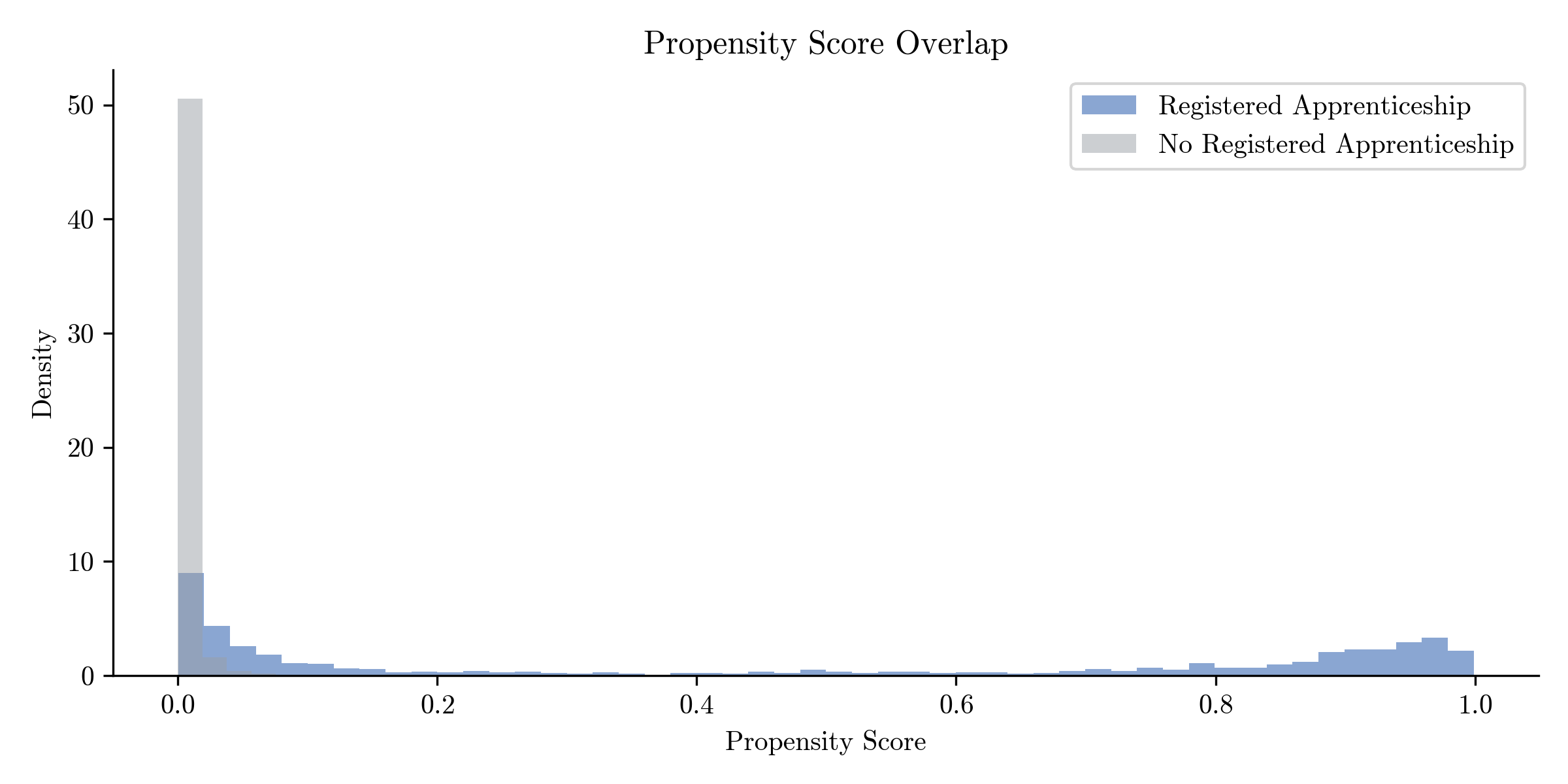}
    \caption{Propensity score overlap for the registered apprenticeship analysis.}
    \label{fig:pscore_apprenticeship}
\end{figure}

Across both propensity models for receiving formal training and participating in registered apprenticeships, the percentage of successful matches is high: $100\%$ and $87.2\%$ respectively. However, there is a high rate of duplicated control matches ($62\%$ and $43\%$ respectively) reflecting the relative scarcity of comparable participation periods that fall in the control group. The fact that in both cases the treatment group appears meaningfully different than the control group is further supported by the meaningful degree of separation between the two groups (as shown in Figures \ref{fig:pscore_training} and \ref{fig:pscore_apprenticeship}). The propensity models for receiving formal training and participating in registered apprenticeships achieve an AUC of $0.83$ and $0.89$ respectively further supporting this claim.

To further characterize selection into each intervention, we examine the logistic regression coefficients from the propensity models. In both cases, local workforce board assignment, pre-program occupation, and industry subsector emerge as important predictors of participation, reinforcing our broader finding that selection into WIOA interventions is heavily geography- and occupation-dependent. Employment status at entry and age also
appear as relevant separating features, consistent with the raw means showing that apprenticeship participants are younger (mean age $28.2$ vs.\ $36.7$) and more likely to be
employed at entry than the broader pool of WIOA participants. We caution that the logistic regression coefficients are not directly comparable across covariate types given the mix of one-hot, target, and standardized encodings used in the propensity model, and therefore
do not draw precise conclusions about the relative importance of individual features.

\begin{figure}[H]
    \centering
    \includegraphics[width=1\textwidth]{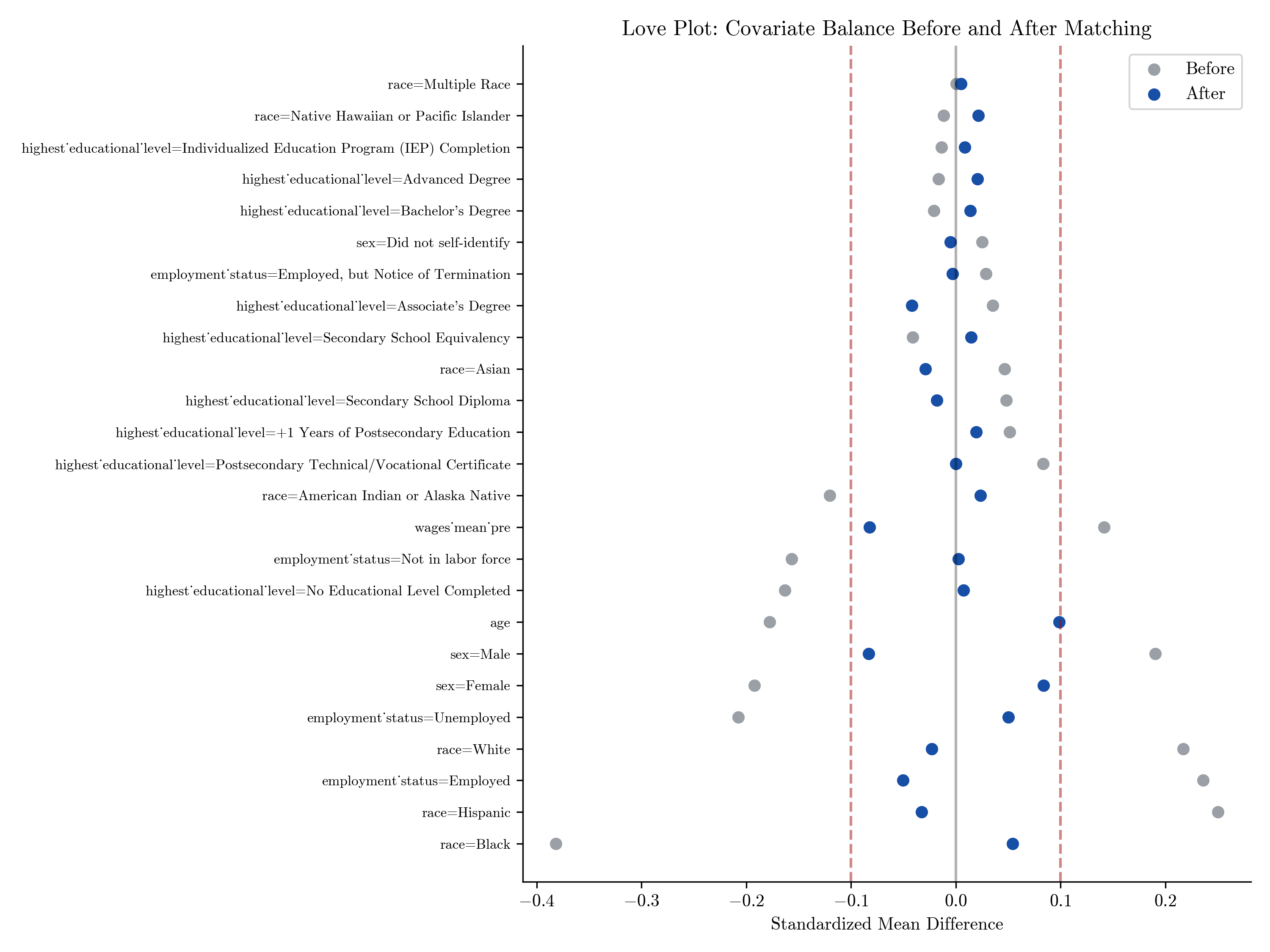}
    \caption{Love plot of covariate balance before and after matching for the received formal
    training analysis. Dashed red lines denote the $|\text{SMD}| = 0.1$ threshold.}
    \label{fig:love_training}
\end{figure}

\begin{figure}[H]
    \centering
    \includegraphics[width=1\textwidth]{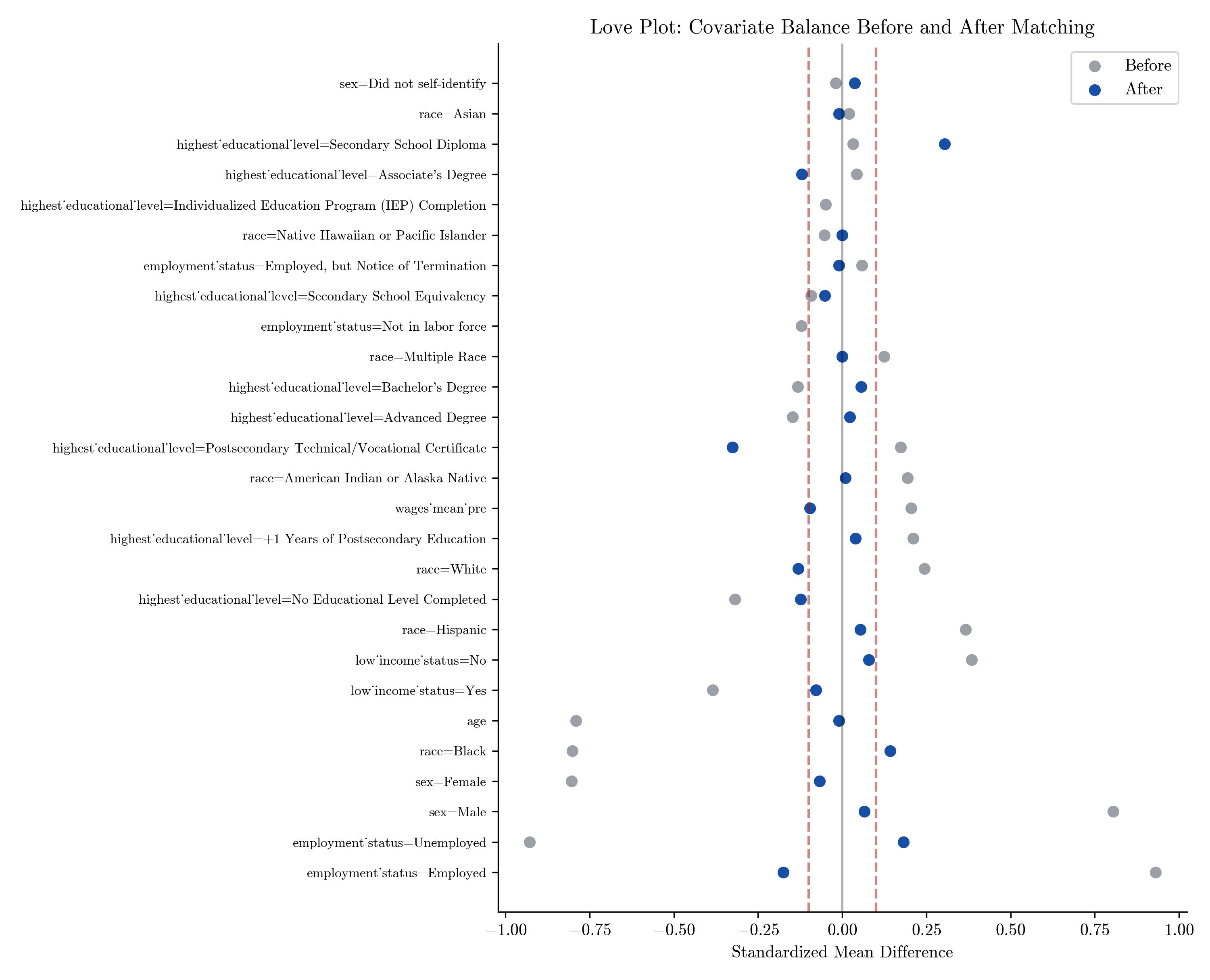}
    \caption{Love plot of covariate balance before and after matching for the registered
    apprenticeship analysis. Dashed red lines denote the $|\text{SMD}| = 0.1$ threshold.}
    \label{fig:love_apprenticeship}
\end{figure}

Despite these limitations, as illustrated in the Love plots in Figures~\ref{fig:love_training} and
\ref{fig:love_apprenticeship}, both models do an adequate job of correcting covariate imbalances after matching. For the formal training model, the majority of covariates fall within the $|\text{SMD}| < 0.1$ threshold after matching, with only age showing meaningful residual imbalance (SMD $= 0.11$). Balance is harder to achieve for the registered apprenticeship model, where several covariates remain outside the threshold after matching (most notably employment status (Employed: SMD $= -0.17$), postsecondary vocational certificate (SMD $= -0.33$), and secondary school diploma (SMD $= 0.32$)) likely reflecting the structural distinctiveness of apprenticeship participants that is difficult to fully account for with the available covariates. We therefore encourage readers to treat the matched estimates as suggestive rather than definitive, and to interpret them in conjunction with the broader observational evidence in Section \ref{section:Characteristics of Success}.

\section{Additional Subgroup Analysis Results Tables}\label{appendix:additional-index-results}

The tables below shows index outcomes aggregated by additional various subgroups. We provide index outcomes computed for both the occupation-level index and industry subsector-level index (as described in \ref{appendix:detailed-definition}). Where applicable, continuous measures are binned into interpretable ranges and bins with less than 100 participation periods with a computable index ($I_n$) are not shown.

\subsection{Received Formal Training}

\TableReceivedTrainingSubsector

\subsection{Employment Status at Entry}

\TableEmploymentStatusSubsector

\subsection{Funding Stream}

\TableFundingStreamOccupationIpw

\TableFundingStreamSubsector


\subsection{Training Service}

\TableTrainingServiceSubsector

\subsection{Demographics}

\TableDemographicsSubsector

\subsection{State}

\TableStateOccupationIpw

\TableStateSubsector

\subsection{Highest Educational Level}

\TableHighestEducationalLevelOccupationIpw

\TableHighestEducationalLevelSubsector

\subsection{Program Year}

\TableProgramYearSubsector

\subsection{Regional Economic Indicators}

\TableRegionalEconomicIndicatorsOccupationIpw

\end{document}